\documentclass[iop]{emulateapj}   

\usepackage{amssymb, amsmath, amsfonts}

\usepackage{graphicx, shadow}

\usepackage{epstopdf}

\usepackage{wasysym}
\usepackage{verbatim}
\usepackage{natbib}

\newcommand{\galfit}{\textsc{Galfit}}
\newcommand{\galapagos}{\textsc{Galapagos}}
\newcommand{\sersic}{{S\'{e}rsic}}
\newcommand{\spitzer}{\emph{Spitzer}}
\newcommand{\scamp}{{\sc Scamp}}
\newcommand{\swarp}{{\sc SWarp}}

\newcommand{\fourteen}{ISCS~J1426.5+3339 ($z=1.163$)}
\newcommand{\oneohthree}{ISCS~J1429.2+3425 ($z=1.162$)}
\newcommand{\twentyfive}{ISCS~J1434.7+3519 ($z=1.372$)}
\newcommand{\eightyfour}{ISCS~J1433.8+3325 ($z=1.369$)}
\newcommand{\threefourtytwo}{ISCS~J1434.5+3427 ($z=1.243$)}
\newcommand{\twentynine}{ISCS~J1432.6+3436 ($z=1.349$)}
\newcommand{\fiftyone}{ISCS~J1429.2+3357 ($z=1.059$)}
\newcommand{\twentytwo}{ISCS~J1438.1+3414 ($z=1.413$)}
\newcommand{\thirtysix}{ISCS~J1432.4+3250 ($z=1.487$)}

\newcommand{\fifty}{ISCS~J1427.9+3430 ($z=1.235$)}

\newcommand{\sfourteen}{ISCS~J1426.5+3339}
\newcommand{\soneohthree}{ISCS~J1429.2+3425}
\newcommand{\stwentyfive}{ISCS~J1434.7+3519}
\newcommand{\seightyfour}{ISCS~J1433.8+3325}
\newcommand{\sthreefourtytwo}{ISCS~J1434.5+3427}
\newcommand{\stwentynine}{ISCS~J1432.6+3436}
\newcommand{\sfiftyone}{ISCS~J1429.2+3357}
\newcommand{\stwentytwo}{ISCS~J1438.1+3414}
\newcommand{\sthirtysix}{ISCS~J1432.4+3250}
\newcommand{\sthirtyfour}{ISCS~J1426.1+3403}
\newcommand{\sfifty}{ISCS~J1427.9+3430}
\newcommand{\sthirty}{ISCS~J1429.3+3437}
\newcommand{\sseventeen}{ISCS~J1432.4+3332}

\newcommand{\thru}{\text{ to }}

\newcommand{\hst}{\textit{HST}}

\newcommand{\rs}{red sequence}
\newcommand{\rss}{red sequences}

\newcommand{\csp}{cSFR}
\newcommand{\ssp}{sbSP}

\shorttitle{Red Sequence in Galaxy Clusters at $1.0 < z < 1.5$}
\shortauthors{G.F. Snyder et al.}

\begin{document}

\title{Assembly of the Red Sequence in Infrared-Selected Galaxy Clusters from the IRAC Shallow Cluster Survey}

\author{Gregory F. Snyder\altaffilmark{1}, Mark Brodwin\altaffilmark{2}, Conor M. Mancone\altaffilmark{3}, Gregory R. Zeimann\altaffilmark{4}, S. A. Stanford\altaffilmark{4,5}, Anthony H. Gonzalez\altaffilmark{3}, Daniel Stern\altaffilmark{6}, Peter R. M. Eisenhardt\altaffilmark{6} , Michael J. I. Brown\altaffilmark{7}, Arjun Dey\altaffilmark{8}, Buell Jannuzi\altaffilmark{8}, Saul Perlmutter\altaffilmark{9,10}}

\altaffiltext{1}{Harvard-Smithsonian Center for Astrophysics, 60 Garden Street, Cambridge, MA 02138}
\altaffiltext{2}{Department of Physics and Astronomy, University of Missouri, Kansas City, MO 64110}
\altaffiltext{3}{Department of Astronomy, University of Florida, Gainesville, FL 32611}
\altaffiltext{4}{Department of Physics, University of California, One Shields Avenue, Davis, CA 95616}
\altaffiltext{5}{Institute of Geophysics and Planetary Physics, Lawrence Livermore National Laboratory, Livermore, CA 94550}
\altaffiltext{6}{Jet Propulsion Laboratory, California Institute of Technology, Pasadena, CA 91109}
\altaffiltext{7}{School of Physics, Monash University, Victoria 3800, Australia}
\altaffiltext{8}{NOAO, 950 North Cherry Avenue, Tucson, AZ 85719}
\altaffiltext{9}{E. O. Lawrence Berkeley National Laboratory, 1 Cyclotron Road, University of California, Berkeley, CA  94720}
\altaffiltext{10}{Department of Physics, University of California, Berkeley, CA 94720}

\begin{abstract}

We present results for the assembly and star formation histories of massive ($\sim L^*$) red sequence galaxies in 11 spectroscopically confirmed, infrared-selected galaxy clusters at $1.0 < z < 1.5$, the precursors to present-day massive clusters with $M \sim 10^{15} M_{\odot}$.  Using rest-frame optical photometry, we investigate evolution in the color and scatter of the red sequence galaxy population, comparing with models of possible star formation histories.  In contrast to studies of central cluster galaxies at lower redshift ($z < 1$), these data are clearly inconsistent with the continued evolution of stars formed and assembled primarily at a single, much-earlier time.  Specifically, we find that the colors of massive cluster galaxies at $z\approx1.5$ imply that the bulk of star formation occurred at $z\sim3$, whereas by $z\approx1$ their colors imply formation at $z\sim2$; therefore these galaxies exhibit approximately the same luminosity-weighted stellar age at $1 < z < 1.5$.  This likely reflects star formation that occurs over an extended period, the effects of significant progenitor bias, or both.  Our results generally indicate that massive cluster galaxy populations began forming a significant mass of stars at $z \gtrsim 4$, contained some red spheroids by $z \approx 1.5$, and were actively assembling much of their final mass during $1 < z < 2$ in the form of younger stars.  Qualitatively, the slopes of the cluster color-magnitude relations are consistent with no significant evolution relative to local clusters.  

\end{abstract}

\keywords{galaxies:\ clusters:\ general -- galaxies:\ elliptical and Lenticular, cD -- galaxies:\ evolution -- galaxies:\ formation -- galaxies:\ photometry}

\journalinfo{Accepted to The Astrophysical Journal}
\slugcomment{Submitted 2012 March 28; Accepted 2012 July 9}

\section{Introduction} \label{s:intro}

\begin{deluxetable*}{cccccccc}
\tablecaption{Summary of \hst\ observations of ISCS clusters.   \label{tab:clusters}}
\tablewidth{1.0\textwidth}
\tablehead{\colhead{Name} & \colhead{R.A.} & \colhead{Dec.} & \colhead{$z$} & \colhead{CMD Area} & \colhead{Instrument/Filter} & \colhead{Depth\tablenotemark{a}} & \colhead{Depth\tablenotemark{a}} \\ 
\colhead{} & \colhead{} & \colhead{} & \colhead{} & \colhead{$ \rm (arcmin^2)$} & \colhead{(optical)} & \colhead{(optical)} & \colhead{(WFC3/F160W)}  } 
\startdata
\sfiftyone\ &  14:29:15.16 &  33:57:08.5 & 1.059 & 3.1 & WFPC2/F814W & 26.1 & 24.7 \\
\sseventeen\tablenotemark{b} & 14:32:29.18 &  33:32:36.0 & 1.112  & 3.6 & ACS/F775W & 26.2 & 24.8 \\
\sthirtyfour\ &  14:26:09.51 &  34:03:41.1 & 1.136 & 3.3 & WFPC2/F814W & 26.1 & 24.7 \\
\soneohthree\tablenotemark{b,c} &14:29:15.16 &  34:25:46.4 & 1.162 & 1.7 & WFPC2/F814W & 26.6 & 24.7 \\
\sfourteen\ &  14:26:30.42 &  33:39:33.2 & 1.163 & 1.9 & WFPC2/F814W & 26.4 & 24.5 \\
\sfifty\tablenotemark{c} &  14:27:54.88 &  34:30:16.3 & 1.235 & 3.3 & WFPC2/F814W & 26.7 & 24.7 \\
\sthreefourtytwo\ &14:34:30.44 &  34:27:12.3 & 1.243  & 4.4 & ACS/F775W & 26.1 & 24.5 \\
\sthirty\ & 14:29:18.51 &  34:37:25.8  & 1.262 & 4.2 & ACS/F850LP & 26.3 & 24.4 \\
\stwentynine\ & 14:32:38.38 &  34:36:49.0  & 1.349 & 4.4 & ACS/F850LP & 26.0 & 24.8 \\
\seightyfour\ & 14:33:51.13 &  33:25:51.1 & 1.369  & 4.5 & ACS/F850LP & 26.1 & 24.8 \\
\stwentyfive\ & 14:34:46.33 &  35:19:33.5  & 1.372 & 4.5 & ACS/F850LP & 26.1 & 24.8 \\
\stwentytwo\ & 14:38:08.71 &  34:14:19.2  & 1.413 & 4.5 & ACS/F850LP & 26.1 & 24.7 \\
\sthirtysix\ &  14:32:24.16 &  32:50:03.7 & 1.487 & 8.0 & ACS/F814W & 26.0 & 25.8 
\enddata
\tablenotetext{a}{5-sigma point source depth in 0.8'' diameter; AB system }
\tablenotetext{b}{\hst\ pointing did not intersect cluster core due to guide star requirements}
\tablenotetext{c}{No single identifiable CMR}
\end{deluxetable*}

A key goal in observational cosmology is to measure the formation and assembly of cosmic structures, a history that can be traced by examining the stars in massive galaxies.  The simplest model consistent with many observations is that the stars in massive early-type galaxies (ETGs) formed in a single short burst at high redshift and evolved passively (i.e., with no further star formation) thereafter.  The dramatic expansion of data from high-quality wide-area galaxy surveys and deep lookback studies has permitted the rigorous testing of this picture, revealing a more complex formation and assembly history.  

Present-day clusters contain a tight red sequence of galaxies, indicating early and uniform star formation histories \citep{bower92}.  As fossil records of cluster assembly, red sequences were found to evolve nearly passively since $z\sim1$ \citep{aragonsalamanca93,sed98,muzzin08}, suggesting negligible ongoing star formation.  Their colors imply formation at $z\sim2\thru3$, and studies at $z \sim 1$ \citep[e.g.,][]{blakeslee03,mei06,lidman08,mei09,strazzullo10} found consistency with this picture.  The single short burst model predicts that the red sequences become bluer and wider nearer their formation epoch, a trend found by \citet{hilton09} and \citet{papovich10} in two clusters at $z \approx 1.5-1.6$.  

This passive view is brought into question by some observations of clusters at $z > 1$.  While \citet{eisenhardt08} found that the colors of cluster galaxies at $z < 1$ are consistent with the passive evolution of a group of stars formed in a short burst at $z_f = 3$, the $z > 1$ candidates favor an earlier formation.  \citet{mancone10} found that cluster galaxies' characteristic $3.6\mu m$ and $4.5\mu m$ magnitudes (from the \spitzer\ Deep, Wide-Field Survey \citep[SDWFS,][]{ashby09}) evolve in a manner consistent with little mass assembly at $z < 1.3$ (see also, \citealt{ytlin06}, \citealt{muzzin08}, \citealt{rettura11}), but are systematically fainter than the extrapolation of passive models at $z > 1.3$ (see also, \citealt{fassbender11}). 

Furthermore, we now know the red sequence is a snapshot of the currently quiescent galaxies, and that galaxies evolve both on and off during formative events.  Such transformations \citep{bo78,dressler80,tran03,mcintosh08} may cause the red sequence's average stellar age to evolve more slowly \citep{vandokkumfranx01} than predicted by the single-burst model.  Therefore, estimates of stellar age from the red sequence color and color dispersion \citep{bower92,demarco10b} can provide important constraints on galaxy assembly in clusters. 

 In this work, we report \textit{Hubble Space Telescope} (\hst) follow-up imaging of clusters drawn from the infrared-selected sample of \citet{eisenhardt08}, the \spitzer/IRAC Shallow Cluster Survey (ISCS).  This survey used near-IR imaging, an excellent proxy for stellar mass, to select structures that are the precursors to today's massive clusters \citep{brodwin07}.  We directly study their histories by examining the red sequence galaxies in 11 spectroscopically confirmed clusters at $1.0 < z < 1.5$.  

In \textsection{\ref{s:data}} we describe the candidate selection and data collection, and in \textsection{\ref{s:cmds}} we quantify the rest-frame optical color-magnitude relations.  We construct simple models for their evolution in \textsection{\ref{s:models}}, and we compare these models with our measements in \textsection{\ref{s:formation}}.  We discuss the implications of these results in \textsection{\ref{s:discussion}} and conclude in \textsection{\ref{s:conclusions}}.  Throughout this work our observed-frame magnitudes are on the AB system, but for convenience in comparing to previous work we also calculate rest-frame \textit{U-V} colors on the Vega system.  We assume a flat cosmology with $\Omega_{\Lambda}=0.72$ and $h=0.71$, consistent with \citet[][WMAP7]{Larson:2011}.


\section{Data, Reduction, and Galaxy Catalogs} \label{s:data}

\subsection{Cluster Catalog}

The \spitzer/IRAC Shallow Cluster Survey \citep[ISCS;][]{eisenhardt08} identified cluster candidates spanning $0.1 < z < 2$ over a 7.25 deg$^2$ area in the Bo\"{o}tes field of the NOAO Deep Wide-Field Survey \citep{ndwfs99}.  The clusters were identified as three-dimensional overdensities using accurate optical/IR photometric redshifts \citep{brodwin06} calculated for the 4.5 $\mu$m flux-limited ($8.8\ \mu \rm Jy$ at $5\sigma$) IRAC Shallow Survey \citep{eisenhardt04} catalog.  Thus, the clusters are selected primarily by stellar mass, largely independent of the presence of a red sequence.  There are 335 ISCS cluster and group candidates, 106 of which are at $z>1$.  Over 20 clusters at $z>1$ have now been spectroscopically confirmed with W.~M.~Keck Telescope and/or \hst\ spectroscopy (\citealt{stanford05,stanford12,elston06,brodwin06,brodwin11xray,eisenhardt08}, Zeimann et al.\ in prep., Brodwin et al. in prep.).  

\subsection{Imaging and Reduction}

In this work we present sensitive, high-resolution \hst\ follow-up imaging obtained from a variety of programs (GO proposal IDs 10496, 11002, 11597, and 11663) for a subset of the confirmed $z>1$ ISCS clusters.  In order to construct a rest--frame optical color magnitude diagram with the most leverage on the galaxies' stellar ages, instrument/filter combinations were chosen to bracket the 4000\AA\ break.  Specifically, each cluster was observed in the near-infrared with the Wide Field Camera 3 \citep[WFC3;][]{kimble08_wfc3} filter F160W.  Optical data were taken with the Advanced Camera for Surveys \citep[ACS;][]{ford98_acs} in filters F775W and F850LP or F814W, or with the Wide Field Planetary Camera 2 \citep[WFPC2;][]{holtzman95_wfpc2} filter F814W, all approximating broad \textit{I} band or \textit{I+z} band coverage.  We provide a summary in Table~\ref{tab:clusters}.

The \hst\ images were reduced using standard procedures with the MultiDrizzle software \citep{koekemoer02,fruchter09}.  The WFC3/F160W imaging was obtained in a single pointing of 4x103s (dithered between each exposure) for all but one of the clusters.  For cluster \thirtysix\, there were two slightly overlapping pointings, each consisting of 700s of integration time in a two-dithered pattern.  The native (undersampled) WFC3 0.13\arcsec\ pixel scale was drizzled down to 0.065\arcsec\ to better match the pixel scale of ACS/WFPC2 and to better recover the point-spread function.  

The regions from which we construct CMDs are limited to the overlap between the optical and WFC3/F160W fields, $\sim 2\thru8 \rm \ arcmin^2$, corresponding to fields of view $\sim 0.7\thru1.5 \rm \ Mpc$ in diameter. Therefore, all results that follow pertain only to the cluster cores, and we will refer to galaxies within $\sim 1\rm\ Mpc$ as ``central''.  The \hst\ fields for these clusters differ in total size, and so we have checked that the conclusions of this paper are unchanged when we consider only sources within a fixed $0.7$ projected physical Mpc radius from the spectroscopic cluster centers.  

The ACS/F814W observation of \thirtysix\ is comprised of 8x564s exposures.  The ACS/F850LP and F775W images for seven clusters were obtained as part of the \hst\ Cluster Supernova Survey \citep[PI Perlmutter, GO-10496][]{dawson09,suzuki12,barbary12a,barbary12b}.  The program consisted of 219 orbits targeting rich clusters at $z\gtrsim 1$, and the images comprise multiple overlapping pointings covering tens of $\rm arcmin^2$.  The single-pointing WFPC2/F814W images were obtained in a joint \spitzer/\hst\ program to survey Luminous Infrared Galaxies (LIRGs) in clusters at $z \gtrsim 1$ (\hst\ proposal ID 11002, \spitzer\ proposal ID 30950).

\scamp\ \citep{bertin06} was applied to align the world coordinate system (WCS) of each image with the reference frame of SDWFS (itself matched to 2MASS). \swarp\ \citep{bertin02} was used to resample the \hst\ optical images to match the pixel scale of the drizzled WFC3 images and to project the images to the tangent plane.  This process aligned the WFC3 and optical images so that pixels in all images for a given cluster covered precisely the same regions of the sky.

\subsection{Photometric Catalogs}

From the reduced and registered images, we measure fixed-aperture photometry using Source Extractor \citep[SE;][]{bertin96} in dual-image mode with sources detected in the WFC3/F160W images.  For comparison we compute colors using aperture diameters of $0.4'' \thru 1.5''$, and choose $0.8''$ for our default color as a reasonable balance between preventing interference in our crowded fields and intersecting a physically representative portion of each galaxy.  We also constructed colors using galaxy-specific apertures scaled to the half-light radius.  We manually inspected the CMRs made with each aperture, and found little change in the appearance and location of the red sequences.  Our results are thus insenstive to our choice of photometric aperture.  We use SE's MAG$\_$AUTO in the F160W band as an approximate total magnitude for our CMRs.  

Differences in ETG color profiles may cause a bias in CMR properties owing to our fixed-aperture color photometry.  The aperture tests described above suggest that the intrinsic widths of the color distributions dominate this bias for most galaxies.  Furthermore, in what follows we assume that the slope of the rest-frame CMR matches that of the Coma cluster, in which the slope was measured \citep{eisenhardt07} in an aperture corresponding to a nearly identical physical scale as the one used herein.  If high-redshift ellipticals form the cores of present-day ETGs \citep[e.g., as suggested by the measurements of][]{trujillo11}, then we expect minimal bias owing to this effect.

Our photometric uncertainties are dominated by sky shot noise, which we estimate in each image from the distribution of fluxes through 5000 randomly-placed apertures with a diameter matched to our catalogs.  The flux distribution is to a good approximation normal except for the sources contaminating the positive half.  We compute the background level $\sigma_{\rm sky}$ by fitting a normal distribution to the negative half.  We verified this procedure by confirming that it produces the correct scaling in photometric scatter of sources detected in the sets of dither images before making the final stack.

We calculated the standard completeness of the WFC3/F160W-selected catalogs by generating artificial objects with known total magnitude and $R^{1/4}$ light profiles where $R_e$ is distributed as a Rayleigh function with $\sigma_{R_e} = 3.5$ pixels ($\sim 0.23$ arcseconds, or $\sim 2$ Kpc at $z=1.5$), matching well the observed SE-derived $R_e$ distribution for sources in these fields.  We found that the WFC3/F160W-detected catalogs for each cluster are more than $90\%$ complete at an AB magnitude of $23.5$, and every field is roughly $97\%$ complete at $H^*(z) + 1.5$, which is approximately $H =$ 22--23 for these redshifts.  $H^*(z) + 1.5$ is the magnitude cut we impose for the CMR analysis presented in this work (Section~\ref{ss:selection}).

      \begin{figure*}[h!]
	\begin{center}
	\begin{tabular}{c}
	\includegraphics[width=5.5in]{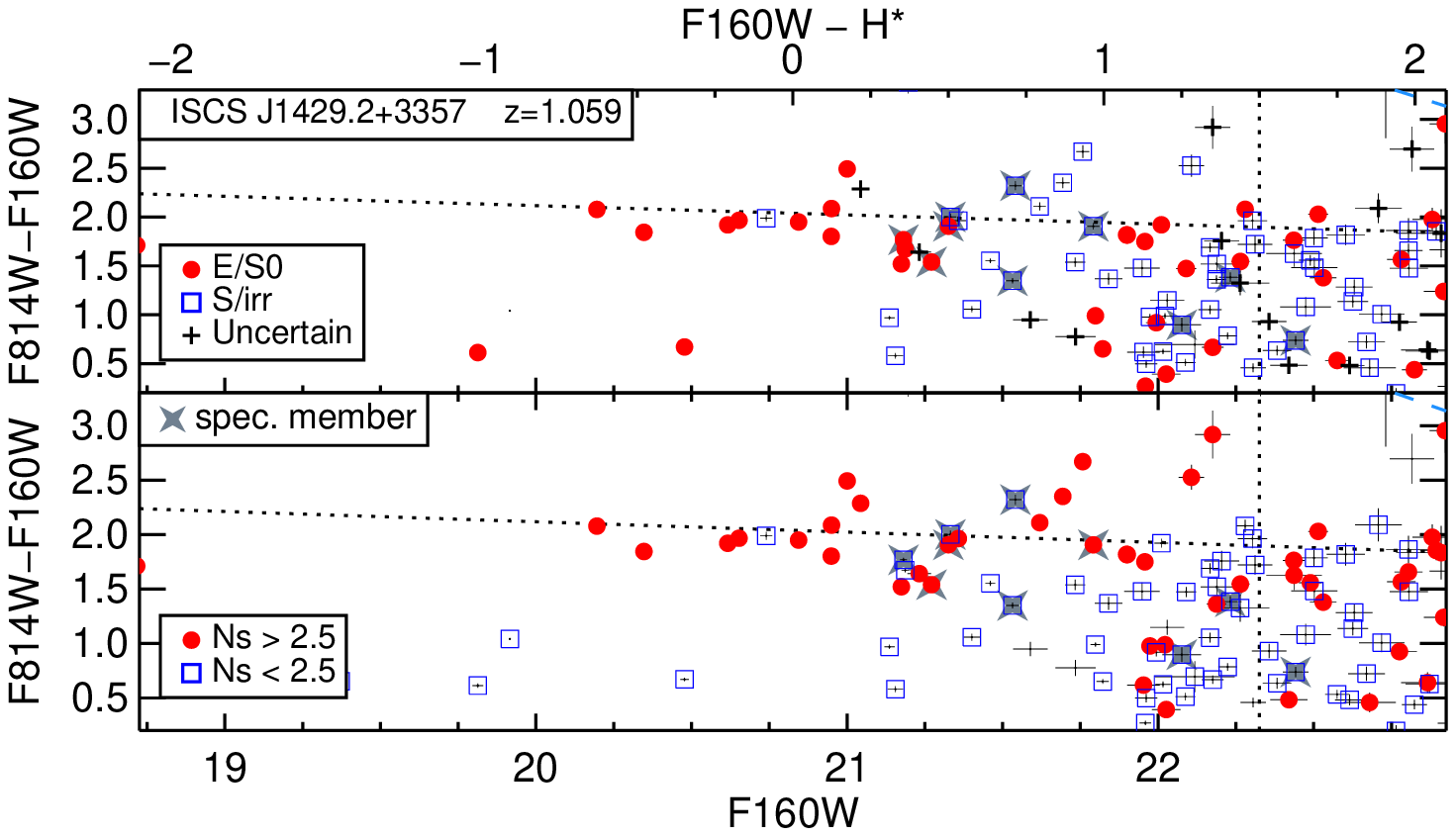} \\
	\includegraphics[width=5.5in]{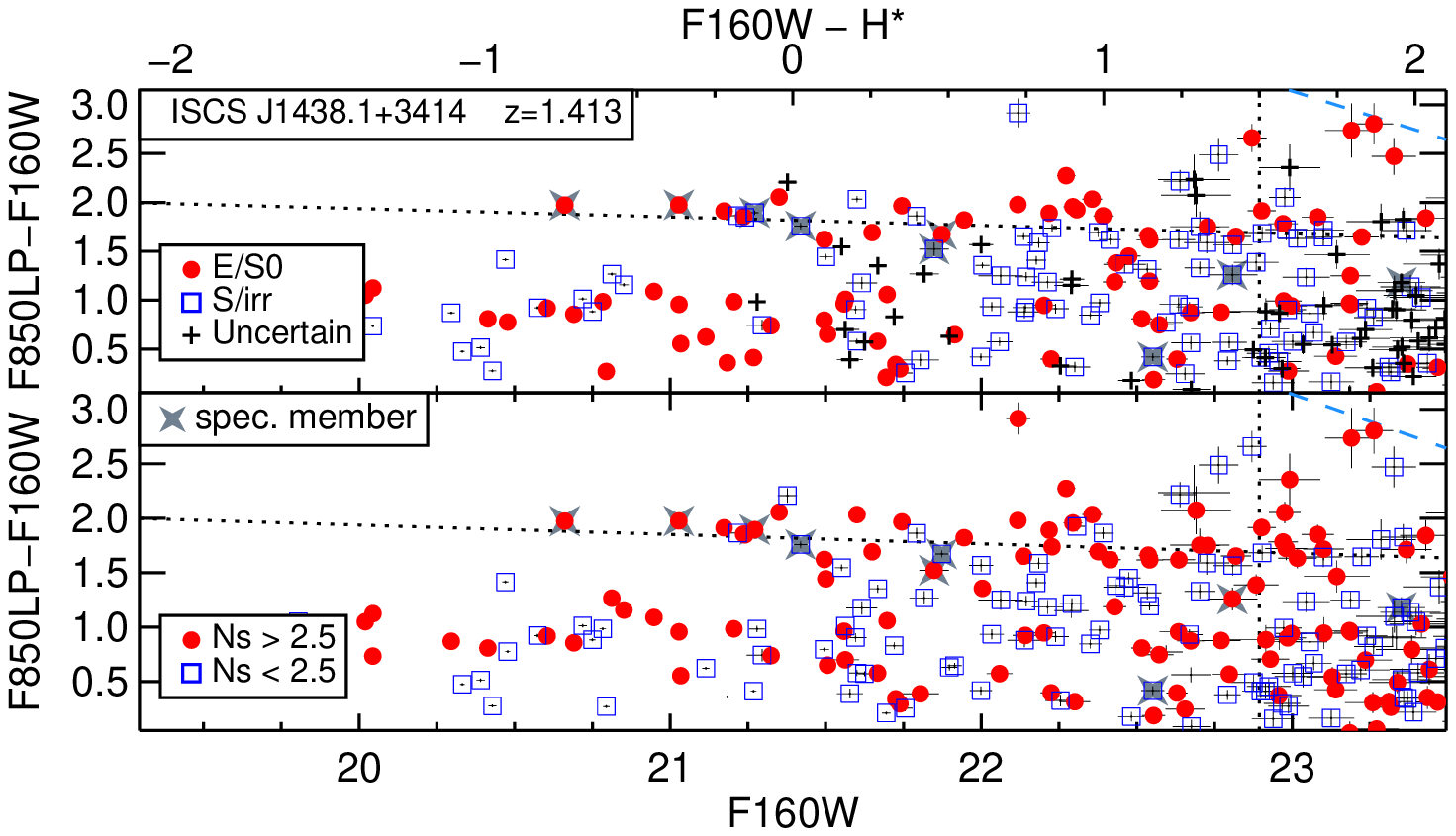}
	\end{tabular}
	\end{center}
      \caption{Comparison between morphological selection techniques for clusters \fiftyone\ and \twentytwo.  We identify objects in the top and bottom panel based on visually-classified and automated morphologies, respectively (\S\ref{ss:morphs}).  The black dotted lines show the CMR calculated from the Coma cluster \citep{eisenhardt07} and evolved (assuming a single short burst at $z_{f}=3$) to the observed redshift.  We plot a vertical dotted line at $H=H^* + 1.5$, representing the magnitude selection limit we apply in Section~\ref{ss:selection}.  Gray crosses represent all spectroscopically confirmed cluster members currently known that are covered by the \hst\ images and that fall within the color-magnitude ranges of the CMD.  An updated description of the spectroscopy will be presented by Zeimann et al. (in prep.) and Brodwin et al. (in prep.).  Where visible in the upper right corners, the diagonal blue dashed lines approximate the boundary of the 5$\sigma$ color and F160W magnitude measurements.  For \twentytwo, we note the presence of many early-type galaxies $\sim 1$ magnitude blueward of the identified cluster \rs.  Several of these objects are part of a foreground structure evident in the photometric redshifts of the ISCS catalog, and are offset from the spectroscopic cluster center by $\sim 1$ arcminute.  Therefore we believe those relatively blue early-type galaxies are interlopers.  We find that both morphology classification schemes yield similar CMRs for all of our clusters, so we present the CMDs in Figure~\ref{fig:raw2}, \ref{fig:rawreject}, and \ref{fig:measure} using only the visual typing, and analyze them in Figure~\ref{fig:measure}.  \label{fig:raw1} \label{fig:morphmethod}}
      \end{figure*} 


\section{Color-Magnitude Relations} \label{s:cmds}

We aim to study the star formation epoch of ETGs on the red sequences for our $z>1$ cluster sample using the color and color scatter of the red sequence CMRs, which are identified using morphological classification and color selection techniques.  The color and scatter of a CMR serve as proxies of the formation history of the galaxies that comprise it.  The simplest physical parameters that they probe is the average formation time or age of the galaxies' stars.  If the galaxies formed separately in short bursts over a period of time, then the color scatter will be inversely proportional to the time since they stopped forming, whereas if they formed in a synchronized fashion, then the scatter between galaxies may be always small.  Since this is true regardless of their average age (and hence average color), then we are motivated to measure both color and scatter in order to better constrain the history of these galaxies.

      \begin{figure*}
	\begin{center}
	\begin{tabular}{c}
	\includegraphics[width=6in,clip=true]{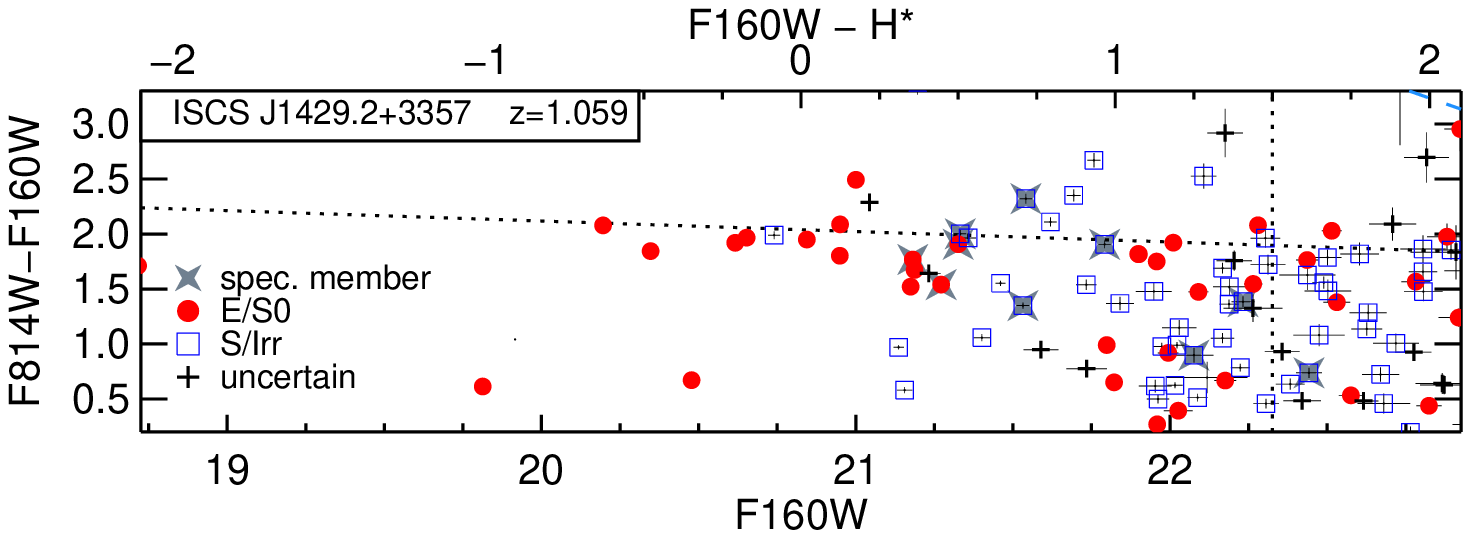} \\
	\includegraphics[width=6in,clip=true]{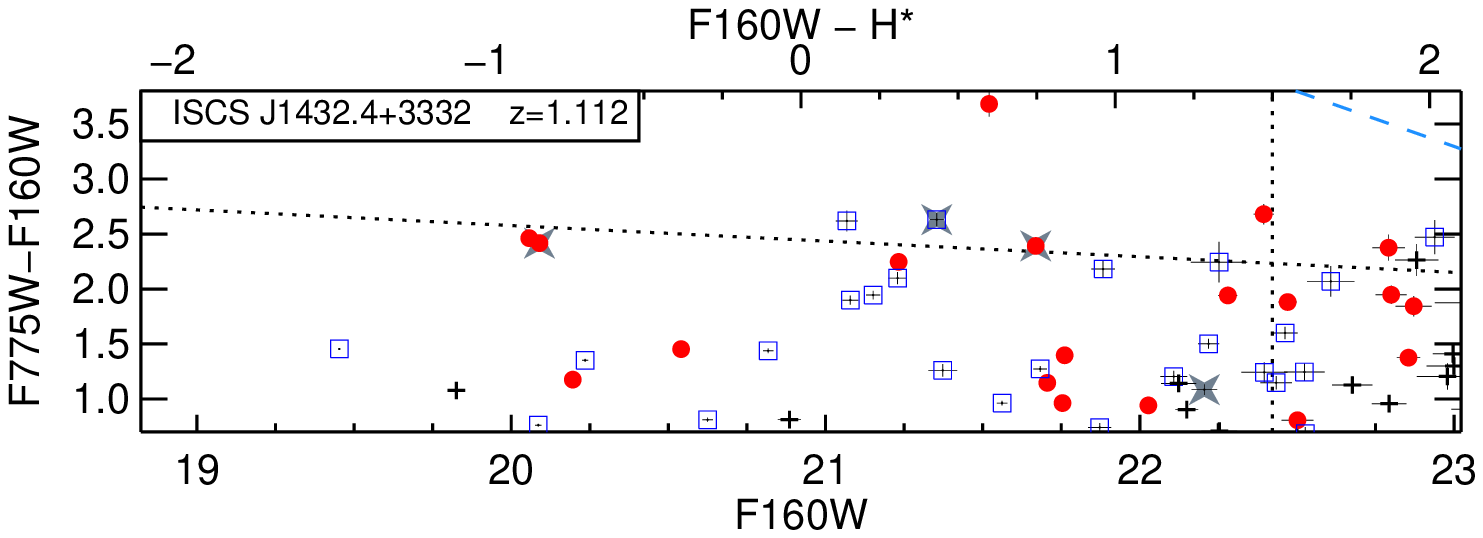} \\
	\includegraphics[width=6in,clip=true]{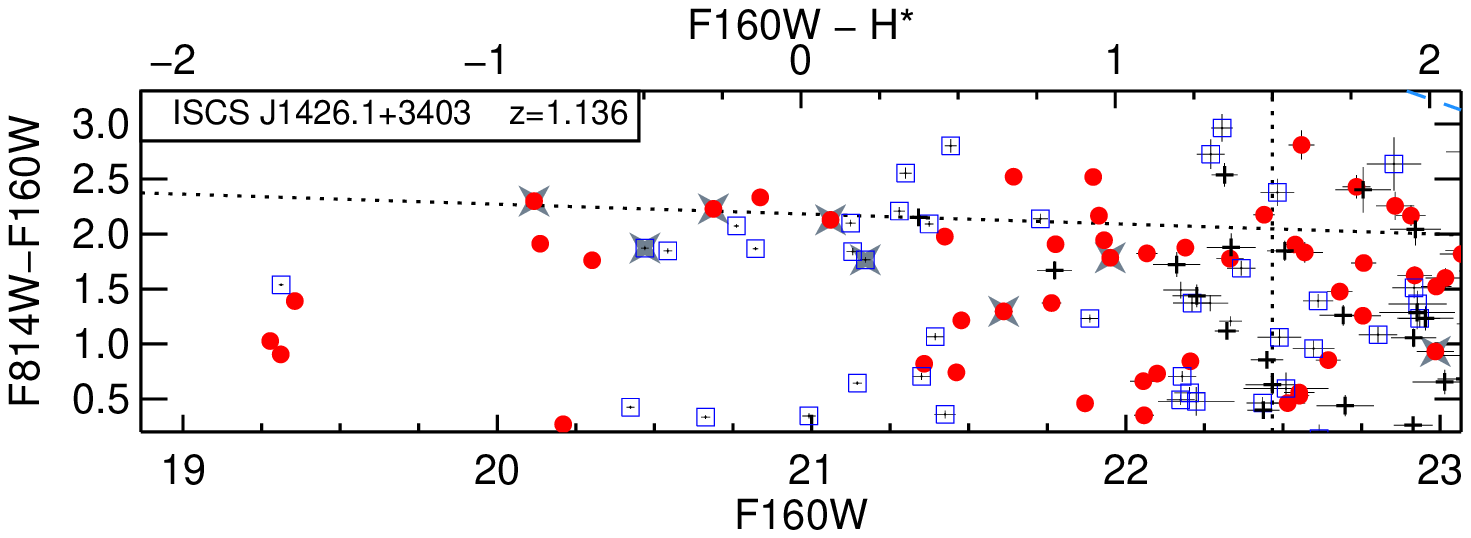}
	\end{tabular}
	\end{center}
      \caption{  CMDs for our cluster sample, in ascending redshift order.  These graphs follow the same formatting as the top panels of the two CMDs in Figure~\ref{fig:raw1}, where plot symbols represent the visual classification system of Section~\ref{ss:morphs}.  \label{fig:raw2}  \label{fig:raw}}
      \end{figure*} 
\setcounter{figure}{1}

      \begin{figure*}
	\begin{center}
	\begin{tabular}{c}	
	\includegraphics[width=6in,clip=true]{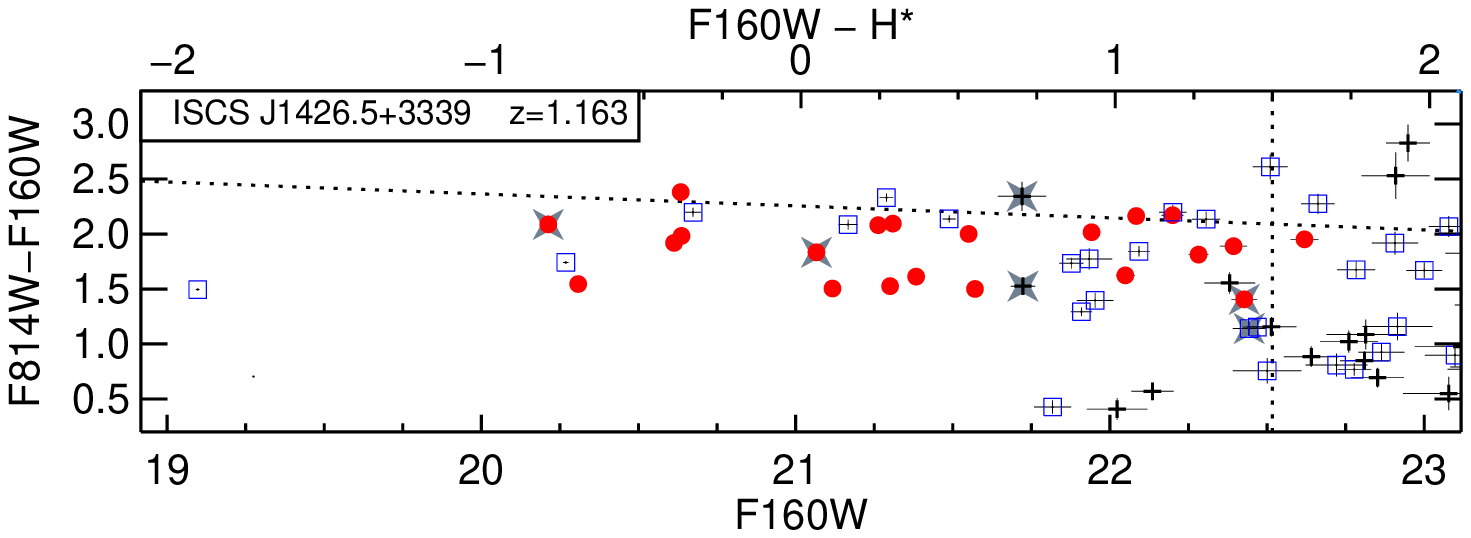} \\
	\includegraphics[width=6in,clip=true]{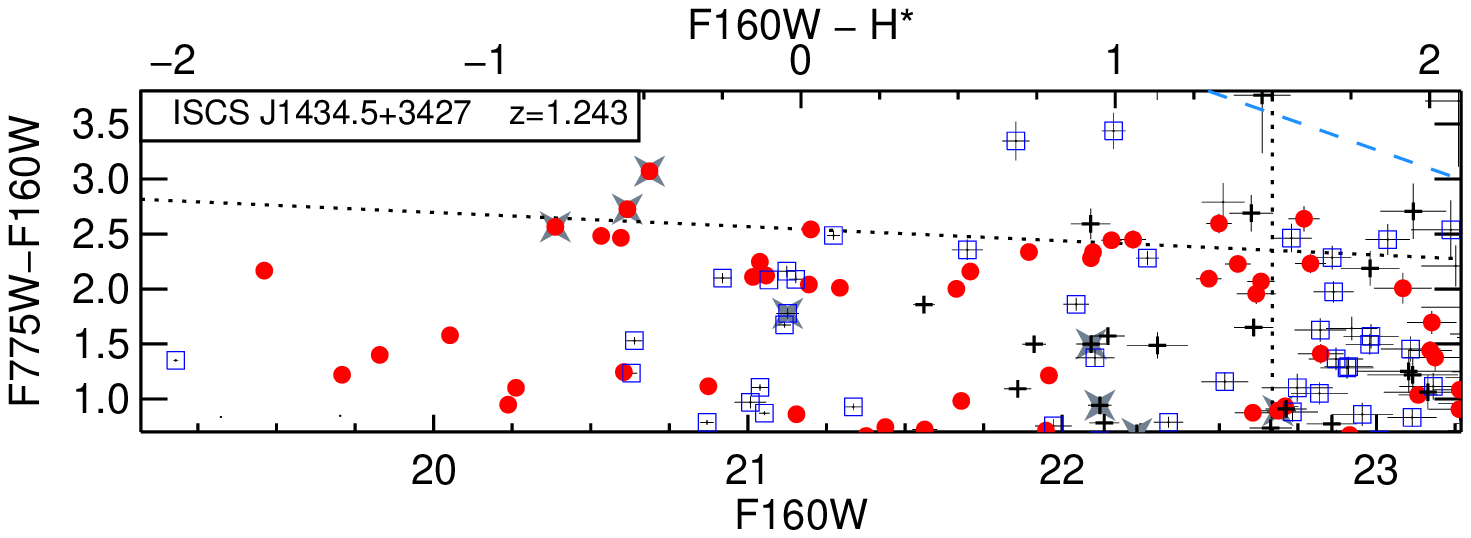} \\
	\includegraphics[width=6in,clip=true]{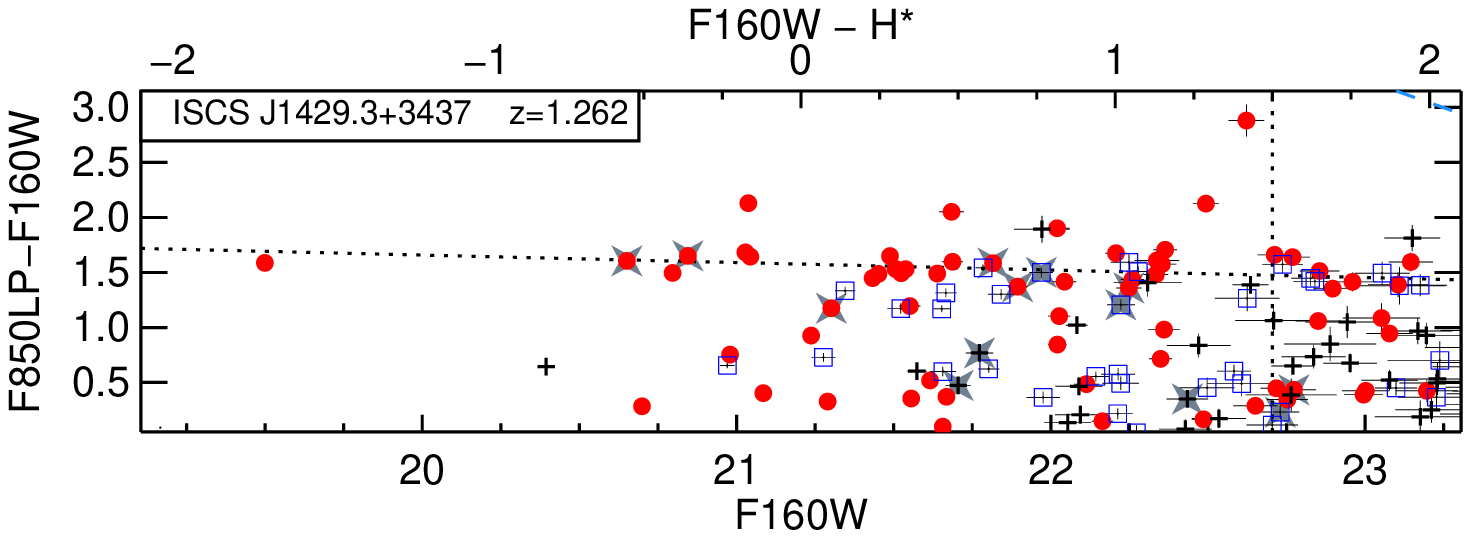}
	\end{tabular}
	\end{center}
      \caption{  Continued from previous page.}
      \end{figure*} 
\setcounter{figure}{1}

      \begin{figure*}
	\begin{center}
	\begin{tabular}{c}
	\includegraphics[width=6in,clip=true]{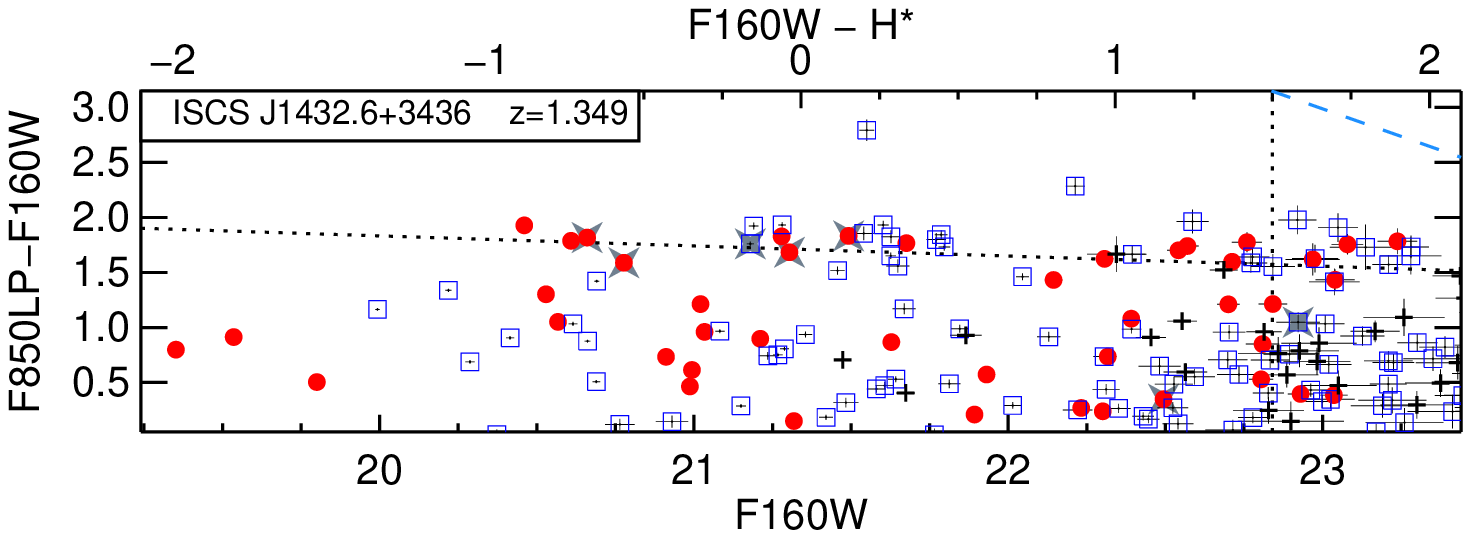} \\
	\includegraphics[width=6in,clip=true]{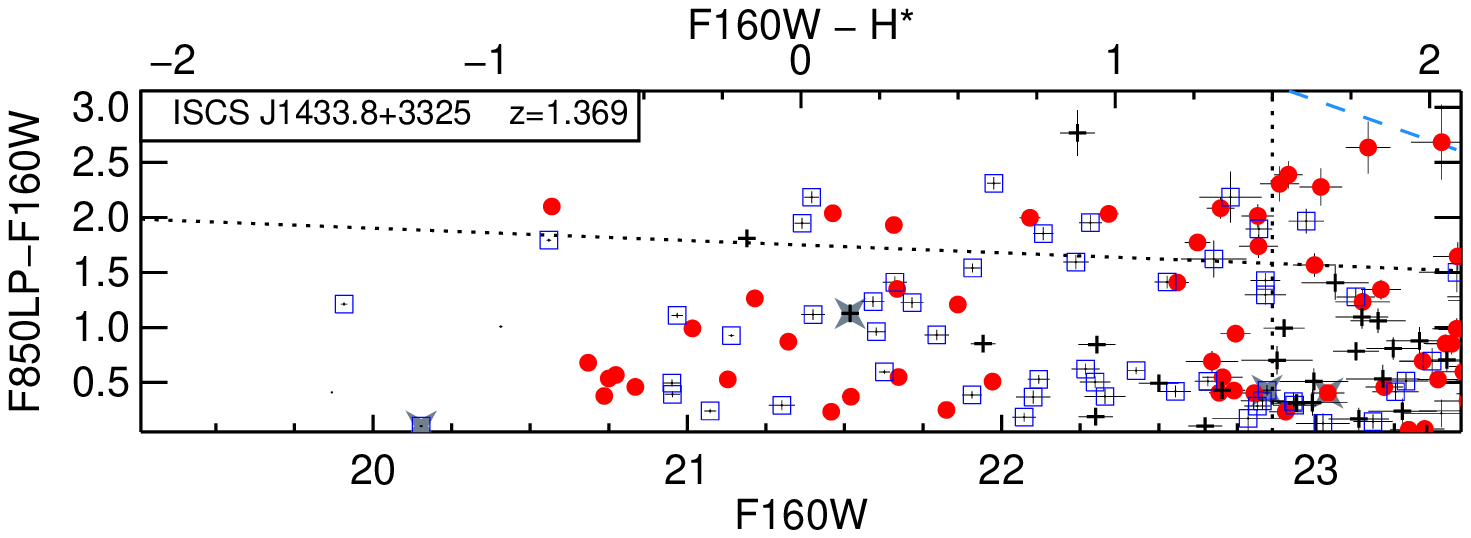} \\
	\includegraphics[width=6in,clip=true]{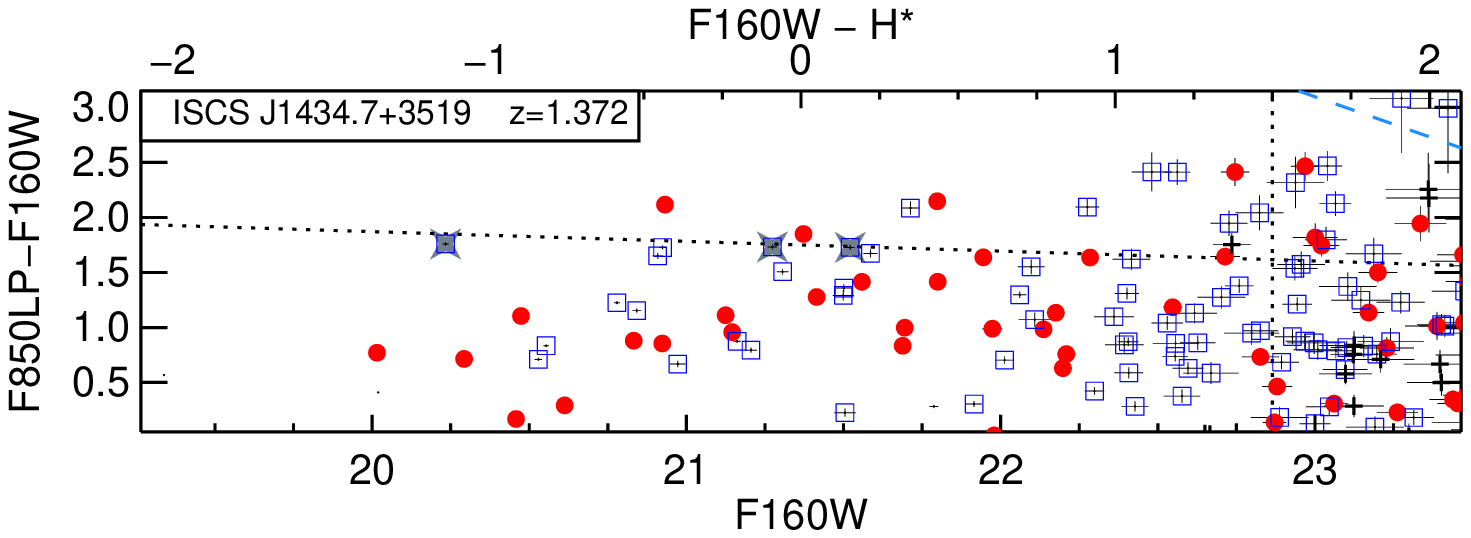}
	\end{tabular}
	\end{center}
      \caption{  Continued from previous page.}
      \end{figure*} 
\setcounter{figure}{1}

      \begin{figure*}
	\begin{center}
	\begin{tabular}{c}
	\includegraphics[width=6in,clip=true]{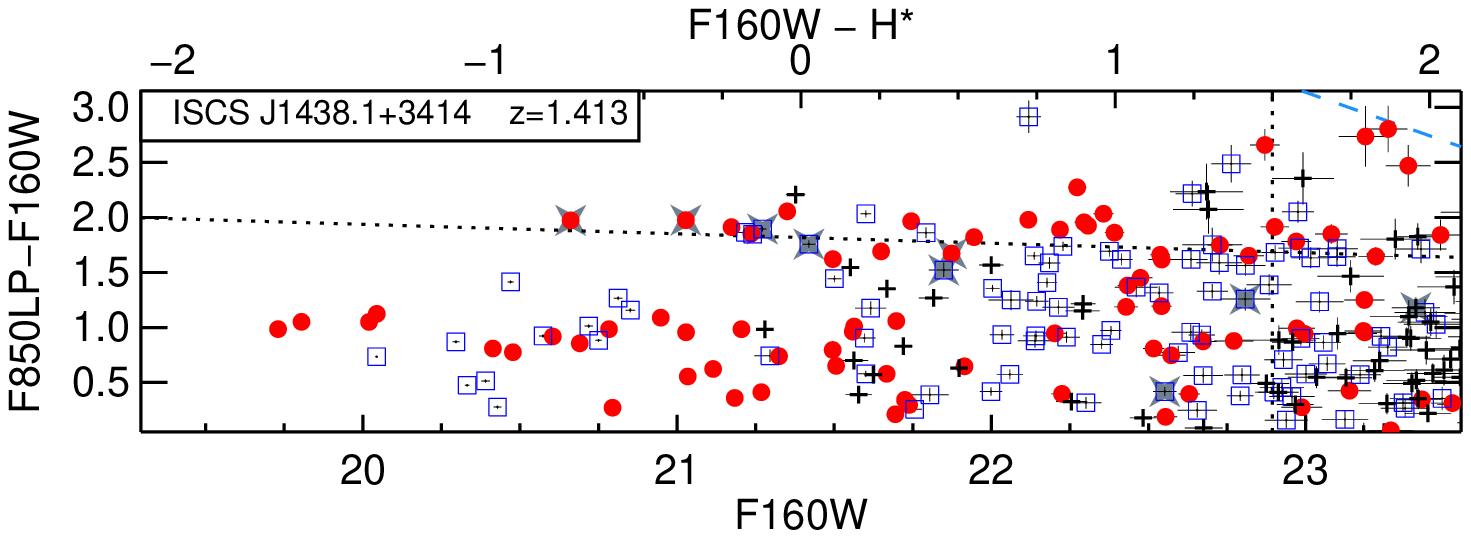} \\
	\includegraphics[width=6in,clip=true]{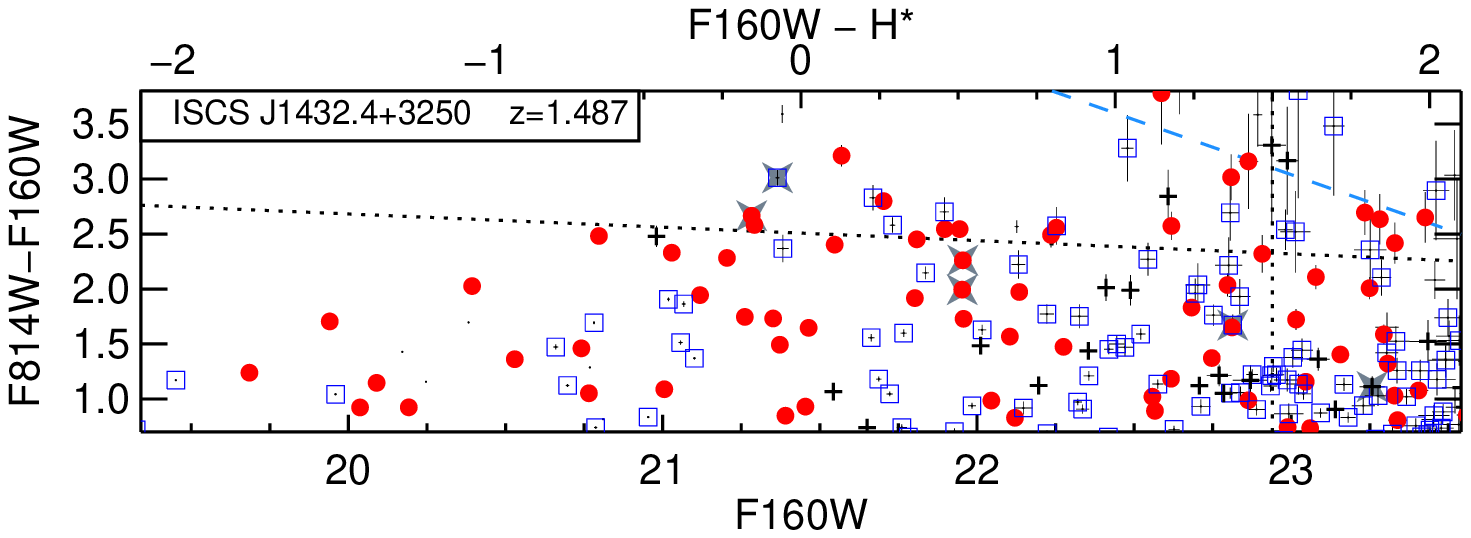}
	\end{tabular}
	\end{center}
      \caption{  Continued from previous page.}
      \end{figure*} 

\subsection{Morphological Classification} \label{ss:morphs}

Visual inspection of the individual galaxies in the F160W-selected catalogs was carried out by one of us (S.~A.~S.) using the WFC3 images.  T-types \citep[RC3;][]{devac_rc3_1991} were assigned where possible to all objects down to $F160W = 23.5$.  If the object was too faint to accurately classify, too close to a bright star, or affected by bad data, no morphology was assigned.  In all other cases an attempt was made to determine the T-type on the scale ranging from $-5$ for a giant elliptical, to 10 for an irregular galaxy.  In addition, information was recorded on the environment of the object, if it was compact or appeared inclined to the line of sight, and if it appeared to be interacting with any neighboring objects.

Quantitative measurements of galaxy shapes were made by fitting a single \sersic\ profile to every object in the F160W-selected catalogs.  The fitting was done with \galfit\ \citep{peng10}, and \galapagos\ \citep{haussler11} was employed to automate the task of running \galfit.  \galfit\ does two dimensional fitting and can fit an arbitrary number of models to any image while accounting for the image's point spread function.  \galapagos\ automates the task of running \galfit\ by generating first--guess \sersic\ parameters from an SE run, measuring the sky for each galaxy, deciding when to simultaneously fit nearby galaxies, and deciding when to subtract off brighter objects that might bias a fit.  It then uses \galfit\ to fit a single \sersic\ profile to every object in the image and builds a catalog of the results.

In Figure~\ref{fig:morphmethod} we present a comparison of the visual and quantitative morphologies in CMDs of two clusters that span the redshift range of our sample. The early--type CMRs resulting from these methods are qualitatively quite similar in these two clusters, as well as in the rest of our sample, and lead to quantitatively similar results.  For clarity we therefore present the remainder of our CMDs using only the visual T-types, but for completeness retain a comparison with the quantitive morphologies in Figures~\ref{fig:morphmethod} and \ref{fig:morphcompare} and in several tables.

\subsection{Red Sequence Selection} \label{ss:selection}

We use the colors, magnitudes, morphological classifications, and available spectroscopy to identify early--type members of the red sequence in each cluster.  Multi--band photometric redshifts, which are not considered in this work, are available for only a small fraction of the galaxies selected in the WFC3/F160W images, because these \hst-based catalogs are dominated by fainter galaxies in fields too crowded for reliable ground-based photometry.

\begin{figure*}
	\begin{center}
	\begin{tabular}{c}
	\includegraphics[width=6in,clip=true]{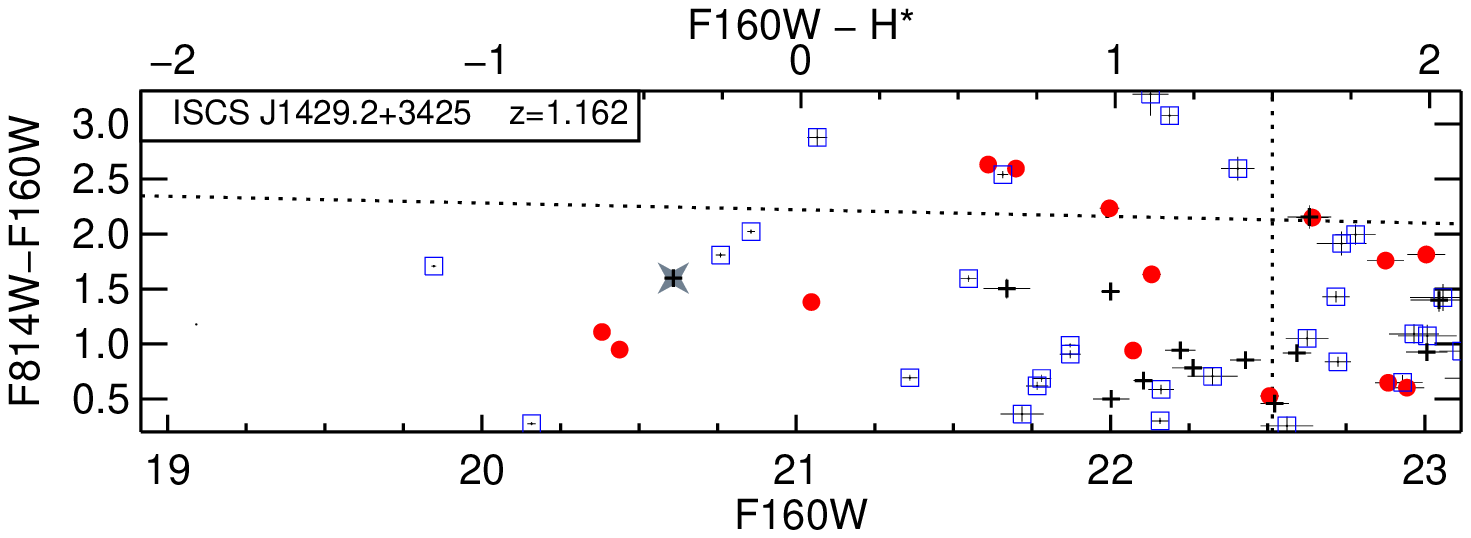} \\
	\includegraphics[width=6in,clip=true]{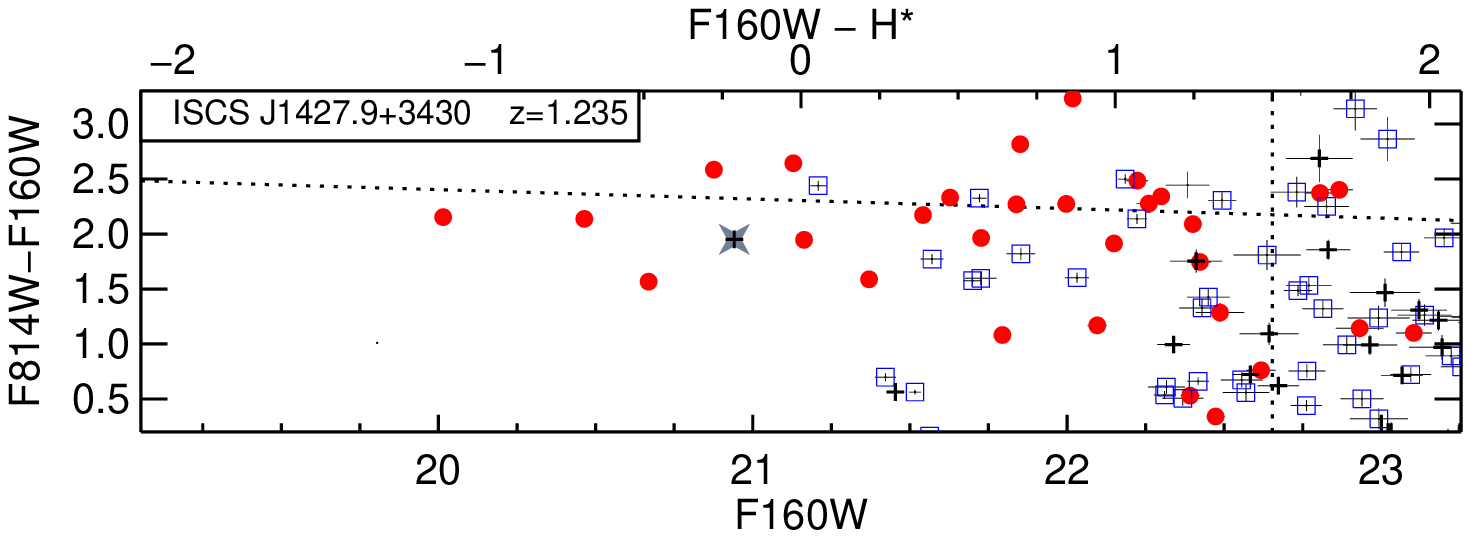} 
	\end{tabular}
	\end{center}
        \caption{CMDs of two clusters for which there are known structures at different but similar ($z \sim 1$) redshifts.  The symbols are the same as in Figure~\ref{fig:raw}.  In order to avoid potential confusion owing to multiple red sequences, these two clusters are not used in the subsequent analyses.  \label{fig:rawreject} }
      \end{figure*}

The basic observables, the CMDs for 13 clusters, are presented in Figures~\ref{fig:raw} and \ref{fig:rawreject}.  In each panel we plot the Coma CMR \citep{eisenhardt07} evolved to the appropriate redshift (see Section~\ref{ss:simplemodels} for details).

In \oneohthree\ and \fifty, shown in Figure~\ref{fig:rawreject}, there exist multiple known structures at different redshifts in the \hst\ field of view.   Moreover, the intersection between the WFC3/F160W image of \oneohthree\ and its WFPC2/F814W image is small and does not include an identifiable CMR.  To avoid a possible bias from including projected large scale structure, as opposed to massive, collapsed clusters, we omit these two clusters from our analysis.  It is possible that one or two other members of the 11 remaining clusters are similar outliers, but lacking such evidence at present, we cannot confidently omit them from our analysis.  

For the clusters with a clear \rs, the rest-frame CMR slope is within $10\%$ of that in Coma, while the color zeropoint and scatter clearly vary.  The shallow slope of the CMR has been found to be nearly constant with redshift \citep{kodama97}, and thus has been interpreted as evidence for a uniformly old sequence in mass and metallicity \citep[e.g.,][]{larson74}, and not one based on galaxy age.  If the tilted CMR is entirely the result of a mass-metallicity relation with a similar slope from $z= 2\thru0$ \citep[as implied by the results of e.g.,][]{erb06,brooks07}, then we do not expect cosmic evolution in the CMR slope so long as we can satisfactorily select galaxies that satisfy this relation.  In what follows we fix the CMR slope in each cluster's rest frame, treating it as independent of age, and consider only measurements of the color zeropoint and scatter.  We will discuss possible deviations from this assumption as needed.

We isolate CMRs in the 11 clusters using the following procedure.  First, we subtract from every galaxy's color our fiducial Coma CMR model (a single stellar population formed in an instantaneous burst at $z_f=3$).  We plot the resulting quantity $\Delta$, in magnitudes, in Figure~\ref{fig:measure1}.  For all clusters except \fourteen, we define a red sequence galaxy (RSG) as having and $ \Delta_{z_f} - 0.25 < \Delta < 0.75$, where $\Delta_{z_f}$ is the color difference between our $z_f = 3$ assumption and a single stellar population formed at redshift $z_f$.  We use $z_f=2$ to set this lower limit on $\Delta$ instead of $z_f=3$ in order to successfully select \rss\ that may have formed more recently.  This $\Delta$ range is chosen as a reasonable balance between including member RSGs and excluding blue interlopers or late-type members that are not on the cluster's \rs.  This range of $\Delta$, within which we select the RSGs, is listed for each cluster in Table~\ref{tab:numbers}; in all cases this is $\gtrsim 1.2$ magnitudes.  Except for \fourteen, which has a very blue \rs, this range cleanly selects the RSGs, and in particular the one composed of ETGs, that are visually identifiable in Figures~\ref{fig:raw1}-\ref{fig:raw}.  For \fourteen, we have manually modified the color selection region to identify the RS: $\Delta_2 - 0.5 < \Delta < 0.75$.  It is unclear why \fourteen\ appears so much younger than other clusters at similar redshifts, though we have verified that our conclusions are unchanged if we omit it from our analysis.  

Next, we define the cluster RSGs as galaxies brighter than $H^*(z) + 1.5$ and selected by the above color criteria, where $H^*(z)$ is calculated by evolving the characteristic brightness for Coma \citep{depropris98} to the cluster's observed redshift, assuming the galaxies formed in a single short burst at $z_f=3$.  Some of the RSGs selected in this way are spectroscopically confirmed cluster members, and we plot these as gray stars in all CMDs.  However, most do not have accompanying spectroscopic or photometric redshift information, and so the RSG samples may suffer from interloper contamination.  In addition to the sample of all RSGs, we analyze two samples drawn from them to identify likely ETGs: visually-selected E's or S0's as defined in Section~\ref{ss:morphs}, and quantitatively-selected ETGs with $n_S > 2.5$.  The latter choice is reasonably consistent with selecting bulge-dominated objects \citep{simard11}, as well as visually-typed E and S0 galaxies \citep{simard09}.  

Finally, to limit the influence of interlopers that are extreme outliers from the cluster's \rs, we drop galaxies from the final ETG or RSG catalog that are further than two median absolute deviations in color from their $\Delta$ zeropoint, as measured below.  Table~\ref{tab:numbers} lists the number of galaxies satisfying our \rs\ selections, as well as the median color uncertainty (for the sample of all RSGs) in the observed bands.

\subsection{Color Zeropoints and Scatters}

For each of the three ETG or RSG samples defined above, we measure the color zeropoint and intrinsic scatter, and their associated uncertainties, applying the biweight estimates of location and scale \citep{mostellertukey77, beers90} for the zeropoint location $\Delta_0$ and scatter about this location $\sigma_{\Delta}$.  These two measurements define our CMR for each subsample.  We calculate the color $C_0$ at $H^*(z)$ by adding back to $\Delta_0$ the previously-subtracted fiducial $z_f=3$ CMR model, and the intrinsic scatter $\sigma_{\rm int}$ by subtracting in quadrature the median photometric error from the biweight scale estimate $\sigma_{\Delta}$ of the sample in question.

Uncertainties are derived from 1000 bootstrap resamplings, from which we measure the scale of the resulting $C_0$ and $\sigma_{\rm int}$ distributions using the median absolute deviation (MAD).  Specifically, $\sigma_C = S_{\rm MAD} = \rm MAD/0.6745$, since a normally-distributed quantity will have $\rm MAD=0.6745 \sigma$.  Where the samples overlap, our measurements of $\sigma_{U-V}$ are consistent with those measured by \citet{meyers12}, except for \twentytwo.  At $z \gtrsim 1.3$, the addition of the \hst\ WFC3/F160W data permits an improved k-correction, and so we believe the present calculation to be reliable.  

For each cluster we present $C_0$, $\sigma_C$, and their uncertainties in Table~\ref{tab:measurements}.  The observed-band $\sigma_{\rm int}$ values are all significantly smaller than $\approx 0.6$ magnitudes, which is the value we would expect to measure if there were no overdensity (e.g., no red sequence) inside the $\Delta$ selection region.  This implies that our measurements are driven by the real red sequences and are not an artifact of our color selection choices (Table~\ref{tab:numbers}).

\begin{deluxetable*}{ccc@{\hspace{1cm}}ccc@{\hspace{1cm}}cc@{\hspace{1cm}}c}
\tablecaption{Number of cluster red sequence galaxies. \label{tab:numbers}}
\tablecolumns{9}
\tablewidth{1.0\textwidth}
\tablehead{
\colhead{}            & \colhead{}        & \colhead{}  \hspace{1cm}  & \multicolumn{3}{c}{Number on CMR}                               \hspace{1cm} & \multicolumn{2}{c}{Spectroscopic Members\tablenotemark{b}}           \hspace{1cm}& \colhead{}    \\
\colhead{}            & \colhead{}        & \colhead{}  \hspace{1cm}  & \colhead{Vis.}    & \colhead{Quant.} & \colhead{All}      \hspace{1cm} & \colhead{WFC3}              & \colhead{All}                     \hspace{1cm}& \colhead{median $\sigma_{\rm phot}$} \\
\colhead{Name} & \colhead{$z$} & \colhead{$\Delta$ width\tablenotemark{a}}  \hspace{1cm}  & \colhead{ETGs} & \colhead{ETGs}  & \colhead{RSGs} \hspace{1cm} & \colhead{Catalog}           & \colhead{RSGs}               \hspace{1cm}& \colhead{All RSGs} 
}
\startdata
   \sfiftyone\          & 1.059 &1.19& 13 & 16 & 26  & 10  & 5 & 0.04 \\ 
   \sseventeen\    & 1.112 &1.25&  5 &  7 & 11    &  6    & 3 & 0.05 \\ 
   \sthirtyfour\        & 1.136 &1.24& 15 & 20 & 30  & 8    & 6 & 0.03 \\ 
   \sfourteen\         & 1.163 &1.76& 18 & 12 & 39  & 6    & 6 & 0.04 \\ 
  \sthreefourtytwo\ & 1.243 &1.25& 19 & 19 & 27 & 10 & 3 & 0.05 \\ 
   \sthirty\              & 1.262 &1.22& 23 & 21 & 33  & 13 & 8 & 0.04 \\ 
   \stwentynine\ & 1.349 &1.24& 12 & 16 & 27  & 7   & 5 & 0.05 \\ 
   \seightyfour\ & 1.369 &1.24&  9 &  9 & 24     & 6   & 0 & 0.07 \\ 
   \stwentyfive\ & 1.372 &1.24& 10 & 14 & 27  & 3   & 3 & 0.04 \\ 
   \stwentytwo\ & 1.413 &1.19& 20 & 25 & 43  & 10 & 7 & 0.04 \\ 
   \sthirtysix\    & 1.487 &1.24& 15 & 13 & 33  & 6   & 4 & 0.10\tablenotemark{c}
\enddata
\tablenotetext{a}{Color width in magnitudes used to define the RSG selection region; see Section~\ref{ss:selection}.}
\tablenotetext{b}{Spectroscopy for parts of this sample were described by \citet{stanford05,elston06,brodwin06,brodwin11xray,eisenhardt08}.  Updated membership identification, including those confirmed with the \hst\ WFC3 grism, will be described by Zeimann et al. (in prep.) and Brodwin et al. (in prep.).  }
\tablenotetext{c}{Although this cluster has the deepest WFC3/F160W image in the sample, its \rs\ galaxies are fainter (and redder), leading to a high average color uncertainty. }
\end{deluxetable*}

      \begin{figure*}
	\begin{center}
	\begin{tabular}{c}
	\includegraphics[width=6in,clip=true]{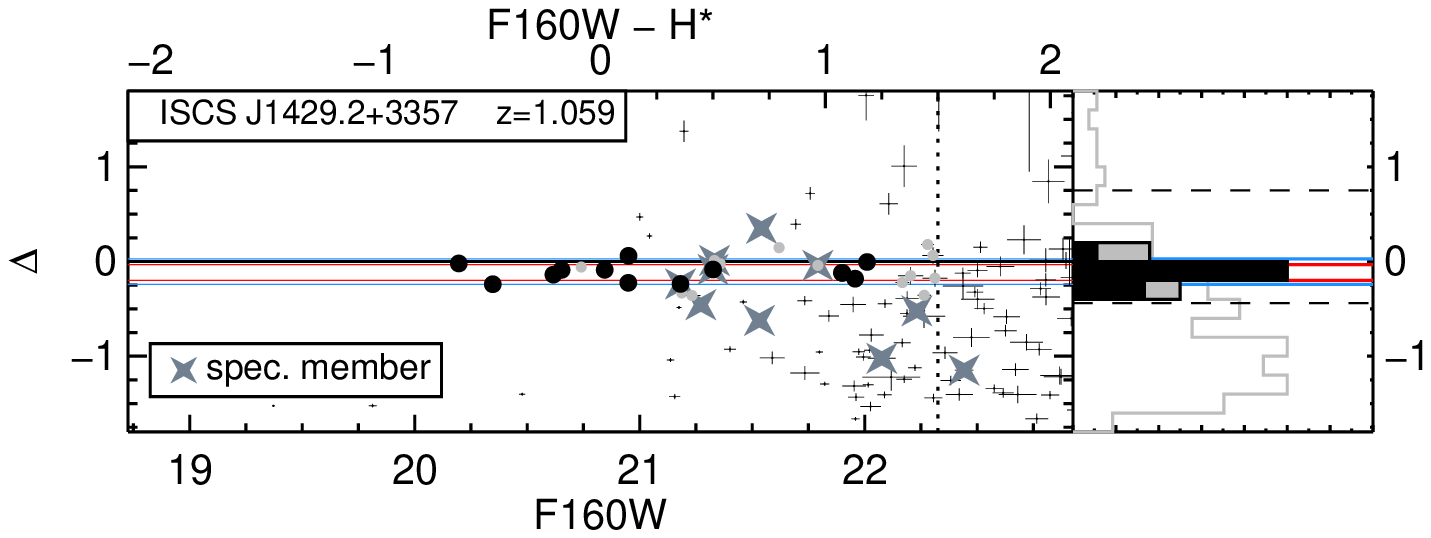} \\
	\includegraphics[width=6in,clip=true]{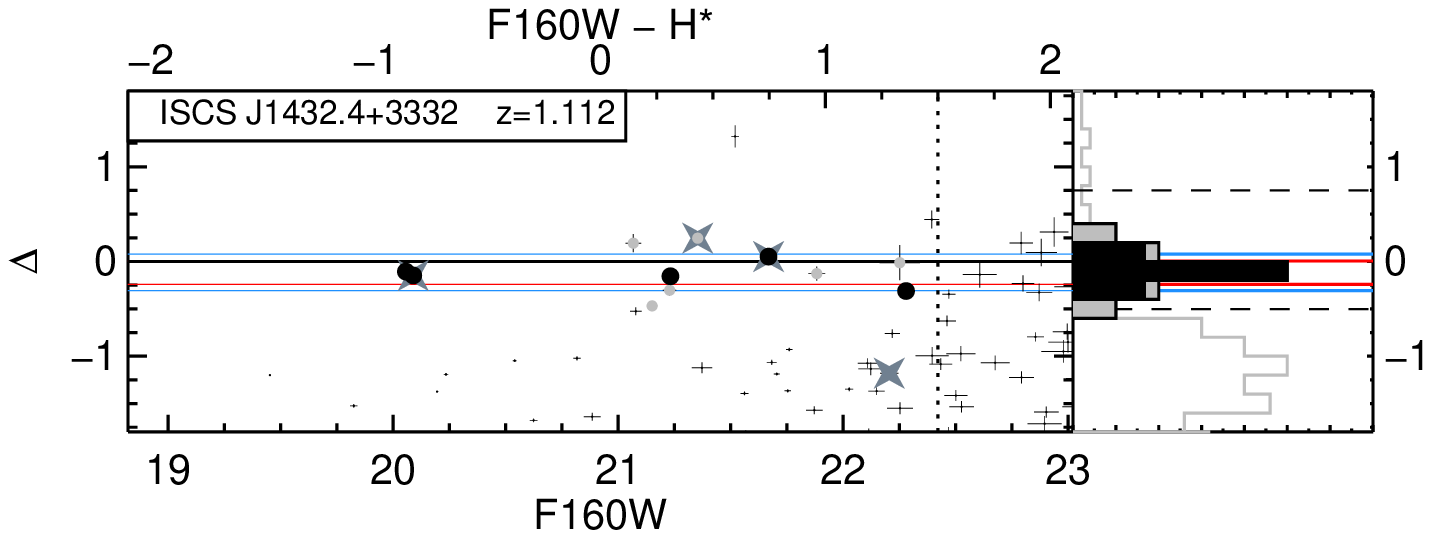} \\
	\includegraphics[width=6in,clip=true]{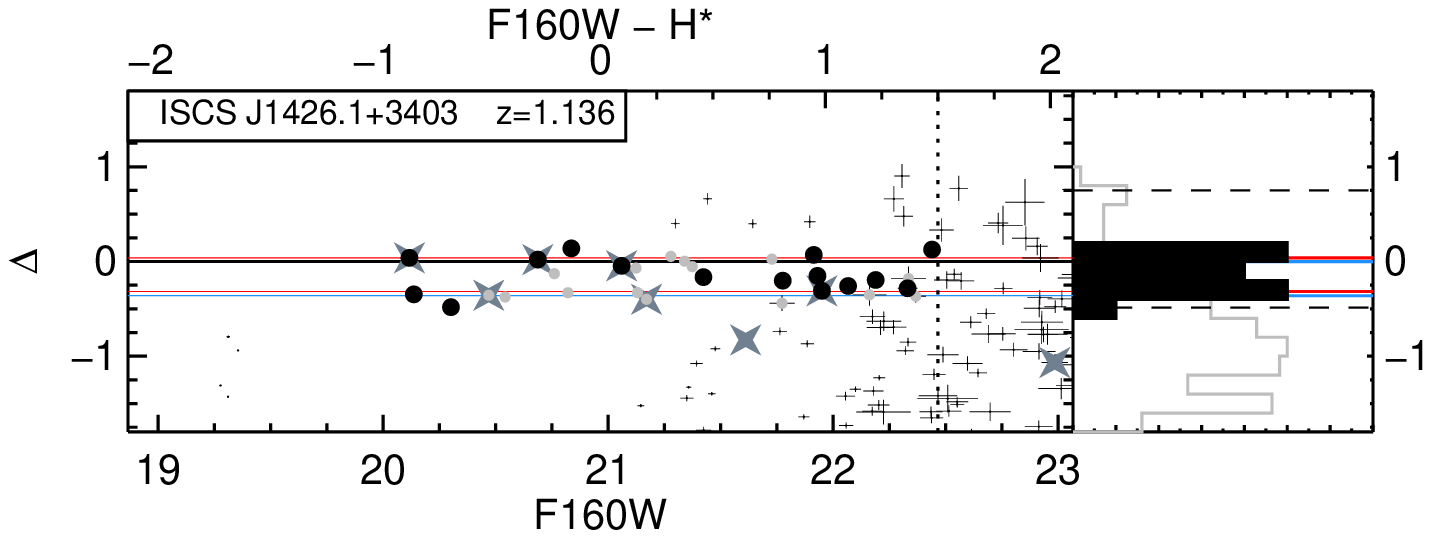} 
	\end{tabular}
	\end{center}
      \caption{ CMD residuals from $z_f=3$ model and subsequent analysis for each cluster, in ascending redshift order.  We restrict attention to elliptical and S0 galaxies satisfying the color and magnitude cuts of Section~\ref{ss:selection}.  All RSGs and the visually-selected ETG subsample are plotted as gray and black filled circles, respectively.  We plot a vertical dotted line at $H=H^* + 1.5$, representing the magnitude selection limit we apply in Section~\ref{ss:selection}.  The location and scale measurements are shown as pairs of horizontal lines centered on the final measured $\Delta_0$ zeropoint value and spaced to surround the one-sigma region defined by the final measured  $\sigma_{\rm int}$ scatter value.  The blue solid lines correspond to the scatter of the subsample containing all RSGs, and the red solid lines correspond to scatter in the visual ETG subsample.  Histograms are tabulated for each subsample on the right of each CMD, and include dashed black lines to show the color selection region set in Section~\ref{ss:selection}.  \label{fig:measure1} \label{fig:measure} \label{fig:measure2}  }
      \end{figure*} 
\setcounter{figure}{3}
      \begin{figure*}
	\begin{center}
	\begin{tabular}{c}
	\includegraphics[width=6in,clip=true]{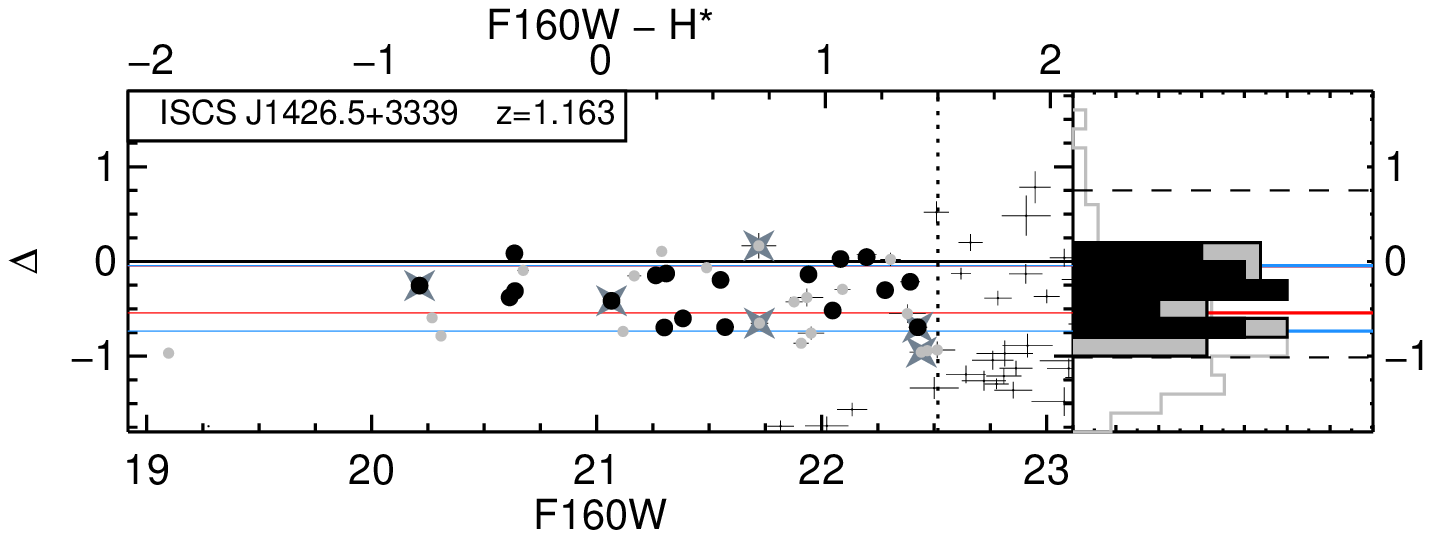} \\
	\includegraphics[width=6in,clip=true]{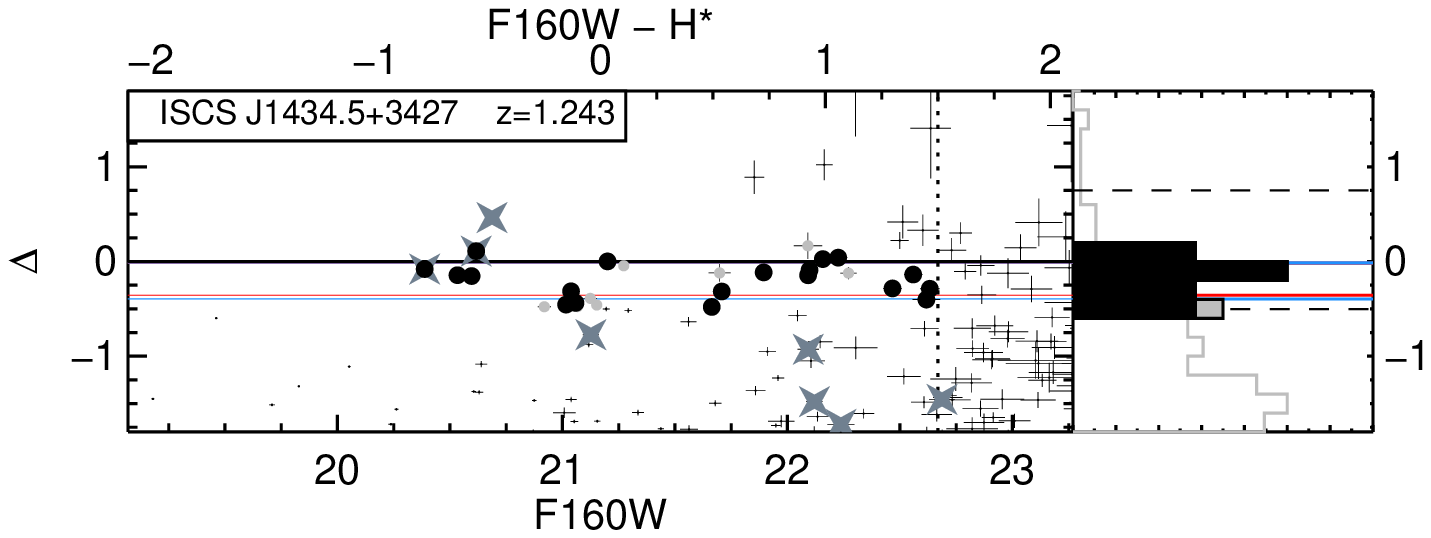} \\ 
	\includegraphics[width=6in,clip=true]{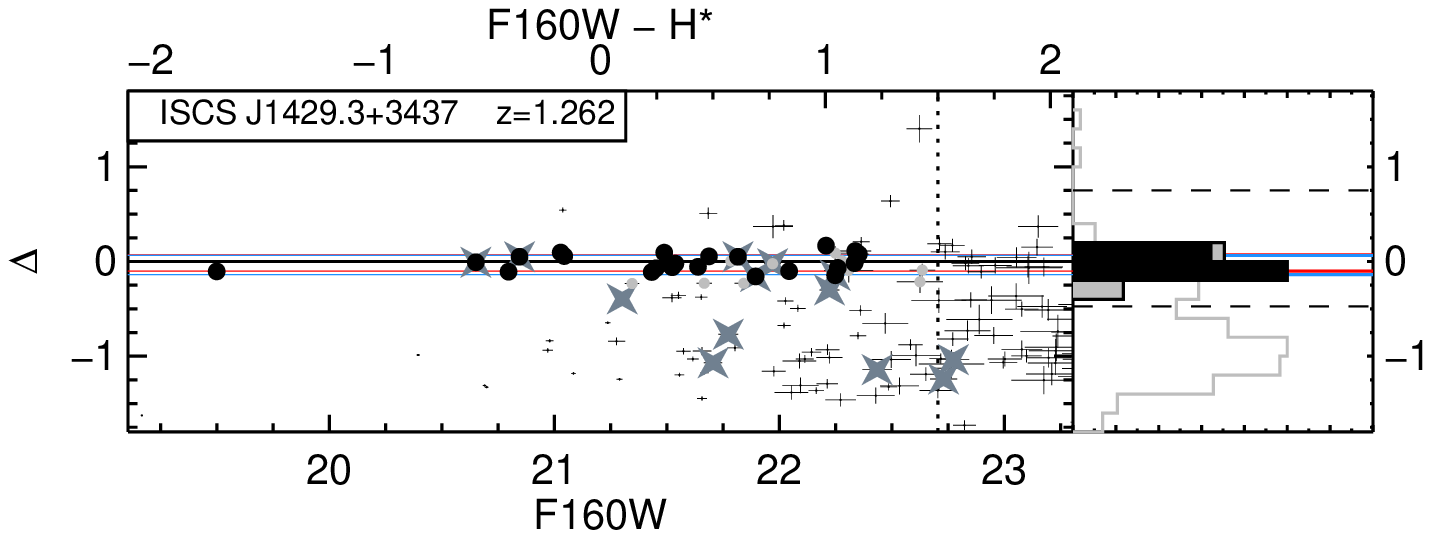} 
	\end{tabular}
	\end{center}
      \caption{ Continued from previous page.}
      \end{figure*} 
\setcounter{figure}{3}
      \begin{figure*}
	\begin{center}
	\begin{tabular}{c}
	\includegraphics[width=6.5in,clip=true]{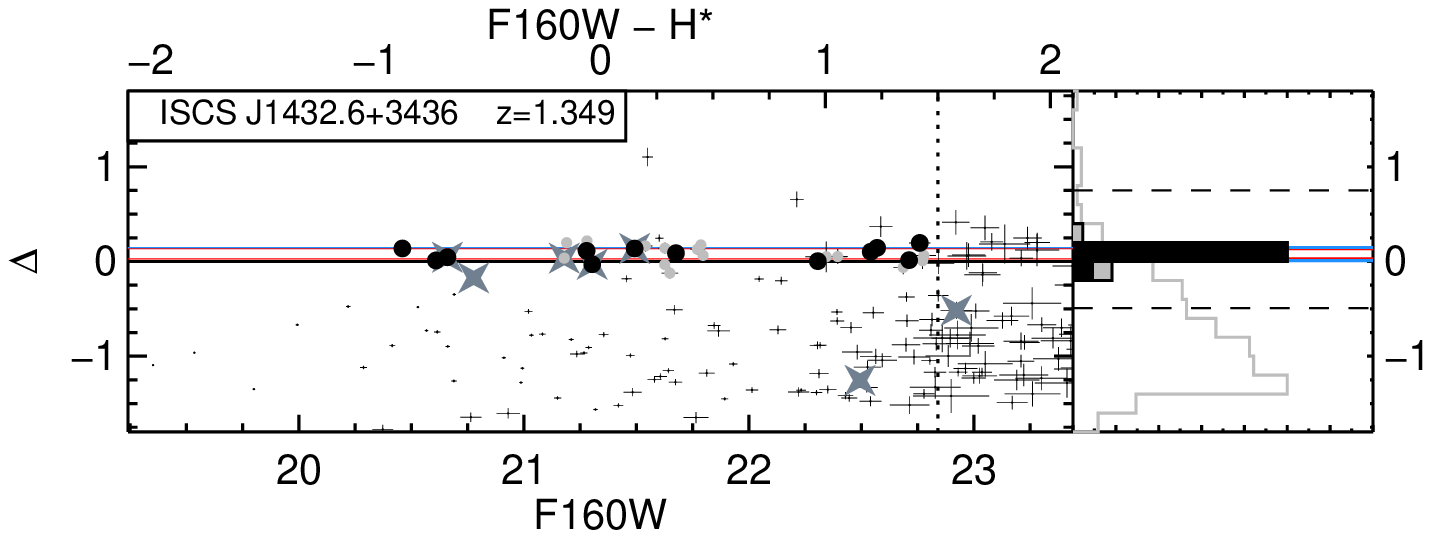} \\
	\includegraphics[width=6.5in,clip=true]{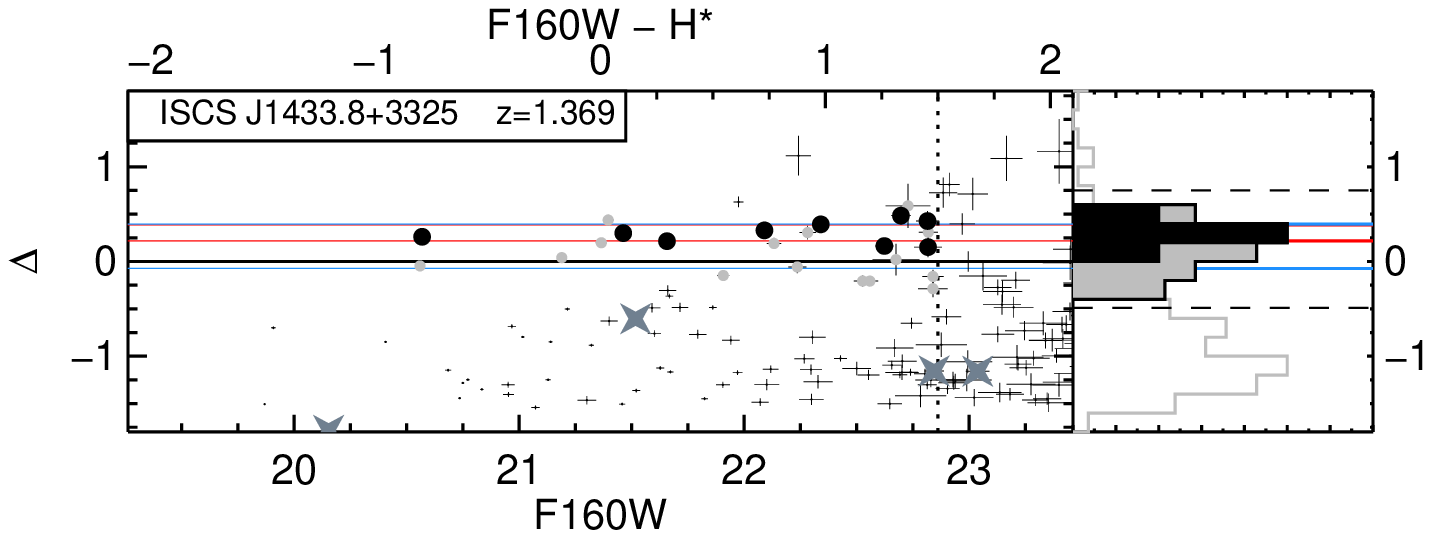} \\
	\includegraphics[width=6.5in,clip=true]{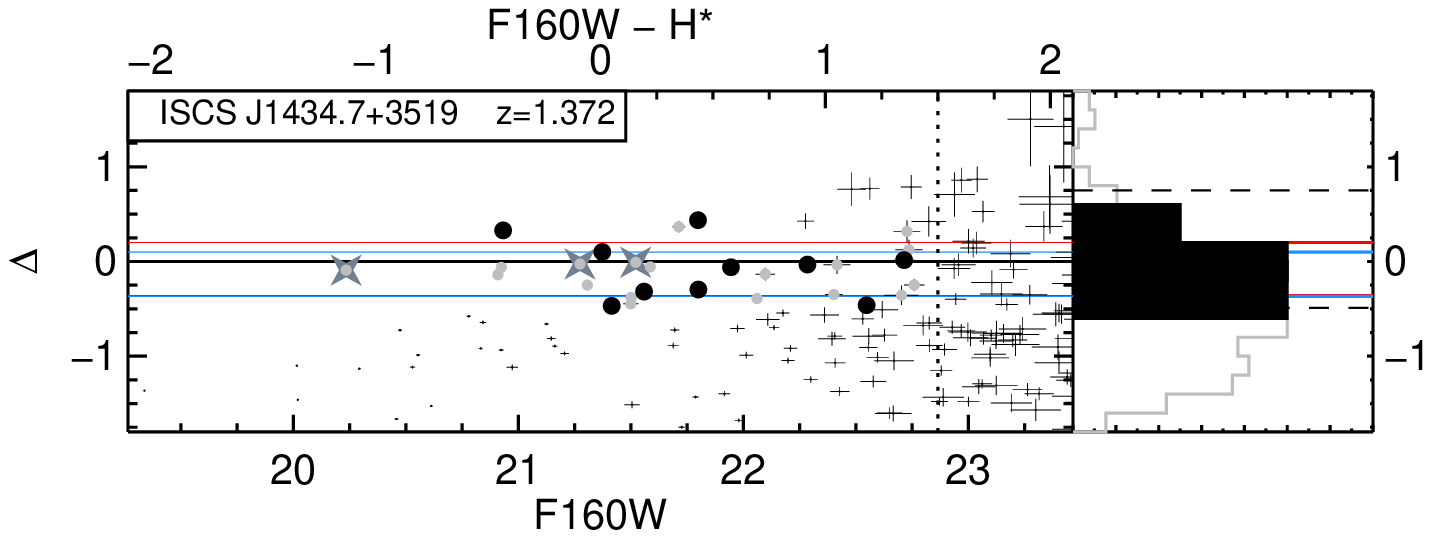} 
	\end{tabular}
	\end{center}
      \caption{ Continued from previous page.}
      \end{figure*} 

\setcounter{figure}{3}
      \begin{figure*}
	\begin{center}
	\begin{tabular}{c}
	\includegraphics[width=6.5in,clip=true]{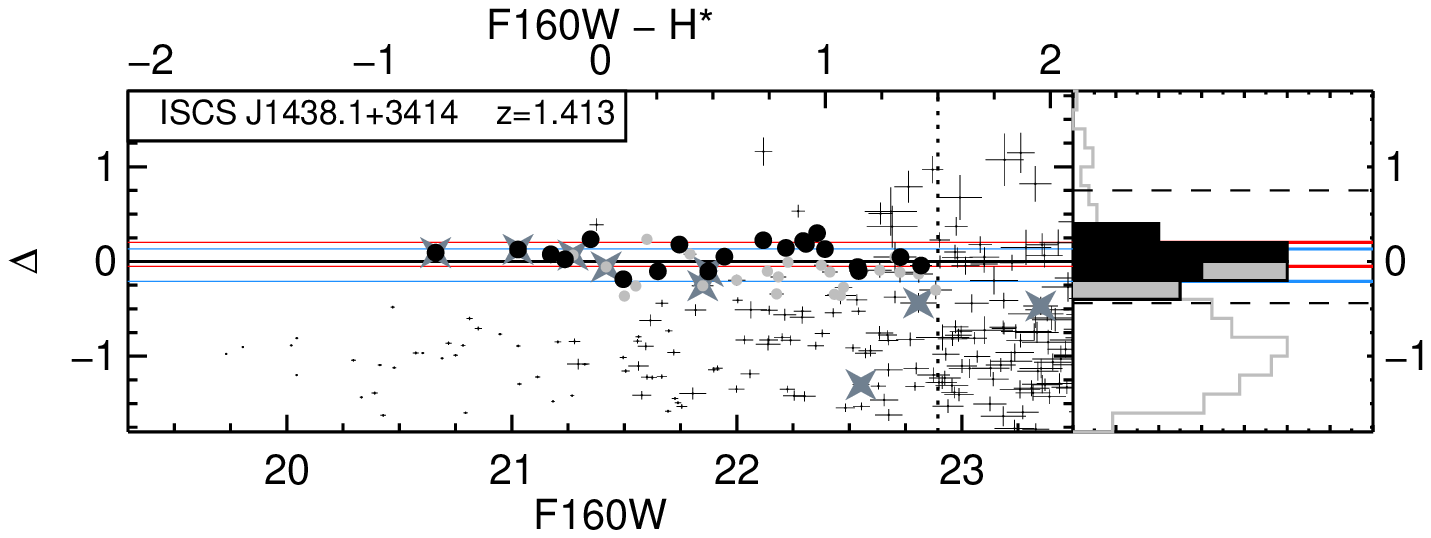} \\
	\includegraphics[width=6.5in,clip=true]{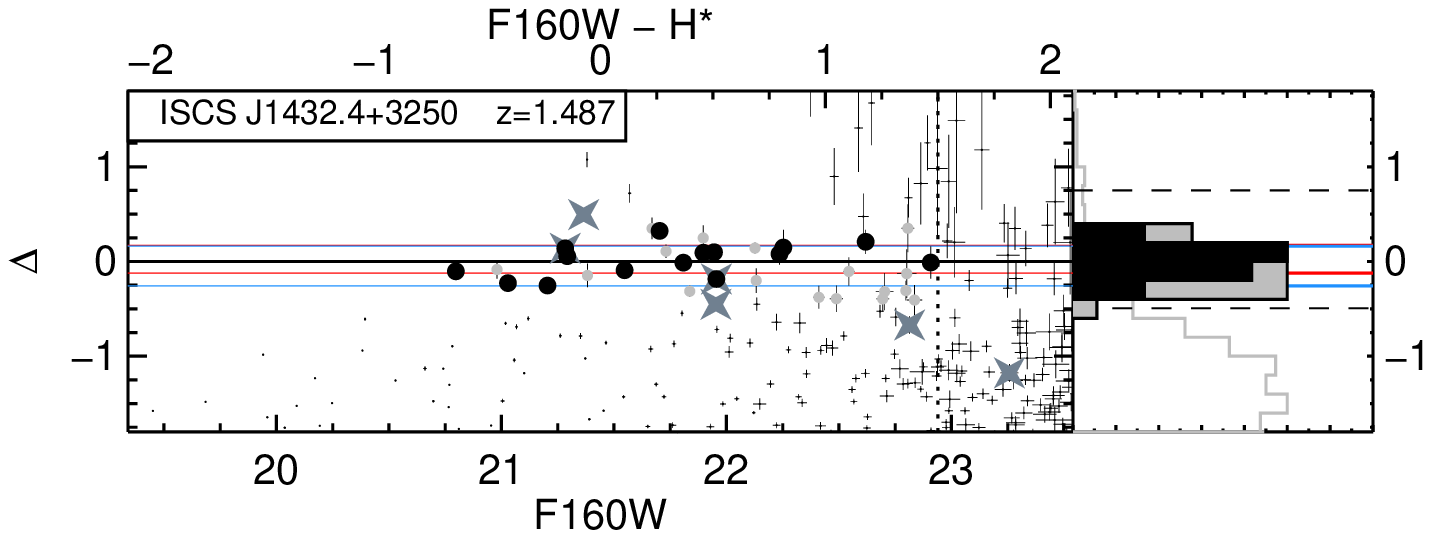} 
	\end{tabular}
	\end{center}
      \caption{ Continued from previous page.}
      \end{figure*} 



\section{Simple Models of Color and Scatter} \label{s:models}  \label{ss:models} \label{ss:simplemodels}

To associate the measured CMRs with formation epochs, we compute grids of model colors in our observed passbands using the models of \citet[][2007 version]{bc03}.  Each model galaxy forms its stars in a single exponentially-decaying burst with a timescale of $\tau=0.1$ Gyr and evolves passively thereafter until it is observed.  We will label these models our ``short-burst stellar population" (\ssp) models, whose evolution after a few hundred Myr is very similar to that of a simple stellar population (SSP).  To construct the grids we use the EZGal code of \citet{mancone12_arxiv}\footnote{http://www.baryons.org/ezgal/}, focusing on solar-metallicity stellar populations drawn from a  \citet{chabrier03} initial mass function (IMF), using formation epochs evenly spaced in time between $z_f=1.8$ and $z_f=10$.

For ease of comparison to other work, we convert our measured quantities to the rest-frame UBV filter set following \citet{mei06} using these models.  First, we evolve a set of \ssp\ models with varying metallicity and formation redshift down to each cluster's redshift.  Then we fit the relation $(i-H)_z = a + b (U-V)_0$ to estimate the parameters a and b.  The slope b is used to derive the rest-frame color scatter given an observed color scatter;  specifically, $\sigma_{UV} = \sigma_{iH}/b$.  We also use these relations to compute the slope of the Coma CMR \citep{eisenhardt07} as it would be observed in the \hst\ bands at these redshifts; these were used for selecting \rss\ in Section~\ref{ss:selection}.  We visually inspected the linear fits and found good agreement in all cases.  We also performed the analysis of Section~\ref{s:formation} using the colors as-observed and using only directly comparable models at the appropriate redshifts.  We found that our results are identical in both cases, confirming the validity of the conclusions based on the k-corrected rest-frame colors.  

We construct model red sequence ETG color scatters following previous studies \citep[][and references therein]{papovich10, hilton09, mei09}.  We consider the simplest models in which the dispersion in optical ETG color for a single model set depends on the redshift at which star formation ceased, $z_{\rm end}$.  Prior to $z_{\rm end}$, our models assume that cluster galaxies form uniformly in time starting at $z_0=10$.  In practice, for each stellar IMF and metallicity we draw $10^5$ individual galaxy models with random formation times between $z_0$ and $z_{\rm end}$ and evolve them to each cluster's observed redshift.  We then compute three quantities: the color dispersion and median color using the biweight scale and location estimators \citep{beers90}, and the luminosity-weighted average formation time, which we label $\langle z_f \rangle$, a typical (star-)formation redshift.  The epochs $z_{\rm end}$ and $\langle z_f \rangle$ describe these galaxies' last major star-forming episode, the nature and timing of which we seek to trace.  In what follows we denote this set of models as our ``constant SFR'' (\csp) models.  

To compare the color zeropoints between the observed clusters and model galaxies, we measure the color of the observed red sequence at the characteristic brightness, $H^*(z_{\rm obs})$, that we evolve from the value in Coma \citep{depropris98}.  We assume each model galaxy has the same mass, and thus refer to all CMR quantities as those measured at $H^*(z_{\rm obs})$.  A galaxy luminosity function evolving by $\sim 0.5$ magnitudes in the fashion of \citet{mancone10} will only mildly affect our interpretation of their stellar populations.  Specifically, the comparison between observed and model color depends only weakly on the magnitude at which we compute numerical values of the colors from the derived CMRs.  This owes to the CMR slopes in our observed bands being shallow ($\approx -0.1$), so a 0.5 magnitude shift in the value of $H^*$ alters our derived $U-V$ colors by only $\approx 0.03$ magnitudes.  To recover the results we obtain in Section~\ref{s:formation} with a fixed formation epoch, $H^*$ must increase by $\gtrsim 3$ magnitudes between $z \approx 1.5$ and $z \approx 1$, a change we believe is too large.  We conclude that this effect may slightly bias our color measurements to the red but it does not change our qualitative results.

The true star formation and assembly histories of ETGs, as observed in cluster cores over time, are likely to be rather complex.  Previous studies showed that these \ssp\ and \csp\ models do not fully explain the evolution of the colors of cluster galaxies.  However, they represent valuable parametrizations of the observed CMRs, allowing for comparisons to previous work and direct, if approximate, estimation of ETG stellar population ages without the need for a complete specification of their varied histories.  Therefore the above models will be applied as the initial framework for analyzing our results.

\begin{deluxetable*}{cc@{\hspace{1.0cm}}ccc@{\hspace{1.0cm}}cccc}
\tablecaption{Cluster red sequence color zeropoints and color scatters. \label{tab:measurements}}
\tablecolumns{8}
\tablewidth{1.0\textwidth}
\tablehead{
\colhead{} & \colhead{}                    \hspace{1cm} & \multicolumn{3}{c}{Observed bands}             & \multicolumn{3}{c}{Rest-frame U-V}  \\
\colhead{Name} & \colhead{$z$}       \hspace{1cm} & \colhead{Vis. ETGs}& \colhead{Quant. ETGs}& \colhead{All RSGs}           \hspace{1cm} & \colhead{Vis. ETGs}& \colhead{Quant. ETGs}& \colhead{All RSGs}
} 
\startdata
\cutinhead{Color Zeropoint, $C_0$}
   \sfiftyone\ & 1.059 & 1.92 $\pm$ 0.03 & 1.93 $\pm$ 0.03 & 1.93 $\pm$ 0.03 & 1.03 $\pm$ 0.03 & 1.04 $\pm$ 0.03 & 1.04 $\pm$ 0.03  \\ 
   \sseventeen\ & 1.112 & 2.33 $\pm$ 0.05 & 2.32 $\pm$ 0.06 & 2.33 $\pm$ 0.06 & 1.05 $\pm$ 0.04 & 1.04 $\pm$ 0.05 & 1.05 $\pm$ 0.04  \\ 
   \sthirtyfour\ & 1.136 & 2.04 $\pm$ 0.05 & 2.00 $\pm$ 0.04 & 2.00 $\pm$ 0.04 & 1.02 $\pm$ 0.05 & 0.99 $\pm$ 0.04 & 0.99 $\pm$ 0.03  \\ 
   \sfourteen\ & 1.163 & 1.95 $\pm$ 0.07 & 1.85 $\pm$ 0.10 & 1.87 $\pm$ 0.06 & 0.91 $\pm$ 0.06 & 0.82 $\pm$ 0.08 & 0.84 $\pm$ 0.05  \\ 
  \sthreefourtytwo\ & 1.243 & 2.35 $\pm$ 0.05 & 2.27 $\pm$ 0.04 & 2.34 $\pm$ 0.04 & 0.94 $\pm$ 0.04 & 0.88 $\pm$ 0.03 & 0.93 $\pm$ 0.03  \\ 
   \sthirty\ & 1.262 & 1.56 $\pm$ 0.02 & 1.53 $\pm$ 0.02 & 1.54 $\pm$ 0.02 & 1.05 $\pm$ 0.02 & 1.01 $\pm$ 0.02 & 1.02 $\pm$ 0.02  \\ 
   \stwentynine\ & 1.349 & 1.79 $\pm$ 0.03 & 1.81 $\pm$ 0.02 & 1.79 $\pm$ 0.02 & 1.12 $\pm$ 0.02 & 1.14 $\pm$ 0.02 & 1.12 $\pm$ 0.02  \\ 
   \seightyfour\ & 1.369 & 2.05 $\pm$ 0.04 & 2.01 $\pm$ 0.05 & 1.91 $\pm$ 0.06 & 1.33 $\pm$ 0.04 & 1.29 $\pm$ 0.05 & 1.20 $\pm$ 0.06  \\ 
   \stwentyfive\ & 1.372 & 1.68 $\pm$ 0.10 & 1.71 $\pm$ 0.08 & 1.62 $\pm$ 0.05 & 0.97 $\pm$ 0.10 & 1.00 $\pm$ 0.07 & 0.91 $\pm$ 0.05  \\ 
   \stwentytwo\ & 1.413 & 1.89 $\pm$ 0.04 & 1.85 $\pm$ 0.03 & 1.78 $\pm$ 0.03 & 1.11 $\pm$ 0.03 & 1.07 $\pm$ 0.03 & 1.00 $\pm$ 0.03  \\ 
   \sthirtysix\ & 1.487 & 2.53 $\pm$ 0.05 & 2.49 $\pm$ 0.06 & 2.46 $\pm$ 0.05 & 1.01 $\pm$ 0.04 & 0.97 $\pm$ 0.04 & 0.95 $\pm$ 0.03  \\ 
\cutinhead{Color Scatter, $\sigma_{\rm int}$}
   \sfiftyone\ & 1.059 & 0.08 $\pm$ 0.02 & 0.12 $\pm$ 0.03 & 0.13 $\pm$ 0.02 & 0.08 $\pm$ 0.02 & 0.11 $\pm$ 0.03 & 0.12 $\pm$ 0.02 \\ 
   \sseventeen\ & 1.112 & 0.12 $\pm$ 0.02 & 0.20 $\pm$ 0.07 & 0.19 $\pm$ 0.05 & 0.09 $\pm$ 0.02 & 0.15 $\pm$ 0.05 & 0.14 $\pm$ 0.03 \\ 
   \sthirtyfour\ & 1.136 & 0.18 $\pm$ 0.03 & 0.17 $\pm$ 0.02 & 0.18 $\pm$ 0.02 & 0.15 $\pm$ 0.02 & 0.14 $\pm$ 0.02 & 0.16 $\pm$ 0.01 \\ 
   \sfourteen\ & 1.163 & 0.25 $\pm$ 0.03 & 0.30 $\pm$ 0.06 & 0.35 $\pm$ 0.03 & 0.21 $\pm$ 0.03 & 0.26 $\pm$ 0.05 & 0.29 $\pm$ 0.02 \\ 
  \sthreefourtytwo\ & 1.243 & 0.17 $\pm$ 0.02 & 0.17 $\pm$ 0.02 & 0.19 $\pm$ 0.02 & 0.13 $\pm$ 0.02 & 0.13 $\pm$ 0.01 & 0.14 $\pm$ 0.02 \\ 
   \sthirty\ & 1.262 & 0.09 $\pm$ 0.01 & 0.09 $\pm$ 0.01 & 0.10 $\pm$ 0.01 & 0.09 $\pm$ 0.01 & 0.09 $\pm$ 0.01 & 0.10 $\pm$ 0.01 \\ 
   \stwentynine\ & 1.349 & 0.05 $\pm$ 0.01 & 0.06 $\pm$ 0.01 & 0.07 $\pm$ 0.01 & 0.05 $\pm$ 0.01 & 0.06 $\pm$ 0.01 & 0.07 $\pm$ 0.01 \\ 
   \seightyfour\ & 1.369 & 0.08 $\pm$ 0.02 & 0.06 $\pm$ 0.04 & 0.24 $\pm$ 0.03 & 0.08 $\pm$ 0.02 & 0.06 $\pm$ 0.03 & 0.22 $\pm$ 0.03 \\ 
   \stwentyfive\ & 1.372 & 0.28 $\pm$ 0.06 & 0.29 $\pm$ 0.04 & 0.23 $\pm$ 0.03 & 0.27 $\pm$ 0.05 & 0.28 $\pm$ 0.04 & 0.22 $\pm$ 0.03 \\ 
   \stwentytwo\ & 1.413 & 0.13 $\pm$ 0.02 & 0.12 $\pm$ 0.02 & 0.17 $\pm$ 0.02 & 0.12 $\pm$ 0.02 & 0.11 $\pm$ 0.01 & 0.16 $\pm$ 0.02 \\ 
   \sthirtysix\ & 1.487 & 0.15 $\pm$ 0.03 & 0.13 $\pm$ 0.04 & 0.21 $\pm$ 0.02 & 0.10 $\pm$ 0.02 & 0.10 $\pm$ 0.03 & 0.15 $\pm$ 0.02
\enddata
\end{deluxetable*}

\section{Results}  \label{s:formation}

\subsection{Color and Scatter Evolution} \label{ss:colorscatters}

\begin{figure}
	\begin{center}
	\begin{tabular}{c}  
	\includegraphics[width=3.3in]{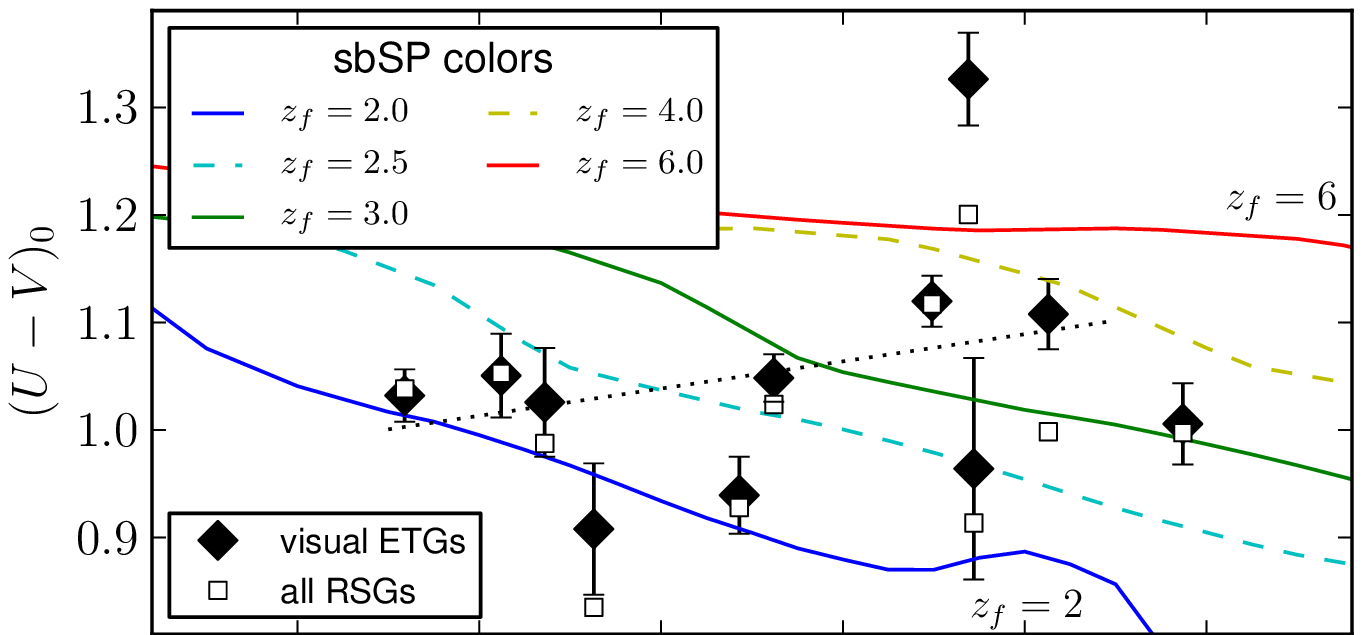}\\
	\includegraphics[width=3.3in]{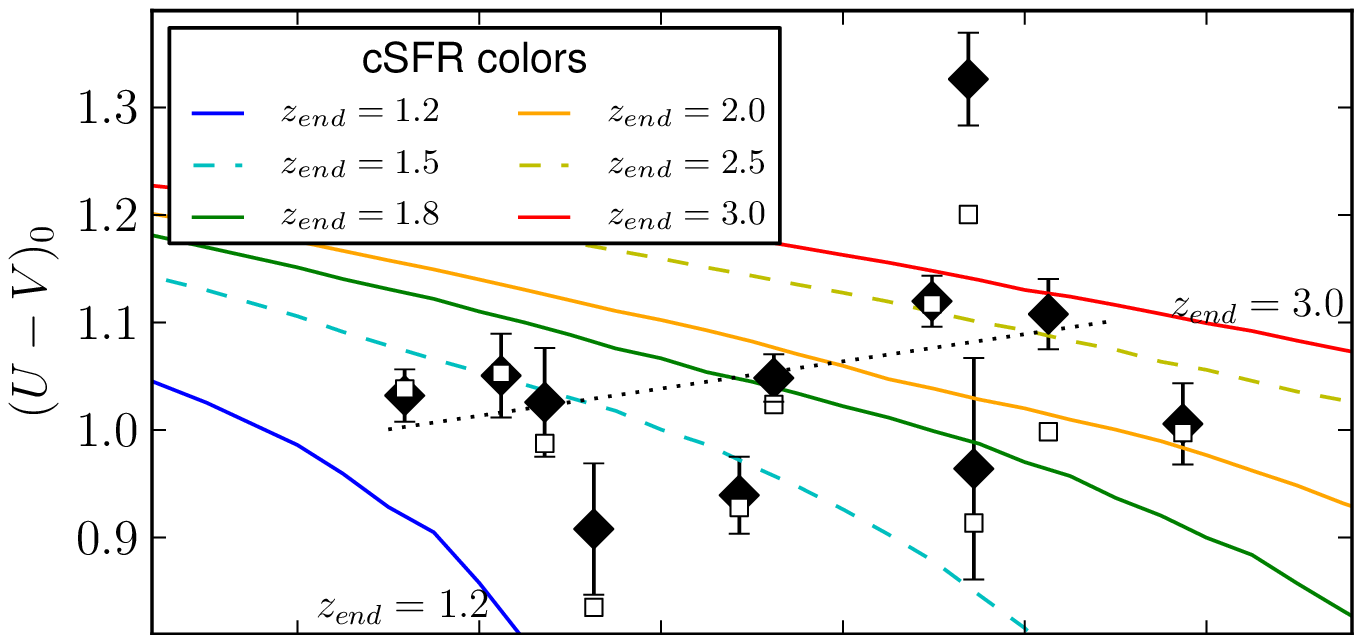}\\
	\includegraphics[width=3.3in]{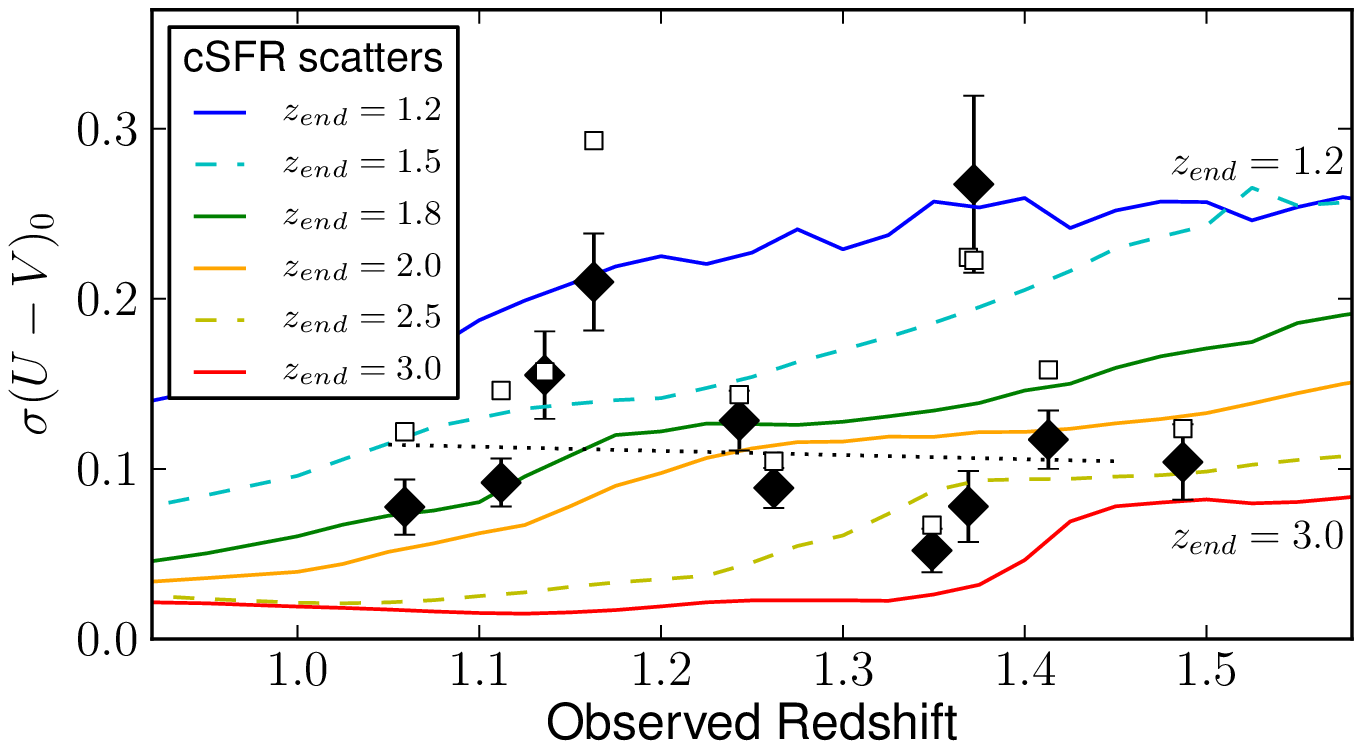} 
	\end{tabular}
	\end{center}
      \caption{ Rest frame colors and scatters of our clusters' CMRs, with associated models.  The top, middle, and bottom figures compare our measurements against the \ssp\ color, \csp\ color, and \csp\ color scatter models, respectively (Section~\ref{ss:simplemodels}).   Here we employ visual morphological classification for the solid black diamonds and show the results for the entire RSG subsample as the open squares.  For clarity, in this and later figures we only plot error bars for the morphologically-selected sample (black diamonds) and not the entire RSG subsample -- they are similar in magnitude and both are listed in Tables~\ref{tab:measurements}-\ref{tab:ages}.  We plot a linear fit to the data in each panel as dotted black lines.  There is a clear disagreement between the continued evolution of any single \ssp\ or \csp\ model and the observed CMR trends.  \label{fig:evolution}}
      \end{figure}

We compile the measured colors and scatters of cluster CMRs in Table~\ref{tab:measurements}, and plot them in Figure~\ref{fig:evolution}.  We also plot the associated \ssp\ and \csp\ models, where for the latter we have utilized the parameter $z_{\rm end}$ rather than a weighted formation redshift $\langle z_f \rangle$ because for a given model $\langle z_f \rangle$ depends on the observation redshift.  Thus the \csp\ models plotted correspond to fixed histories, where star formation continues uniformly in time and ceases at $z_{\rm end}$.  

In the top two rows of Figure~\ref{fig:evolution}, we see that the rest-frame $U-V$ colors of cluster RSGs are nearly constant across the redshift range 1 to 1.5, in contrast to any of the model stellar populations that redden monotonically with time.  The color scatters in the bottom row of Figure~\ref{fig:evolution} reflect the same physical scenario: the rest-frame $\sigma (U-V)$ is nearly constant (or slightly declining) with increasing redshift, while the scatter increases monotonically in \csp\ models.  

We checked several other stellar population models \citep{Maraston:2005, conroy09_fsps, conroy10_fsps} and parameter choices (metallicity, IMF, SFH) and found similar results.  Models with $\sim 0.4 Z_{\odot}$ can reproduce the flat $(U-V)_0 \sim 1$ trend at these redshifts, but \textit{cannot} account for a flat or declining color scatter trend, nor for the colors of $z < 1$ giant cluster galaxies \citep{aragonsalamanca93, sed98}.

Several clusters are outliers to the median trends observed in these figures, and will be discussed in more detail in Sections~\ref{ss:outliers} and \ref{s:discussion}.  These general trends imply that we are not approaching the primary epoch of star formation in the centers of these clusters, and/or that their formation histories were not as simple as those assumed by the simplest models.   Colors of the CMRs are redder than, and the color scatters smaller than the simplest \ssp\ or \csp\ estimates in which the CMRs get bluer and wider as one approaches their primary formation epoch.

\subsection{Effect of Morphology Selection} \label{ss:progenitorbias}

\begin{figure}
	\begin{center}
	\begin{tabular}{c}
	\includegraphics[width=3.3in]{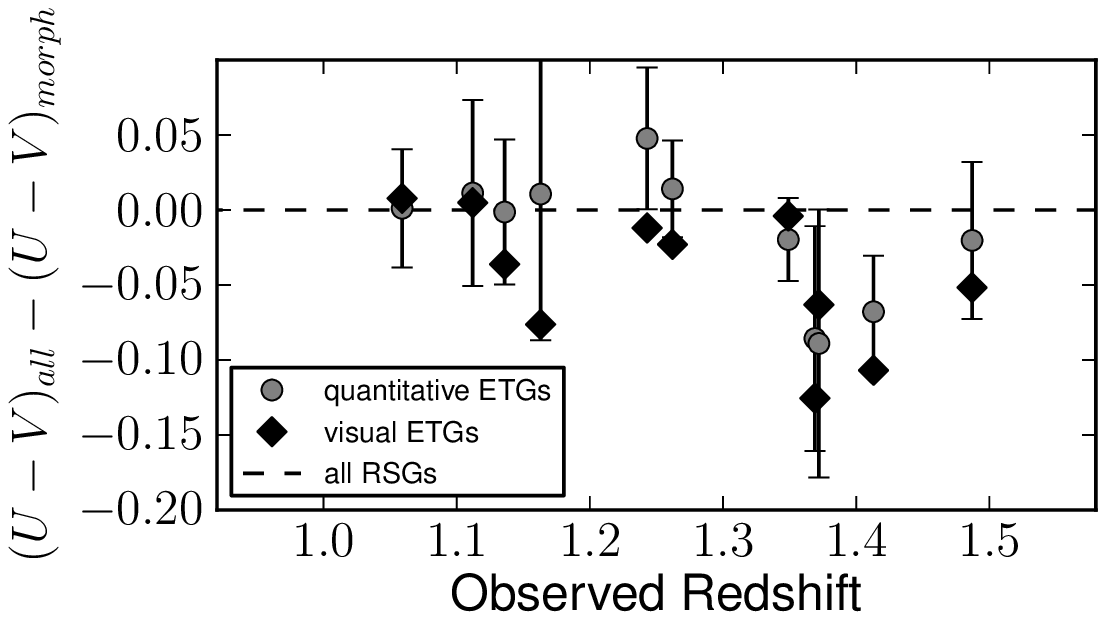}  \\
	\includegraphics[width=3.3in]{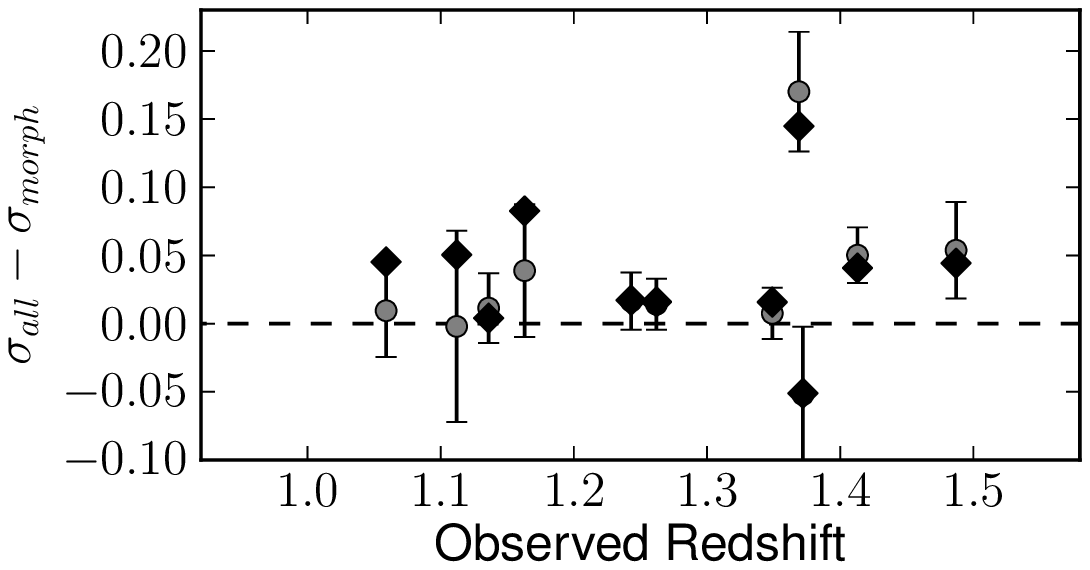}
	\end{tabular}
	\end{center}
      \caption{ Demonstration of the effect of our morphological selection, in terms of colors (top) and scatters (bottom).  Here we plot the change in measured CMR quantities when we turn off our morphology cut.  In the top panel, a negative number occurs when the CMR of ETGs is on average redder than CMR of all RSGs in the cluster.   In the bottom panel, a positive number occurs when the CMR of ETGs  is on average narrower than the CMR of all RSGs in the cluster.  Although weakly, the data imply the existence of bluer/younger non-ETGs on the red sequence in clusters above $z\sim 1.3$.  One interpretation of this trend is as evidence of ongoing morphological transformations of relatively young late-type galaxies into cluster ellipticals.  However, these in-situ color differences do not account for all evolution in the median formation epoch seen in Figure~\ref{fig:evolution}.  Therefore, a significant fraction of ETG luminosity at $z\sim 1$ was assembled from outside the cluster cores (or at least, outside the ETGs' progenitors) since $z \sim 1.5$, and composed of relatively younger stellar populations.  \label{fig:morphcompare}}
      \end{figure}

 We demonstrate the effect of the morphology cut on both colors and scatters in Figure~\ref{fig:morphcompare}.  At $z > 1.3$, the colors of morphologically-selected ETGs are slightly redder than the full red sequences at the same redshift.  This might reflect the canonical in-situ progenitor bias as described in \citet{vandokkumfranx01}, where galaxies drop out of the ETG sample at higher redshift because they are continually being transformed and as such are not necessarily identified as ETGs in a cluster's progenitor.  Progenitor bias, or a similar effect, may provide a natural explanation for the seemingly unintuitive way that the \rss\ remain narrow and red, a scenario we consider further in Section~\ref{s:discussion}.  The rate of transformations inside the cluster center appears to have slowed by $z < 1.3$, lending additional evidence that above $z\approx 1.3$ we are beginning to see the late stages of evolution by the giant cluster galaxies.  
 
The change in U-V color owing to this effect is approximately $0.1$ magnitudes, and is therefore insufficient to explain the entirety of the deviation we see from any one \ssp\ in Figure~\ref{fig:evolution}, leaving the colors and scatters roughly constant across the redshift range.  In this series of figures (Figures~\ref{fig:evolution}, \ref{fig:morphcompare}, \ref{fig:zforms}, and \ref{fig:ages}) we plot the entire RSG sample as unfilled black squares to demonstrate the (typically inconsequential) difference from the morphologically-selected ETG sample.  We conclude that the cluster ETG samples are growing across the entire range $1 < z < 1.5$ by the incorporation of younger RSGs that are missed in the samples at higher redshift, not only because they were star forming (and hence still growing) or disk-like in the centers at $z > 1.3$, but also because RSGs/ETGs that grow or enter the central region have younger stars than those previously present.  

\subsection{Inferred (Simple) Star Formation Epochs} \label{ss:epochs}   \label{ss:ages}

Although the simple models of Section~\ref{ss:simplemodels} may not reflect the \emph{evolving} stellar populations of the ISCS clusters, we can nevertheless gain some insight by fitting the individual formation redshift for each cluster.  Using the \ssp\ and \csp\ models, we interpret the derived ages as the average time elapsed since the last major epoch of star formation in the galaxies currently making up the red sequence.  

\begin{figure}
	\begin{center}
	\begin{tabular}{c}
	\includegraphics[width=3.3in]{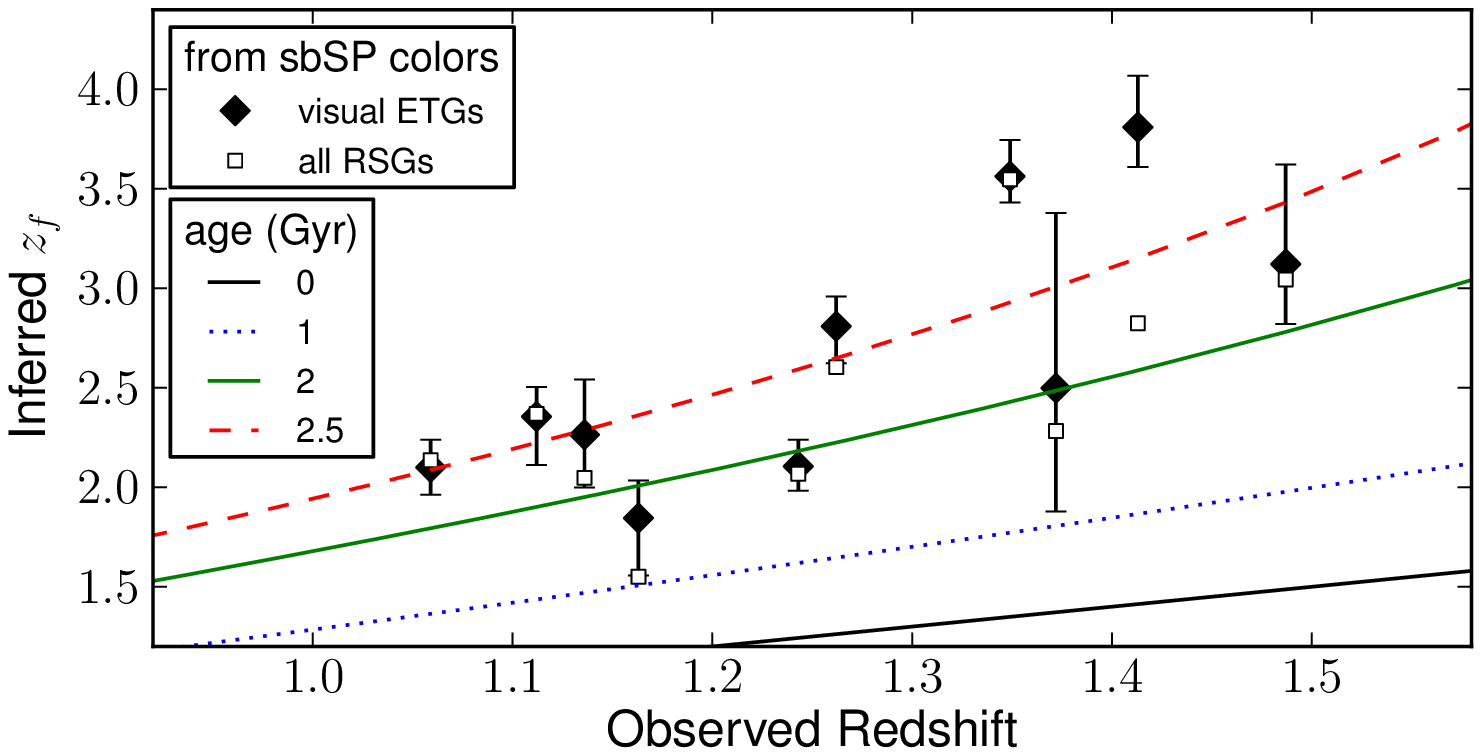} \\
	\includegraphics[width=3.3in]{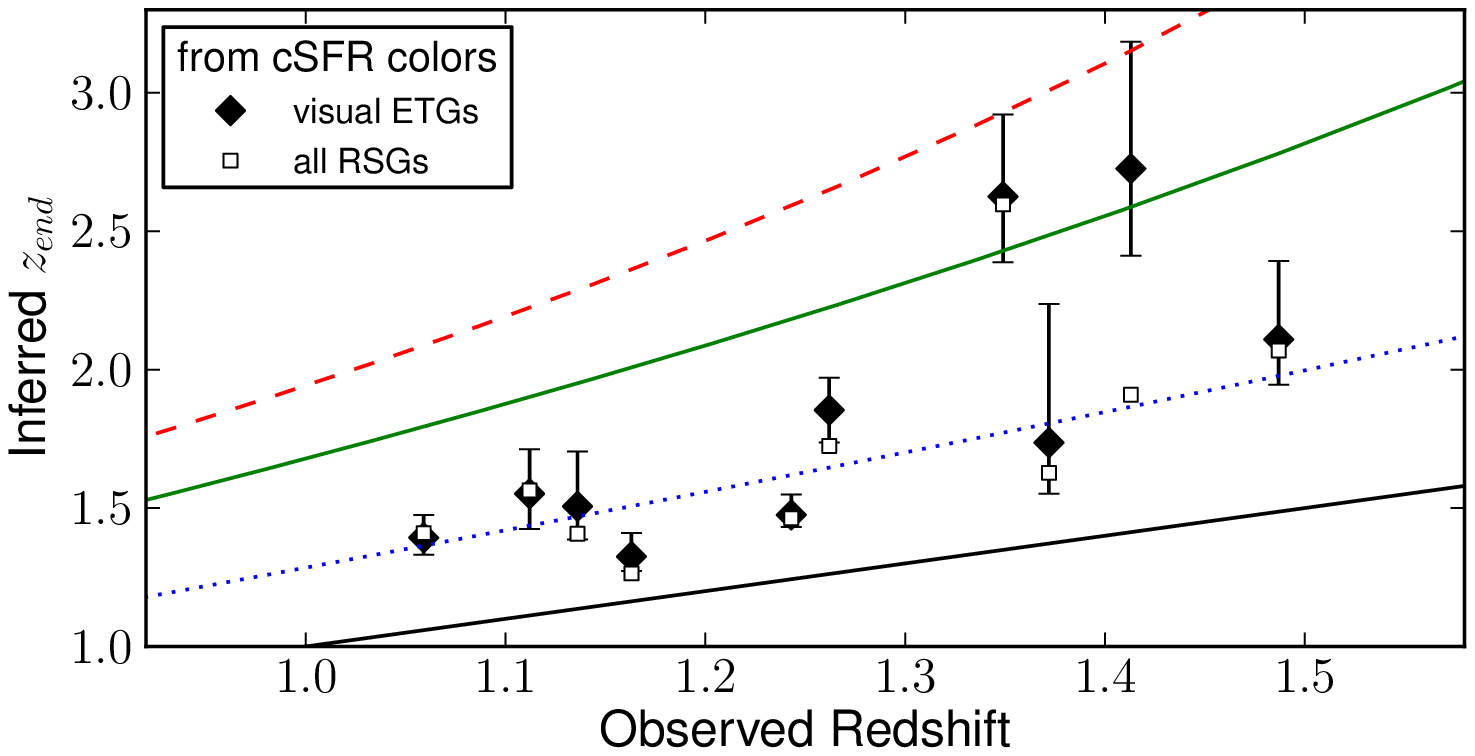} \\
	\includegraphics[width=3.3in]{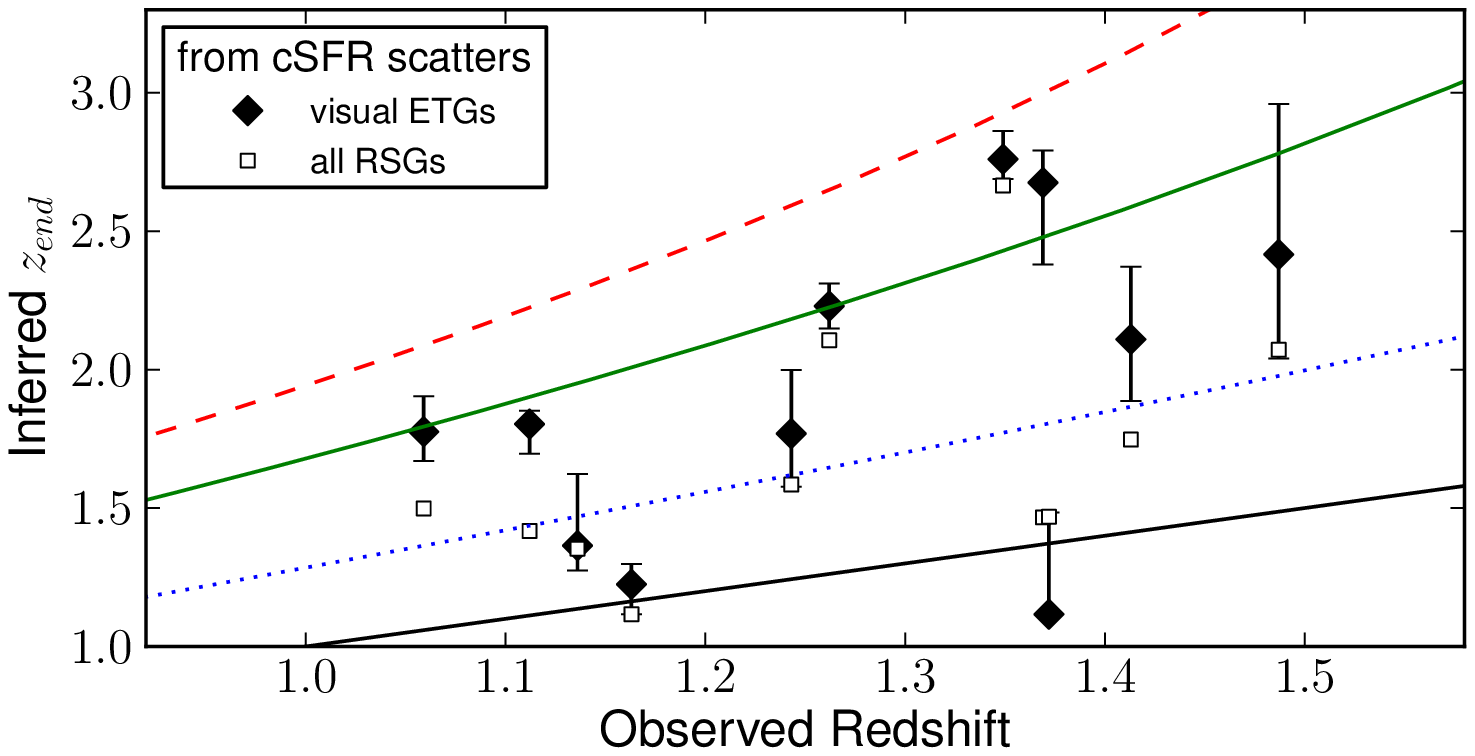}  
	\end{tabular}
	\end{center}
      \caption{ Inferred formation redshifts, corresponding to the same rows as Figure~\ref{fig:evolution}.  The values in the top, middle, and bottom figures are inferred from our \ssp\ color, \csp\ color, and \csp\ color scatter models, respectively (Section~\ref{ss:simplemodels}).  Here we employ visual morphological classification for the solid black diamonds and show the results for the entire RSG subsample as the open squares.  Cluster \eightyfour\ does not appear in the top two panels (nor in the visual ETG sample in the bottom panel) because its CMR color indicates $z_f > 6$ ($z_{\rm end} > 4$).  In all cases, the inferred \rs\ formation epoch is later in lower-redshift clusters, suggesting a combination of ongoing accretion of younger stars by the massive cluster galaxies during this epoch and/or formation of younger cluster ETGs.  This extends the results found by, e.g., \citet{mei09} and \citet{jaffe11} that the formation redshift inferred by SSPs increases with redshift.   \label{fig:zforms}}
      \end{figure}

The trends of \S\ref{ss:colorscatters} imply a tight positive correlation between a cluster's observed redshift and its inferred star formation epoch.  Below $z \approx 1.3$, these cluster RSGs resemble a single burst with $\langle z_f \rangle \approx 2.0$ (or $z_{\rm end} \approx 1.5 $), while higher-redshift clusters have $\langle z_f \rangle \gtrsim 3.0$ ($z_{\rm end} \gtrsim 2.0$).  We plot these fitted values in Figure~\ref{fig:zforms}.   Error bars are calculated by fitting the endpoints of the one-sigma color and scatter ranges in the same manner.  These data are compiled in Table~\ref{tab:zforms}.  \emph{In the lower-redshift subsample, no clusters exist that are the evolutionary continuation of clusters at higher z along the \csp\ or \ssp\ tracks.}

\begin{figure}
	\begin{center}
	\begin{tabular}{cc}
	\includegraphics[width=3.3in]{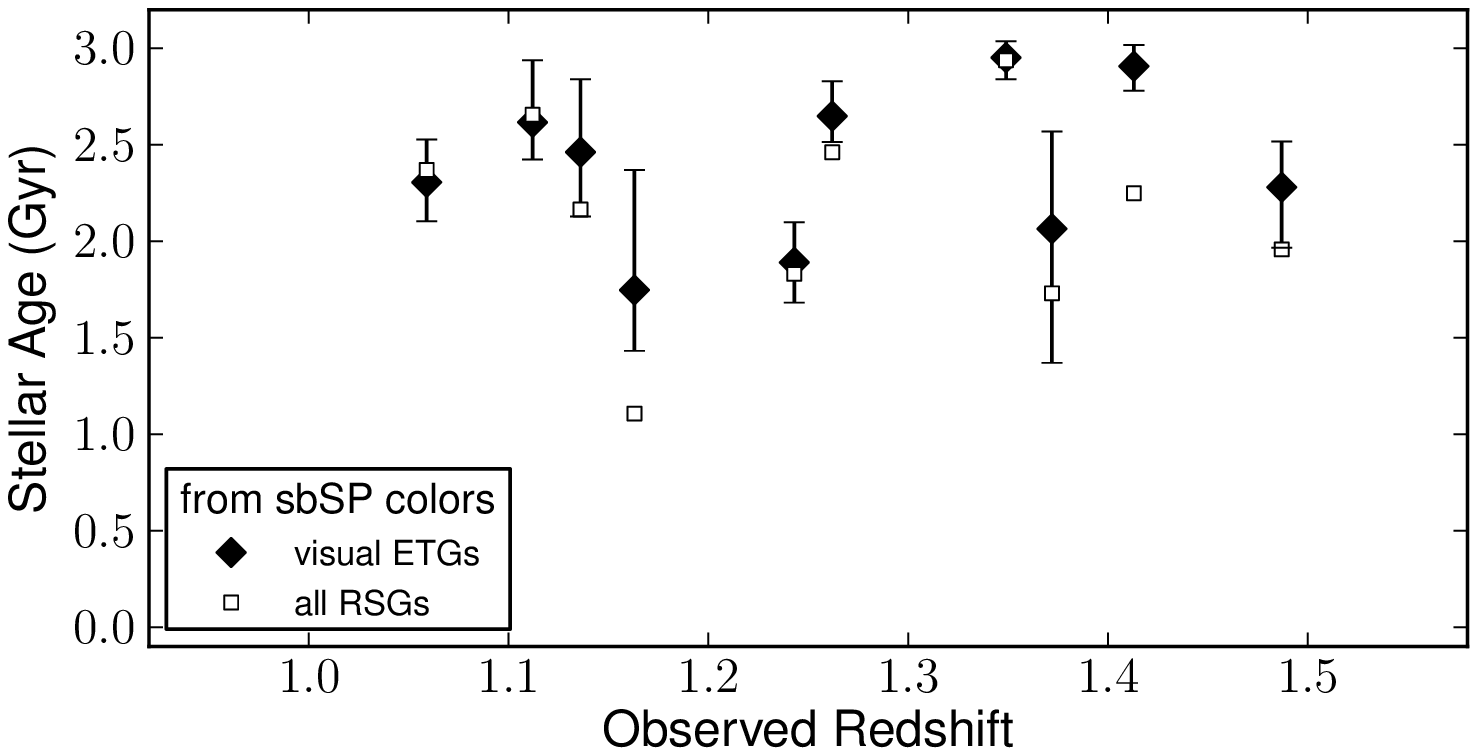} \\
	\includegraphics[width=3.3in]{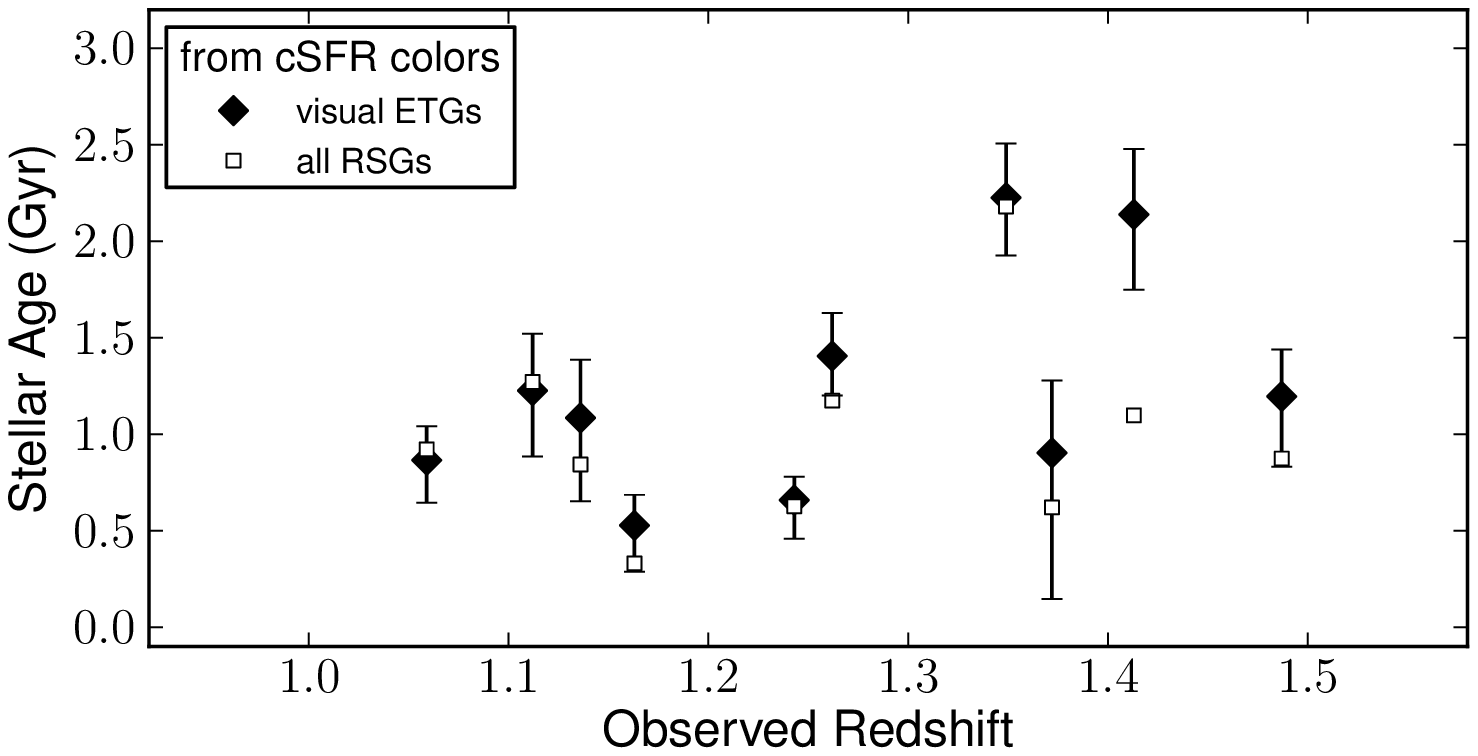} \\
	\includegraphics[width=3.3in]{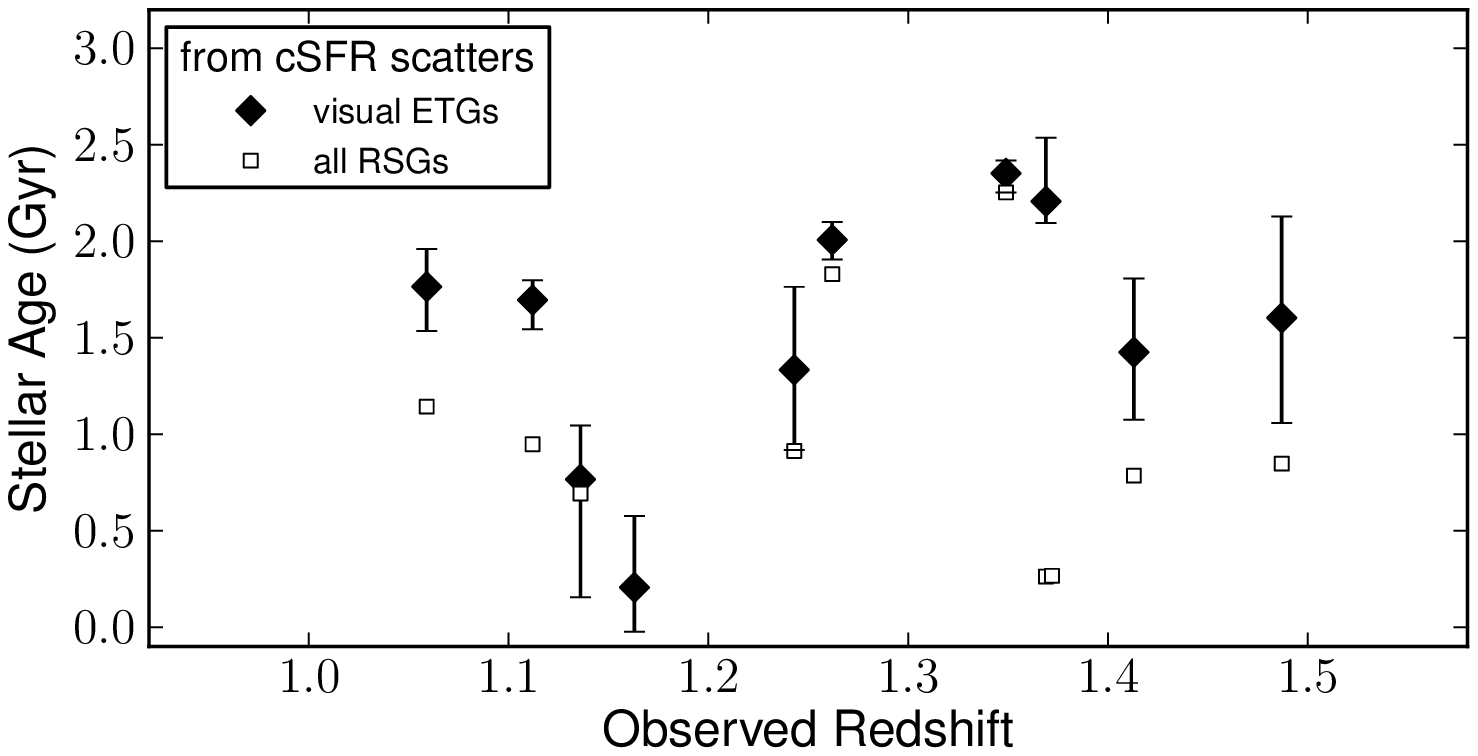} 
	\end{tabular}
	\end{center}
      \caption{ Inferred ages, corresponding to the same rows as Figure~\ref{fig:evolution}.  The values in the top, middle, and bottom figures are inferred from our \ssp\ color, \csp\ color, and \csp\ color scatter models, respectively (Section~\ref{ss:simplemodels}).  Here we employ visual morphological classification for the solid black diamonds and show the results for the entire RSG subsample as the open squares.    Cluster \eightyfour\ does not appear in the top two panels (nor in the visual ETG sample in the bottom panel) because its CMR color indicates an age $\gtrsim 4 \rm\ Gyr$.  \label{fig:ages}}
      \end{figure}

Figure~\ref{fig:ages} and Table~\ref{tab:ages} show the same information, but converted into stellar ages, which we interpret as the average time since the last major star-formation epoch for galaxies on the red sequence.  These trends are qualitatively similar regardless of the property-model combination applied, \ssp\ color, \csp\ color, or \csp\ scatter.  The scatter measurements in the bottom panel of Figures \ref{fig:zforms} and \ref{fig:ages} show the same general trend with redshift as the median color measurements of the top two panels: clusters in this sample at higher redshift indicate an earlier formation epoch.  Therefore the time elapsed since the last major star forming eposide appears roughly constant, yielding a luminosity-weighted average age of $\approx 2.3\pm0.6$ Gyr, consistent with the timescales inferred from rest-frame optical color scatters by \citet{meyers12} using the models of \citet{vandokkumfranx01} for seven of these clusters.

Numerically, the best-fit ages show a large scatter, $\sim 0.5\thru1$ Gyr, between when they are based on the scatter and when based on median color.  It may be that the simple star formation histories we assumed were inadequate.  For example, at a fixed \rs\ color (and hence ``age''), the color scatter between \rs\ members can be made arbitrarily small by increasing the extent to which their star formation histories are correlated.  Thus one possibility is that the formation histories of galaxies in some clusters are more uniform than others.

\begin{deluxetable*}{cc@{\hspace{0.2cm}}ccc@{\hspace{0.2cm}}ccc}
\tablecaption{Inferred CMR formation epochs. \label{tab:zforms}}
\tablecolumns{8}
\tablehead{
\colhead{} & \colhead{}                      \hspace{0.2cm} & \multicolumn{3}{c}{$\langle z_f \rangle$}                          \hspace{0.2cm}    & \multicolumn{3}{c}{$z_{\rm end}$}  \\
\colhead{Name} & \colhead{$z$}           \hspace{0.2cm} & \colhead{Vis. ETGs}& \colhead{Quant. ETGs}& \colhead{All RSGs}        \hspace{0.2cm}   & \colhead{Vis. ETGs}& \colhead{Quant. ETGs}& \colhead{All RSGs}
} 
\startdata
\cutinhead{From \ssp\ Colors}
      \sfiftyone & 1.059 & $2.10_{-0.14}^{+0.14}$ & $2.12_{-0.18}^{+0.17}$ & $2.15_{-0.15}^{+0.14}$ & \nodata & \nodata & \nodata  \\ 
     \sseventeen & 1.112 & $2.36_{-0.25}^{+0.15}$ & $2.28_{-0.27}^{+0.21}$ & $2.38_{-0.28}^{+0.15}$ & \nodata & \nodata & \nodata  \\ 
    \sthirtyfour & 1.136 & $2.24_{-0.24}^{+0.28}$ & $2.05_{-0.13}^{+0.20}$ & $2.04_{-0.12}^{+0.18}$ & \nodata & \nodata & \nodata  \\ 
      \sfourteen & 1.163 & $1.85_{-0.29}^{+0.19}$ & $1.54_{-0.04}^{+0.29}$ & $1.55_{-0.02}^{+0.23}$ & \nodata & \nodata & \nodata  \\ 
\sthreefourtytwo & 1.243 & $2.10_{-0.12}^{+0.14}$ & $1.89_{-0.22}^{+0.13}$ & $2.07_{-0.11}^{+0.11}$ & \nodata & \nodata & \nodata  \\ 
        \sthirty & 1.262 & $2.81_{-0.19}^{+0.15}$ & $2.50_{-0.15}^{+0.19}$ & $2.62_{-0.17}^{+0.17}$ & \nodata & \nodata & \nodata  \\ 
    \stwentynine & 1.349 & $3.57_{-0.14}^{+0.21}$ & $3.67_{-0.13}^{+0.25}$ & $3.55_{-0.10}^{+0.12}$ & \nodata & \nodata & \nodata  \\ 
    \seightyfour\tablenotemark{a} & 1.369 & $ > 6$ & $ > 6$ & $8.86_{-5.03}^{+0.91}$ & \nodata & \nodata & \nodata  \\ 
    \stwentyfive & 1.372 & $2.54_{-0.64}^{+0.88}$ & $2.71_{-0.39}^{+0.72}$ & $2.27_{-0.39}^{+0.21}$ & \nodata & \nodata & \nodata  \\ 
     \stwentytwo & 1.413 & $3.79_{-0.19}^{+0.23}$ & $3.56_{-0.27}^{+0.15}$ & $2.82_{-0.19}^{+0.30}$ & \nodata & \nodata & \nodata  \\ 
     \sthirtysix & 1.487 & $3.14_{-0.30}^{+0.49}$ & $2.84_{-0.22}^{+0.37}$ & $2.73_{-0.16}^{+0.20}$ & \nodata & \nodata & \nodata  \\
\cutinhead{From \csp\ Colors}
      \sfiftyone & 1.059 & $2.17_{-0.10}^{+0.16}$ & $2.19_{-0.15}^{+0.21}$ & $2.21_{-0.11}^{+0.19}$ & $1.39_{-0.06}^{+0.09}$ & $1.41_{-0.09}^{+0.11}$ & $1.42_{-0.07}^{+0.10}$ \\ 
     \sseventeen & 1.112 & $2.44_{-0.24}^{+0.25}$ & $2.33_{-0.20}^{+0.34}$ & $2.46_{-0.26}^{+0.31}$ & $1.56_{-0.14}^{+0.16}$ & $1.50_{-0.12}^{+0.20}$ & $1.57_{-0.15}^{+0.19}$ \\ 
    \sthirtyfour & 1.136 & $2.31_{-0.17}^{+0.32}$ & $2.17_{-0.10}^{+0.15}$ & $2.17_{-0.09}^{+0.13}$ & $1.49_{-0.11}^{+0.19}$ & $1.41_{-0.06}^{+0.09}$ & $1.41_{-0.05}^{+0.08}$ \\ 
      \sfourteen & 1.163 & $2.01_{-0.10}^{+0.17}$ & $1.88_{-0.08}^{+0.11}$ & $1.89_{-0.06}^{+0.09}$ & $1.33_{-0.06}^{+0.09}$ & $1.25_{-0.03}^{+0.07}$ & $1.26_{-0.03}^{+0.04}$ \\ 
\sthreefourtytwo & 1.243 & $2.22_{-0.08}^{+0.13}$ & $2.11_{-0.08}^{+0.07}$ & $2.21_{-0.08}^{+0.08}$ & $1.48_{-0.05}^{+0.07}$ & $1.40_{-0.04}^{+0.03}$ & $1.46_{-0.05}^{+0.06}$ \\ 
        \sthirty & 1.262 & $2.84_{-0.19}^{+0.20}$ & $2.53_{-0.10}^{+0.19}$ & $2.64_{-0.13}^{+0.18}$ & $1.84_{-0.11}^{+0.13}$ & $1.68_{-0.06}^{+0.10}$ & $1.73_{-0.07}^{+0.10}$ \\ 
    \stwentynine & 1.349 & $3.89_{-0.30}^{+0.45}$ & $4.16_{-0.34}^{+0.39}$ & $3.85_{-0.22}^{+0.30}$ & $2.62_{-0.25}^{+0.34}$ & $2.83_{-0.26}^{+0.33}$ & $2.59_{-0.18}^{+0.23}$ \\ 
    \seightyfour\tablenotemark{a}  & 1.369 & $> 6$ & $> 6$ & $8.87_{-4.50}^{+0.23}$ & $> 6$ & $> 6$ & $9.22_{-6.23}^{+0.55}$ \\ 
    \stwentyfive & 1.372 & $2.63_{-0.33}^{+0.79}$ & $2.75_{-0.32}^{+0.70}$ & $2.40_{-0.12}^{+0.19}$ & $1.76_{-0.19}^{+0.51}$ & $1.84_{-0.19}^{+0.46}$ & $1.62_{-0.07}^{+0.11}$ \\ 
     \stwentytwo & 1.413 & $3.98_{-0.39}^{+0.54}$ & $3.52_{-0.26}^{+0.29}$ & $2.82_{-0.13}^{+0.30}$ & $2.69_{-0.30}^{+0.42}$ & $2.34_{-0.17}^{+0.22}$ & $1.91_{-0.09}^{+0.17}$ \\ 
     \sthirtysix & 1.487 & $3.16_{-0.30}^{+0.36}$ & $2.87_{-0.17}^{+0.34}$ & $2.80_{-0.14}^{+0.14}$ & $2.13_{-0.18}^{+0.26}$ & $1.96_{-0.12}^{+0.21}$ & $1.90_{-0.09}^{+0.11}$ \\
\cutinhead{From \csp\ Scatters}
      \sfiftyone & 1.059 & $2.81_{-0.14}^{+0.17}$ & $2.44_{-0.23}^{+0.26}$ & $2.39_{-0.19}^{+0.18}$ & $1.77_{-0.09}^{+0.12}$ & $1.54_{-0.12}^{+0.17}$ & $1.51_{-0.10}^{+0.10}$ \\ 
     \sseventeen & 1.112 & $2.79_{-0.12}^{+0.12}$ & $2.27_{-0.49}^{+0.54}$ & $2.21_{-0.31}^{+0.43}$ & $1.77_{-0.07}^{+0.09}$ & $1.47_{-0.28}^{+0.32}$ & $1.43_{-0.19}^{+0.25}$ \\ 
    \sthirtyfour & 1.136 & $2.13_{-0.17}^{+0.41}$ & $2.27_{-0.21}^{+0.33}$ & $2.12_{-0.10}^{+0.19}$ & $1.38_{-0.10}^{+0.24}$ & $1.47_{-0.14}^{+0.19}$ & $1.37_{-0.05}^{+0.12}$ \\ 
      \sfourteen\tablenotemark{a}  & 1.163 & $1.73_{-0.55}^{+0.22}$ & $ < 2$ & $ < 2$ & $1.20_{-0.08}^{+0.09}$ & $ < 1.2$ & $< 1.2$ \\ 
\sthreefourtytwo & 1.243 & $2.75_{-0.34}^{+0.33}$ & $2.75_{-0.32}^{+0.28}$ & $2.41_{-0.12}^{+0.26}$ & $1.79_{-0.20}^{+0.20}$ & $1.79_{-0.19}^{+0.19}$ & $1.59_{-0.08}^{+0.16}$ \\ 
        \sthirty & 1.262 & $3.40_{-0.13}^{+0.08}$ & $3.39_{-0.15}^{+0.10}$ & $3.24_{-0.26}^{+0.14}$ & $2.23_{-0.08}^{+0.07}$ & $2.22_{-0.10}^{+0.08}$ & $2.12_{-0.18}^{+0.10}$ \\ 
    \stwentynine & 1.349 & $4.09_{-0.11}^{+0.12}$ & $4.03_{-0.15}^{+0.09}$ & $3.97_{-0.13}^{+0.12}$ & $2.76_{-0.09}^{+0.11}$ & $2.70_{-0.09}^{+0.09}$ & $2.67_{-0.09}^{+0.09}$ \\ 
    \seightyfour & 1.369 & $3.99_{-0.43}^{+0.16}$ & $4.16_{-0.34}^{+0.37}$ & $2.09_{-0.25}^{+0.17}$ & $2.68_{-0.31}^{+0.12}$ & $2.81_{-0.26}^{+0.33}$ & $1.46_{-0.35}^{+0.07}$ \\ 
    \stwentyfive\tablenotemark{a}  & 1.372 & $< 2$ & $< 2$ & $2.12_{-0.27}^{+0.19}$ & $< 1.5$ & $< 1.5$ & $1.47_{-0.36}^{+0.09}$ \\ 
     \stwentytwo & 1.413 & $3.20_{-0.39}^{+0.38}$ & $3.37_{-0.27}^{+0.42}$ & $2.61_{-0.12}^{+0.11}$ & $2.13_{-0.23}^{+0.25}$ & $2.25_{-0.20}^{+0.29}$ & $1.76_{-0.07}^{+0.07}$ \\ 
     \sthirtysix & 1.487 & $3.60_{-0.60}^{+0.77}$ & $3.81_{-0.73}^{+0.98}$ & $2.79_{-0.14}^{+0.14}$ & $2.43_{-0.39}^{+0.58}$ & $2.58_{-0.48}^{+0.77}$ & $1.89_{-0.08}^{+0.11}$
\enddata
\tablenotetext{a}{In some cases, fitting resulted in seemingly unphysical parameters --- e.g.\ galaxies formed earlier than our code permitted ($\sim 5\times10^8$ years) or have a negative age --- in these cases we give a one-sided estimate.  }
\end{deluxetable*}

\begin{deluxetable*}{cc@{\hspace{1.0cm}}ccc}
\tablecaption{Inferred CMR ages, in Gyr. \label{tab:ages}}
\tablecolumns{5}
\tablehead{\colhead{Name} & \colhead{z}           \hspace{1.0cm} & \colhead{Vis. ETGs}& \colhead{Quant. ETGs}& \colhead{All RSGs}   }
\startdata
&& \multicolumn{3}{c}{Stellar Age, $t_{\rm obs} - \langle t_f \rangle$, from \ssp\ Colors} \\
      \sfiftyone & 1.059 & $2.30_{-0.23}^{+0.21}$ & $2.33_{-0.29}^{+0.23}$ & $2.38_{-0.23}^{+0.19}$ \\ 
     \sseventeen & 1.112 & $2.66_{-0.34}^{+0.17}$ & $2.57_{-0.39}^{+0.25}$ & $2.69_{-0.37}^{+0.17}$ \\ 
    \sthirtyfour & 1.136 & $2.44_{-0.36}^{+0.34}$ & $2.16_{-0.21}^{+0.30}$ & $2.15_{-0.19}^{+0.26}$ \\ 
      \sfourteen & 1.163 & $1.73_{-0.61}^{+0.33}$ & $1.09_{-0.11}^{+0.61}$ & $1.10_{-0.06}^{+0.49}$ \\ 
\sthreefourtytwo & 1.243 & $1.89_{-0.19}^{+0.19}$ & $1.55_{-0.43}^{+0.22}$ & $1.83_{-0.17}^{+0.16}$ \\ 
        \sthirty & 1.262 & $2.66_{-0.18}^{+0.13}$ & $2.34_{-0.18}^{+0.20}$ & $2.47_{-0.19}^{+0.17}$ \\ 
    \stwentynine & 1.349 & $2.95_{-0.09}^{+0.11}$ & $3.01_{-0.08}^{+0.13}$ & $2.94_{-0.06}^{+0.07}$ \\ 
    \seightyfour\tablenotemark{a}  & 1.369 & $> 3$ & $> 3$ & $4.13_{-1.09}^{+0.07}$ \\ 
    \stwentyfive & 1.372 & $2.06_{-0.89}^{+0.74}$ & $2.24_{-0.43}^{+0.56}$ & $1.73_{-0.60}^{+0.26}$ \\ 
     \stwentytwo & 1.413 & $2.90_{-0.10}^{+0.11}$ & $2.77_{-0.17}^{+0.09}$ & $2.23_{-0.18}^{+0.25}$ \\ 
     \sthirtysix & 1.487 & $2.30_{-0.25}^{+0.32}$ & $2.06_{-0.21}^{+0.29}$ & $1.95_{-0.16}^{+0.18}$ \\ 
&& \multicolumn{3}{c}{Stellar Age, $t_{\rm obs} -  t_{\rm end} $, from \csp\ Colors} \\
      \sfiftyone & 1.059 & $0.86_{-0.18}^{+0.23}$ & $0.90_{-0.26}^{+0.29}$ & $0.93_{-0.20}^{+0.25}$ \\ 
     \sseventeen & 1.112 & $1.30_{-0.35}^{+0.35}$ & $1.16_{-0.32}^{+0.45}$ & $1.32_{-0.38}^{+0.40}$ \\ 
    \sthirtyfour & 1.136 & $1.05_{-0.28}^{+0.43}$ & $0.84_{-0.17}^{+0.23}$ & $0.83_{-0.15}^{+0.21}$ \\ 
      \sfourteen & 1.163 & $0.52_{-0.17}^{+0.25}$ & $0.30_{-0.10}^{+0.20}$ & $0.32_{-0.08}^{+0.13}$ \\ 
\sthreefourtytwo & 1.243 & $0.66_{-0.14}^{+0.17}$ & $0.47_{-0.10}^{+0.09}$ & $0.63_{-0.12}^{+0.14}$ \\ 
        \sthirty & 1.262 & $1.39_{-0.20}^{+0.22}$ & $1.07_{-0.14}^{+0.20}$ & $1.19_{-0.15}^{+0.19}$ \\ 
    \stwentynine & 1.349 & $2.21_{-0.28}^{+0.32}$ & $2.41_{-0.26}^{+0.27}$ & $2.18_{-0.21}^{+0.23}$ \\ 
    \seightyfour\tablenotemark{a}  & 1.369 & $> 3$ & $ > 3$ & $4.16_{-1.66}^{+0.04}$ \\ 
    \stwentyfive & 1.372 & $0.90_{-0.41}^{+0.83}$ & $1.06_{-0.39}^{+0.71}$ & $0.63_{-0.16}^{+0.22}$ \\ 
     \stwentytwo & 1.413 & $2.11_{-0.32}^{+0.36}$ & $1.72_{-0.23}^{+0.26}$ & $1.07_{-0.16}^{+0.28}$ \\ 
     \sthirtysix & 1.487 & $1.24_{-0.28}^{+0.34}$ & $0.96_{-0.21}^{+0.32}$ & $0.87_{-0.16}^{+0.18}$ \\ 
 && \multicolumn{3}{c}{Stellar Age, $t_{\rm obs} - t_{\rm end}$, from \csp\ Scatters} \\
      \sfiftyone & 1.059 & $1.75_{-0.18}^{+0.22}$ & $1.24_{-0.31}^{+0.38}$ & $1.18_{-0.25}^{+0.24}$  \\ 
     \sseventeen & 1.112 & $1.75_{-0.14}^{+0.16}$ & $1.07_{-0.82}^{+0.71}$ & $0.97_{-0.54}^{+0.60}$  \\ 
    \sthirtyfour & 1.136 & $0.76_{-0.29}^{+0.60}$ & $1.00_{-0.37}^{+0.45}$ & $0.74_{-0.16}^{+0.32}$  \\ 
      \sfourteen\tablenotemark{a}  & 1.163 & $0.12_{-0.28}^{+0.28}$ & $ < 0.1$ & $ < 0.1$  \\ 
\sthreefourtytwo & 1.243 & $1.36_{-0.43}^{+0.36}$ & $1.35_{-0.40}^{+0.33}$ & $0.93_{-0.18}^{+0.34}$  \\ 
        \sthirty & 1.262 & $2.01_{-0.11}^{+0.09}$ & $1.99_{-0.14}^{+0.11}$ & $1.85_{-0.29}^{+0.14}$  \\ 
    \stwentynine & 1.349 & $2.35_{-0.08}^{+0.10}$ & $2.29_{-0.09}^{+0.09}$ & $2.26_{-0.09}^{+0.09}$  \\ 
    \seightyfour & 1.369 & $2.21_{-0.34}^{+0.12}$ & $2.34_{-0.26}^{+0.27}$ & $0.25_{-1.06}^{+0.18}$  \\ 
    \stwentyfive\tablenotemark{a} & 1.372 & $< 0.1$ & $< 0.1$ & $0.27_{-1.08}^{+0.23}$  \\ 
     \stwentytwo & 1.413 & $1.42_{-0.38}^{+0.34}$ & $1.60_{-0.28}^{+0.35}$ & $0.80_{-0.14}^{+0.13}$  \\ 
     \sthirtysix & 1.487 & $1.64_{-0.53}^{+0.57}$ & $1.80_{-0.61}^{+0.65}$ & $0.85_{-0.16}^{+0.18}$
\enddata
\tablenotetext{a}{In some cases, fitting resulted in seemingly unphysical parameters --- e.g.\ galaxies formed earlier than our code permitted ($\sim 5\times10^8$ years) or have a negative age --- in these cases we give a one-sided estimate.  }
\end{deluxetable*}


\subsection{Qualitative Discussion of CMR Assumption} \label{ss:cmrassumption} \label{ss:outliers}

Several additional features pertaining to cluster galaxy evolution are seen in these data.  Specifically, since the cluster selection technique does not depend directly on the presence of a canonical red sequence, these CMRs have a wide variety of properties.  

In nearly all cases, the cluster red sequence spans a factor of at least $\sim 10$ in brightness, indicating that cluster ETGs spanning the range $L \approx 0.3 \thru 3\ L^*$ are present by $z \sim 1.5$.  On the other hand, we measure several clusters with very tenuous or wide RSs.  \fourteen, \twentyfive, \eightyfour, and \thirtysix\ all contain an over-density of red galaxies that are near the expected \rs\ color but do not define the canonical locus in color-magnitude space consistent with a single CMR of old galaxies.

Moreover, clusters \fiftyone, \twentynine, and \threefourtytwo\ appear to have red sequences where the \rs\ is deficient in ETGs at $L < 0.3\ L^*$.  Furthermore, \threefourtytwo\ and \twentytwo\ have $5\thru10$ RSGs at $L < L^*$ that are bluer by up to $\sim 0.4$ magnitudes than the most apparent CMR, comprising $\approx 30\%$ of the \rs\ galaxies.  We also notice a large number of early-type galaxies that are as bright and brighter than the \rs\ we identified for \twentytwo, and that are $\sim 1$ magnitude bluer.  However, they are likely interlopers based on their projected spatial offset ($> 0.5\ \rm Mpc$) from the cluster center and association with a known foreground structure.  

Taken together with the quantitative results described above, these qualitative properties outline a picture whereby some massive cluster ETGs (ellipticals and S0s) are in place in many clusters, but perhaps not all, by $z \approx 1.5$, with stellar populations formed at $\langle z_f \rangle > 3$.  At $z \sim 1$, the stellar populations of the ETGs in this sample appear more recently formed, on average by $\approx 1 \rm\ Gyr$, having $\langle z_f \rangle \sim 2$.  


\section{Discussion} \label{s:discussion}

We analyzed the rest-frame optical CMRs of \rs\ galaxies in 11 spectroscopically confirmed, infrared-selected massive galaxy clusters at $1.0 < z < 1.5$.  Clustering \citep{brodwin07}, extended X-ray emission \citep{brodwin11xray}, weak lensing \citep{jee11}, and stellar masses \citep{eisenhardt08} support the view that these are massive ($M \gtrsim 10^{14} M_{\odot}$) clusters and are the precursors to present-day massive ($M \sim 10^{15} M_{\odot}$) clusters.  These observations thus provide a valuable window into the evolution of galaxies within the most massive halos in the universe by revealing the time during which most of their stellar mass formed and assembled from smaller pieces.

\subsection{Star Formation Histories of Cluster Ellipticals}

The present data (Figure~\ref{fig:zforms}) indicate that massive, red ETGs existed in the centers of many rich clusters by at least $z = 1.5$.  Other observations suggest this is true at even higher redshifts, both in massive, mature clusters \citep{stanford12} and assembling structures or ``proto-clusters'' \citep[e.g.,][]{kurk09}.  The red colors of these galaxies at $z \approx 1.5$ indicate that their luminosity-weighted age is $\approx 2\thru3\ \rm Gyr$ (a formation epoch of $z \sim 4$), and/or that the formation of stars then present ceased by $\approx 1\ \rm Gyr$ prior to their observed state ($z \sim 2.5$).  

This story is reminiscent of a ``monolithic collapse'' \citep{eggen62}, whereby the stars in these galaxies formed in a single short burst at a much earlier time.  Massive cluster ETGs at $z \lesssim 1$ evolve in a manner consistent with this passively-evolving model \citep{bower92, ellis97, sed98, mancone10}, although their rest-frame optical colors are consistent with a wide variety of potential formation epochs (e.g., the first panel of Figure~\ref{fig:evolution}).

The analysis presented here connects the low-redshift cluster galaxy population to their pasts by quantifying the stellar populations in cluster galaxies over a span of $1.5\ \rm Gyr$ following the end of the peak star formation activity in the universe at $z \sim 2$.  The measurements of Section~\ref{s:formation} suggest that the characteristic ages of the total stellar populations in massive RSGs are roughly the same across this cluster sample at $1 < z < 1.5$ (Figure~\ref{fig:ages}).  This manifests as an inferred star formation epoch $z_f$ that continues to increase with the observed cluster redshift, as found previously in samples presented by, e.g., \citet{mei09} and \citet{jaffe11}.  

This is explained naturally if they experience significant stellar mass assembly during this epoch, a scenario we discuss in Sections~\ref{ss:luminosity}--\ref{ss:toymodels} below.  Alternatively, these observations could reflect drastic changes during this epoch in the nature of the galaxies/overdensities selected for this study, a possibility we consider in Sections~\ref{ss:mass}--\ref{ss:dust}.  

\subsection{The Impact of Cluster Mass} \label{ss:mass}

One factor that might confuse our analysis is if the nature of the clusters in the ISCS sample changes significantly over the redshift range we consider, or if they are fundamentally different structures than the progenitors of today's massive clusters, as our interpretation assumes.  The correlation function of ISCS cluster candidates \citep{brodwin07} implies that they are predominately $M > 10^{14} M_{\odot}$ structures and likely the precursors to today's massive ($\sim 10^{15}M_{\odot}$) clusters.  Furthermore, for the spectroscopically confirmed subsample considered herein, clusters with sufficiently-deep X-ray observations \citep{brodwin11xray} are detected at a significance consistent with their being $\gtrsim 10^{14} M_{\odot}$ objects \citep[see also,][]{stanford12}.  Finally, shear measurements by \citet{jee11} for {\it all} of these clusters with very deep imaging from the \hst\ Cluster Supernova Survey \citep[PI Perlmutter, GO-10496; see][]{dawson09} yield individual (i.e., not stacked) weak--lensing masses comfortably above $10^{14} M_{\odot}$.

This does not preclude our spectroscopic subsample from containing a small number of structures that are somehow different than the remainder of the sample (for instance see \eightyfour\ and \oneohthree).  Furthermore, the expected rapid assembly of cluster-mass halos at these redshifts implies significant changes to the cluster galaxy population resulting from the incorporation of external groups via mergers, changes that might occur stochastically and lead to the outliers we observe in particular cases.  Therefore the stellar-mass selection employed for this sample is effective at selecting cluster-size objects whose galaxy populations are not required to be fully evolved, permitting the analysis of a more diverse set of evolutionary paths in massive clusters.  The fact that these clusters, which have similar stellar masses owing to their selection, exhibit similar colors implies that they had similar rates of star formation in their recent pasts.  This matches our conclusion that massive cluster galaxies star formation over an extended period at $1 < z < 1.5$.

\subsection{The Effect of Reddening} \label{ss:dust}

Significant dust attenuation in cluster galaxies would bias our measurements toward a redder color, although we have no reason to believe that these ETGs contain an unusual amount of dust.  If we posit that galaxies follow an evolutionary path from star-forming disks, through quenching, and ultimately to red-and-dead galaxies, then it is not clear that ETGs at higher redshift will have significantly more dust obscuration than those at $z \ll 1$.  \citet{meyers12} analyzed the ETGs in a set of clusters that includes seven members of the ISCS, and found that these galaxies, some of which are on the \rss\ in the present work, likely experience little dust reddening: $E(B-V) \lesssim 0.06$.  

If cluster ETGs at $z\gtrsim 1.3$ are more highly obscured than their lower-redshift counterparts, but reflect the same general star formation histories (suggesting $z_f \approx 2$), then cluster galaxies would appear redder \citep{eisenhardt08} and fainter \citep{mancone10}.  Line emission may also bias the present measurements to the red, since the WFC3/F160W filter includes the $H\alpha$ line for clusters at $1.14 < z < 1.55$.  However, WFC3/G141 spectra (Zeimann et al.\ in prep.) indicate weak $H\alpha$ emission in most of these RSGs.  Furthermore, if either of these effects cause the red colors, we would expect significant variation between objects, and so the red sequence color scatters would systematically \emph{increase} with redshift, contrary to the constant or decreasing scatters we found in Section~\ref{ss:epochs}.  

This does not mean that dust plays no role at these redshifts.  At least one cluster's CMD (\threefourtytwo) contains a spectroscopically confirmed member \citep{wagg12_dog} satisfying the dust-obscured galaxy (DOG) criteria of \citet{dey08}.   In addition, some objects (spectroscopic members or otherwise) are very red members of the clusters' red sequences.  If this reflects contamination of the \rss\ by dusty galaxies then our colors and/or color scatters may be biased high in some cases, an effect we expect to get larger at higher redshifts owing to the increasing presence of dusty starbursts.  This may also contribute to several of the outliers mentioned in Section~\ref{ss:cmrassumption}.

Importantly, if dust or line emission do play a role in contaminating the \rss\ or reddening cluster ETGs at $z\sim1.5$, then it can only strengthen our conclusions that we are beginning to witness a period during which significant star formation activity recently occurred in massive central cluster galaxies.  

\subsection{Luminosity Evolution} \label{ss:luminosity}

Star formation over an extended period implies that cluster \rss\ should be growing in stellar mass, and this implies evolution in the galaxy luminosity function shape or normalization, scenarios that have been explored in the ISCS by exploiting the rest-frame near-IR data obtained with \spitzer.  \citet{mancone10} found that above $z \sim 1.3$, cluster galaxies are $\approx 0.5$ magnitudes fainter at $3.6$ and $4.5\mu m$ than one would expect from the extrapolation of their lower-redshift evolution, positing significant ongoing assembly at this epoch, leading to a rapid increase in luminosity with time.  This conclusion, as well as the one consistent with the present work by \citet{eisenhardt08}, is based on a sample of cluster galaxies that does not apply a color or morphological selection, lending more evidence to the notion that stellar mass must be actively increasing in cluster member progenitors.  Rather than the entirety of such evolution being caused by progenitor bias, we are likely witnessing the effect of real growth in the stellar mass of some cluster galaxies.  

The results by \citet{mancone10} and \citet{eisenhardt08} were based on photometric cluster members within $1.5\rm\ and\ 1.0\ projected\ Mpc$, respectively, of the cluster candidate's center.  Like we posit for the present work, those studies may track a galaxy sample that misses stellar mass that formed outside of these regions, or in galaxies not bright enough to be selected.  Thus we suggest the luminosity growth in massive cluster galaxies at these redshifts is due in part to star formation that occurred over an extended period, rather than arising entirely from the merging of equally old progenitors.  

\subsection{Red Sequence Growth}  \label{ss:growth}

Then what is the star formation and assembly history of massive cluster galaxies?  The data in Figure~\ref{fig:evolution} are consistent with each cluster's ETGs forming at a different time, as the \ssp\ models with $z_f = 2\thru6$ roughly bracket the observed colors.  We might reasonably expect that the star formation epoch of the massive galaxies in one cluster differs from another based on the assembly history or halo mass, and that differences in this history are small enough to have damped out by $z \lesssim 1$.  

However, cluster galaxies in the present sample at $1.0 < z < 1.3$ tend to be bluer than those implied by the continued passive evolution of the clusters at $z > 1.3$:  any \ssp\ ($z_f \gtrsim 3$) that fits the high-redshift clusters is too red for the lower-redshift ones, suggesting that the colors of massive cluster galaxies might evolve in a manner altogether different from the commonly-used simple \ssp\ and time-uniform composite models considered in Section~\ref{ss:simplemodels}.  Such evolution has been observed for some time \citep{bell04,brown07} in surveys of massive field galaxies.  Additional significant growth of the massive red galaxy population occurs at $z>1$ \citep{kriek08, vandokkum08, taylor09,whitaker10}, and this may be likely true in all environments.  We therefore consider the possibility that our cluster sample selects, at least in part, a continuous evolutionary sequence that experiences star formation over an extended period leading to an increase in stellar mass on the red sequence.

Importantly, for the ETGs to reside on the red sequence and maintain a nearly-constant luminosity-weighted age as we observe, the \rs\ must accrete newer stars in some fashion, making the CMR bluer than one which evolves passively.  Moreover, the newer stars or galaxies must have ceased forming some time ago ($\gtrsim 100$ Myr) in order to be considered RSGs, and must comprise a body of stellar mass that was experiencing an \emph{increasing} star formation rate prior to its quenching.  Otherwise the observed luminosity-weighted age would increase with time.  Thus, these stars likely formed in gas-rich, rapidly star-forming objects that are not found in large numbers on the red sequence in the centers of their progenitor clusters.  In this picture, the halting of star formation in massive cluster ETGs does not occur suddenly and absolutely in every galaxy of the final cluster, but continuously, during or after periods of significant gas accretion in objects that eventually make up the final, present-day ETGs.  

With this interpretation, the CMRs of these clusters may reflect in part the ongoing growth of galaxies in the cluster centers as they accrete or form younger stars not present at a previous time, as well as the transformation from star-forming to quiescent galaxies. The latter is the canonical in-situ progenitor bias of \citet{vandokkumfranx01}, who described this effect at $z < 1$ for satellite galaxies, which exhibit remarkable constancy in their optical colors and color scatters over time.  We now see this same result in massive cluster ETGs (Figure~\ref{fig:evolution}), and we showed in Section~\ref{ss:progenitorbias} that the transformations of late-type RSGs into ETGs are not sufficient to account for the entirety of the evolution we observe in the CMR formation epochs.  Therefore another form of progenitor bias likely contributes, in which the red galaxy samples in the cluster centers gain stellar mass over time from outside the RSG sample in order to achieve roughly constant colors and scatters.

Contrast this result with a mass-selected field galaxy sample at comparable redshifts such as that considered by \citet{whitaker10}, who found that the color scatter of field RSGs increases rapidly from $z=1$ to $z=1.5$.  This likely owes to the way our sample of massive cluster galaxies were selected: we require them to exist in the center of a cluster, the bottom of a deep potential well that continuously collects significant amounts of stellar mass.    We see only those galaxies and stars present there at the time we observe them, but may miss a significant amount of stellar mass that formed in massive galaxies \emph{outside} the central ETGs at $z \lesssim 2\thru3$.  

\subsection{Models of Ongoing Assembly}  \label{ss:assemblymodels}

We now seek an understanding of the impact of stellar mass assembly on the appearance of cluster cores over this redshift range.  Previous work \citep[e.g.,][]{vandokkumfranx01} highlighted the effect of ``progenitor bias'', whereby galaxies identified as ETGs or RSGs at one redshift are not selected in comparable higher or lower redshift samples.  Potential causes of this effect include the continued formation or growth of these galaxies, bursts of star formation in RSGs, and/or the transformation of star-forming disk galaxies into RSGs/ETGs via merging or other processes.  Therefore we expect this and similar effects to be apparent in this sample.  

As galaxies form stars, possibly merge, and eventually experience quenching as they fall into the center of the assembling halo, significant amounts of time may pass before they become incorporated into what we see as galaxies in the cluster center.  Also, massive galaxies are thought to accrete a significant fraction of their stellar mass from their environment, through numerous minor mergers or smooth accretion \citep[e.g.,][]{keres05,white07,dekel09,oser10}.  In this way massive galaxies experience assembly of structures that alter their average stellar population with time.  If the stellar ages of the new material are different enough from those already assembled, then arbitrary paths through the space of inferred ages (through a simple proxy such as color) are possible, depending on the precise history of assembly.

In any case, the salient result of this structure growth is that the cluster cores grow in stellar mass density, a generic process for which we seek an intuitive description in terms of the quantities we measure here.  A complete specification of the different star formation histories in massive structures is beyond the scope this work, yet simple models can demonstrate the key characteristics expected of galaxies that evolve in this fashion.  For this purpose we apply three such models inspired by \citet{vandokkumfranx01} that differ only in the time delay between when a galaxy's stars are formed and when they are incorporated into the ETG sample.  Model galaxies that have formed within $\Delta t$ of the observation time are masked from the corresponding CMR model, mimicking the time before a galaxy would be identified as an ETG or RSG in the center of a cluster.  This procedure has been used previously \citep[for example,][]{jaffe11} to analyze the evolution of cluster galaxies using their \rs\ colors and scatters.  

This delay, and its time evolution, could represent various physical scenarios, including galaxies that experience a small burst that removes them from the \rs, galaxies that have not yet entered the center of the cluster, and galaxies that are not bright enough to be detected in the sample.  We do not apply any color cuts on these model CMRs: all model galaxies are assumed to be on the \rs.  If a particular choice of $\Delta t$ leads to very blue galaxies, then the CMR model would be relatively blue on average.  We could test an infinite number of choices for $\Delta t$, but our data can only reasonably constrain a small number of representative scenarios.  Therefore we will test three such toy models in the hope of gaining some additional physical insight into the formation of massive cluster galaxies.  

For the first two models we choose to demonstrate a constant $\Delta t$ of  $1.0$ and $2.3$ Gyr, roughly consistent with the $t_{\rm delay} \sim 2\rm\ Gyr$ value found by \citet{meyers12} after utilizing models by \citet{vandokkumfranx01} to the optical color scatter in a subset of seven ISCS clusters.  Our third model, ``sudden infall'', corresponds to sudden mass growth by assuming that $\Delta t$ is a function of redshift, with a minimum of $200 \rm\ Myr$ occurring at $z=1.2$ but steeply rising to $2.3 \rm\ Gyr$ before and after.  Because galaxies that are younger than the existing RSGs are being added in large numbers to the CMR at $1.2 \lesssim z \lesssim 1.8$, this corresponds to a model \rs\ that becomes bluer with time, where the amount of color variation depends on the suddenness of the assembly, the minimum delay time, as well as the closeness of the infall time to the time of peak SFR in the galaxies that eventually make up the cluster.  In this case, because we track an estimate of the universal SFR, this peak occurs at $z \sim 2\thru3$.   

To implement these models, we assume that each galaxy is formed in a single short burst and enters the cluster ETG sample in a first-made, first-in fashion exactly $\Delta t$ after its formation.  We simulate $10^5$ \ssp\ tracks that form at a rate such that the final red sequence experienced the same star formation history in all three cases.  To reduce the number of free parameters, we use models with the universal star formation history estimated by \citet{hopkinsbeacom06}.  For simplicity we simulate only the galaxy color distribution and not the entire CMR.  

These models have the beneficial features that a) the colors of the galaxy populations in higher redshift clusters are redder than one would expect given a monolithic collapse scenario with a single formation redshift, b) these colors are consistent with an earlier inferred star-formation epoch without necessarily leaving a redder descendent, and c) the final model ``red sequences'' have identical stellar populations, regardless of the details of their assembly history.  Therefore they only reflect the consequences of (highly idealized) structure growth and do not depend on specific assumptions for the physics of star formation quenching in massive halos.  

\subsection{Colors and Scatters of Assembling Red Sequences}  \label{ss:toymodels} \label{ss:discmodels}

\begin{figure*}
	\begin{center}
	\begin{tabular}{c} 
	\includegraphics[width=6in]{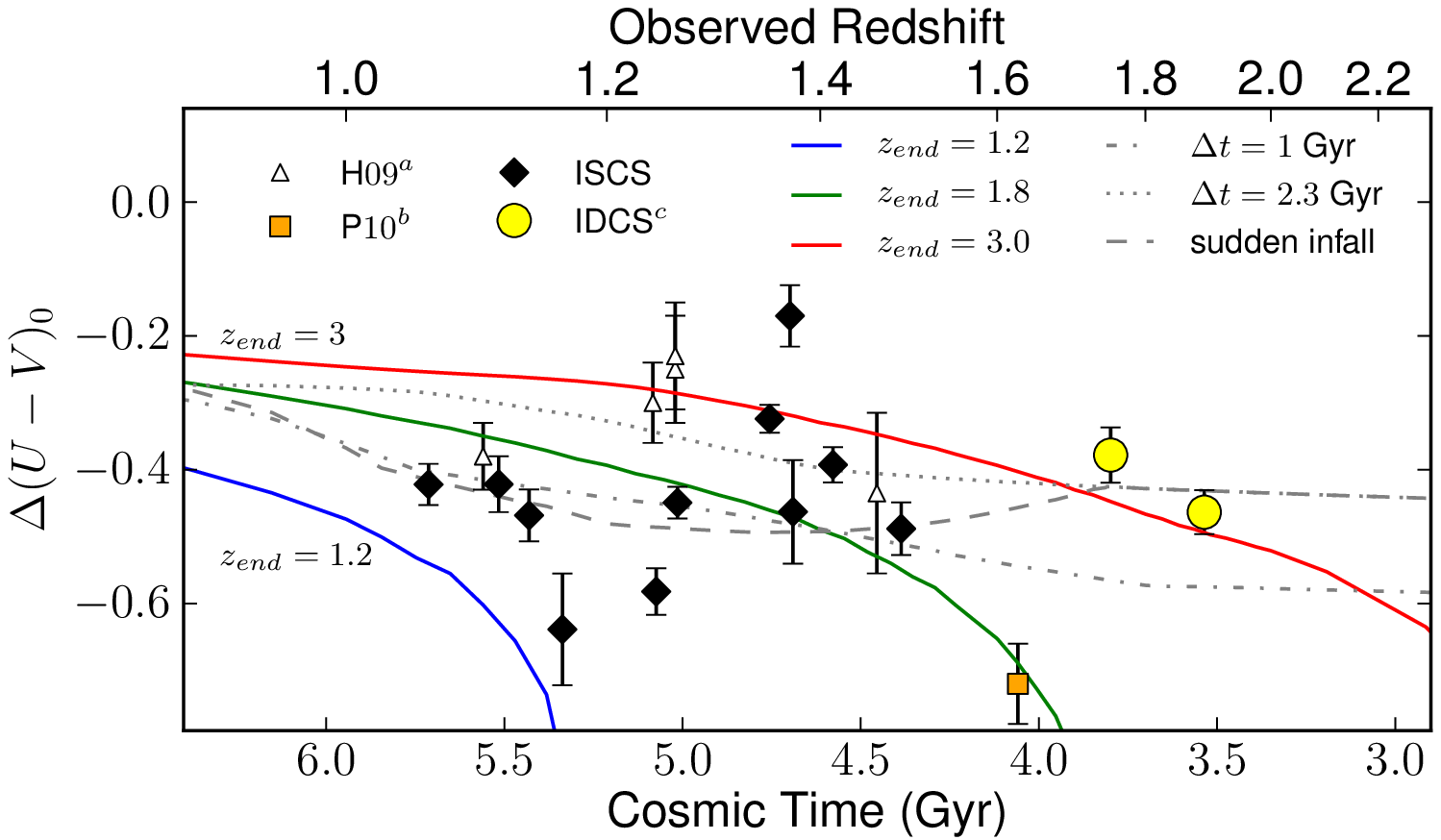} \\
	\includegraphics[width=6in]{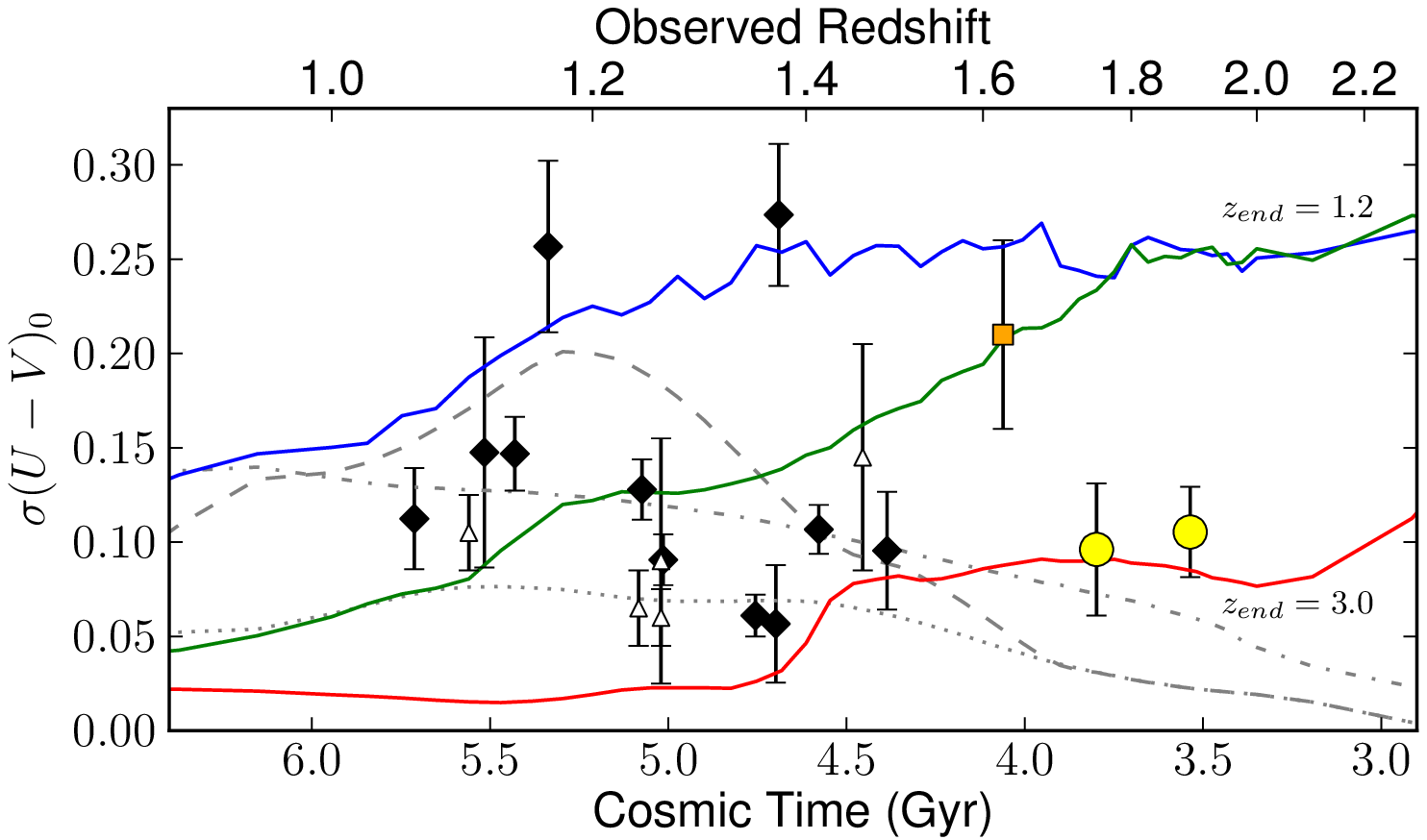}
	\end{tabular}
	\end{center}
         \caption{Top: Evolution of the rest-frame \textit{U-V} colors of $z > 1$ cluster galaxies, relative to Coma.   Bottom: Rest-frame \textit{U-V} scatters of cluster galaxies.  We plot cluster data from the present work, and data compiled by \citet{hilton09}$^{a}$, including clusters from \citet{blakeslee03} and \citet{mei06a,mei06}.  We also include clusters CIG J0218.3-0510 \citep{papovich10}$^b$, and two new clusters from the IRAC Deep Cluster Survey$^c$, IDCS 1426+3508 \citep[$z=1.75$;][]{stanford12} and IDCS 1433+3306 \citep[$z=1.89$;][]{zeimann12_biden}.   The curves are different toy models for the stellar populations in massive ETGs at the centers of clusters.  The solid red, green, and blue curves are the \csp\ models corresponding to $z_{\rm end}=3$, $z_{\rm end}=1.8$, and $z_{\rm end}=1.2$, which we described in Section~\ref{ss:models}.  We show the simple alternatives of Section~\ref{ss:assemblymodels} in the gray curves: these models experience the same overall star formation histories \citep{hopkinsbeacom06}, but differ in the time delay between the formation of stars in a galaxy and the time at which the galaxy becomes apparent in the cluster red sequence.  \label{fig:delta}}
      \end{figure*}

We plot the color and color scatter evolution of these assembly-inspired models as the gray curves in Figure \ref{fig:delta}.  In the top panel we summarize the intrinsic U-V colors of our sample and compare it to data from the literature on clusters at $1.0 < z < 1.9$.  In this panel, colors are computed relative to the Coma cluster and denoted $\Delta (U-V)_0$. The color scatters of cluster ETGs (Figure~\ref{fig:delta}, bottom) agree largely with inferences made from their colors only, as seen in Figures \ref{fig:evolution}, \ref{fig:zforms}, and \ref{fig:morphcompare}.  

%

As seen in previous work, the immediate result of the delay between the galaxies' formation and their settling on the red sequence is that cluster \rss\ continue to remain redder and narrower than any single \ssp\ or \csp\ model implies, even while the \rs\ is sparse or tenuous as observed in some cases (Section~\ref{ss:cmrassumption}), presumably before they are fully formed.  

The feature that inferred ages appear more or less constant at $z>1$ is a natural consequence of the stars entering high-redshift ETG samples having been drawn from a history with an \emph{increasing} star formation rate.  This scenario was observed in distant galaxies ($z \gtrsim 3$) by, for example, \citet{lee11_sfh} and \citet{papovich11_sfh}, suggesting that the galaxies in assembling clusters may be imprinted with this rising-SFR signature from their gas accretion epoch.  In that case, for a given delay $\Delta t$, the dominant stellar population in the optical at any time resembles an \ssp\ formed a time $\Delta t$ ago.  Our third such model experiences a rapid decrease in the SF-assembly delay at $z=1.2$, possibly mimicking the result of a dramatic enhancement to star formation in galaxies falling into the cluster, such as a merger with a star-forming group or a temporary increase in smooth gas accretion.  This demonstrates one plausible effect that can make assembling CMRs bluer (or wider) for a time (cf.\ cluster \fourteen), which may be necessary or expected during this epoch.  We do not claim that our data are yet able to constrain this type of model in detail, but these models do provide some evidence that such scenarios, and perhaps combinations thereof, might naturally occur in a sample like the ISCS.

The color scatter measurements provide an additional constraint, because they reflect the extent to which the star formation episodes of the cluster ETGs (at fixed color) were correlated.  We demonstrate specific variants of this in the bottom panel of Figure~\ref{fig:delta}; the assembly-inspired models can yield the necessary slowly-varying (or increasing) color scatter as time progresses for clusters at $z  > 1$.  In these models, each ``galaxy'' is an uncorrelated \ssp\ formed at a random time such that the correct average SFH is borne out in a large sample.  Therefore the correlation between model ETG star formation episodes reflects only the correlation implied by the underlying SFH (e.g., when it peaks, galaxies forming at that time will have correlated SFHs), and so any such scatter models could be considered as upper limits.  

The model tracks we show in this work are $\approx0.05$ magnitudes bluer than clusters at $z < 1$, reflecting either a possible systematic offset in the photometric systems between the model and sample (here we merely quote the numeric values from the cited works), that cluster galaxies experience an earlier peak in their star formation than the SFH we assumed, or that massive cluster ETGs at low redshift have a slightly higher than solar metallicity.  

While order-unity departures from roughly solar metallicity in cluster ETG stellar populations are inconsistent with their evolution at $z < 1$, the metallicity evolution in cluster ETG progenitors at $z \gg 1$ is unclear.  The color and scatter trends we have observed here, at a fixed formation epoch ($z_f\sim3$) using our \ssp\ or \csp\ models, would imply that the luminosity-weighted average stellar metallicities decrease by a factor of $\sim 5$ during approximately $1 \text{ Gyr}$ at $1 < z < 1.5$.  This by itself is perhaps unphysical, but if it does contribute, it again implies stellar mass growth, in this idealized case owing to the incorporation of relatively low-mass (low-metallicity) galaxies.  Like our toy models at fixed metallicity but evolving average age, this scenario would suggest significant ongoing star formation near cluster cores, and assembly into cluster cores, over an extended period at $z \sim 1.5$.  In reality both effects might contribute to the observed trends at this epoch, when assembling clusters contain massive galaxies (high-Z, early $z_f$), but grow their \rss\ rapidly by incorporating or growing lower-mass (lower-Z, late $z_f$) galaxies \citep{raichoor11}.

In this work, we have not separated the elliptical versus S0 galaxy samples, and therefore the apparent evolution of the ETG population may owe to an increasing population of S0 galaxies (and not necessarily ellipticals) with time, extending the results of, e.g., \citet{poggianti09}.  Subsequent analyses will determine the evolution of the cluster galaxies' morphologies.  

\section{Conclusions} \label{s:conclusions}

We studied sensitive, high-resolution, rest-frame optical imaging of the central regions in 13 spectroscopically confirmed infrared-selected galaxy clusters at $1 < z < 1.5$ that are plausible progenitors of today's massive clusters with $M \sim 10^{15} M_{\odot}$.  In 11 of these clusters we measured the CMR of red galaxies at a color consistent with the cluster's known redshift, and subsequently we analyzed the stellar populations implied by the location and dispersion of the red sequence.  We find the following:

\begin{enumerate}
\item{ Red sequences of bright galaxies are present in the centers of many clusters by $z=1.5$.   However, the properties of their CMRs indicate large cluster-to-cluster variations in the formation histories of cluster galaxies observed at a given redshift.}
\item{ As a function of redshift from $z=1.5\thru1.0$, the median rest-frame $U-V$ color of cluster galaxies in this sample is nearly constant at $(U-V)_0 \sim 1.0$, and the scatter about the median color is nearly constant at $\sigma_{\rm int} \sim 0.1 \text{ magnitudes}$.  Some clusters deviate widely from these trends, and outliers may represent contamination by dust-obscured galaxies, the evolution of galaxies through the ``green valley'', or red sequences that are not yet fully formed.} 
\item{The flatness of the color and scatter trends at $1 < z < 1.5$ requires star formation histories more diverse than a single short burst in the past. These average trends are plausibly in part the result of progenitor bias, where the progenitors of lower-redshift cluster ETGs are excluded by the sample selection at higher redshifts. }
\item{ Across this redshift range, these average colors and scatters both imply a roughly constant luminosity-weighted stellar age in the cluster RSGs, where the time since the bulk of stars formed is roughly $2\thru2.5$ Gyr, and the time since star formation halted is roughly $1\thru2$ Gyr.}
\item{A constant stellar age is a natural consequence of observing the center of an assembling halo whose final stellar population is experiencing ongoing or increasing star formation at the observed redshift.  If the red sequence stars are nearly completely in place, they would reflect a small continuing star formation rate and their luminosity-weighted ages should increase with time.  By contrast, at $z\sim1.5$ our observations imply that significant current star formation is occuring that will contribute to the stellar populations of central cluster galaxies by $z\sim1$.}
\item{Assuming the star formation of cluster galaxies follows that of other systems, the color and scatter trends we find are consistent with a factor of two growth in the cluster red sequences from $z=1.5\thru1.0$.}  
\item{ These observations lend credence to the idea that the red sequence in the centers of clusters was growing rapidly at $z\sim1.5$, where this mass assembly was likely in the form of relatively younger, smaller, and/or metal-poorer galaxies that were actively growing in their recent past. }

\end{enumerate}

\acknowledgements

This work is based on observations made with the NASA/ESA \emph{Hubble Space Telescope}, obtained at the Space Telescope Science Institute (STScI), which is operated by the Association of Universities for Research in Astronomy, Inc., (AURA) under NASA contract NAS 5-26555. These observations are associated with programs 10496, 11002, 11597, and 11663.  Support for these programs was provided by NASA through grants from the STScI, which is operated by AURA under NASA contract NAS 5-26555.  This work is also based in part on observations made with the \emph{Spitzer Space Telescope}, which is operated by the Jet Propulsion Laboratory, California Institute of Technology under a contract with NASA.  This work is also based in part on data obtained at the W.M. Keck Observatory, which is operated as a scientific partnership among the California Institute of Technology, the University of California and NASA. The Observatory was made possible by the generous financial support of the W.M. Keck Foundation.  This work makes use of image data from the NOAO Deep Wide-Field Survey (NDWFS) as distributed by the NOAO Science Archive.  The research activities of AD and BTJ are also supported by NOAO, which is operated by the AURA under a cooperative agreement with the National Science Foundation.  MultiDrizzle is a component of the STSDAS and PyRAF software, products of the STScI, which is operated by AURA for NASA.  

This work relies on considerable efforts by the \emph{Spitzer}, \hst, and Keck Observatory support staffs and instrument teams, and we thank the many hundreds of people who have contributed to these facilities over the years.  We thank the anonymous referee for a number of useful suggestions, and Kyle Dawson for a helpful review of the manuscript.  This research has made use of NASA's Astrophysics Data System.  Support for M.B. was provided by the W.M. Keck Foundation.  The work by SAS at LLNL was performed under the auspices of the U. S. Department of Energy under Contract No. W-7405-ENG-48.


\end{document}